\documentclass[aps,prd,groupedaddress,preprintnumbers,
               floatfix, nofootinbib,longbibliography]{revtex4-1}
\usepackage[utf8]{inputenc}
\usepackage[pagebackref=false,colorlinks=true,citecolor=blue,filecolor=blue,linkcolor=blue,anchorcolor=blue,urlcolor=blue]{hyperref}
\usepackage{slashed}
\usepackage{natbib}
\usepackage{graphicx}
\usepackage{epsfig}
\usepackage{amsmath}
\input{epsf}
\usepackage{psfrag}
\usepackage{subcaption}
\usepackage{ulem}
\usepackage{gensymb}
\usepackage[percent]{overpic}

\usepackage[usenames,dvipsnames]{color}

\newcommand{\beq}{\begin{eqnarray}}
\newcommand{\eeq}{\end{eqnarray}}

\newcommand{\nn}{\nonumber \\}

\usepackage{amsmath,color}
\usepackage{amssymb}
\usepackage{bm}
\usepackage{graphicx}
\usepackage{multirow}

\usepackage{soul} 
\newcounter{RSQ}

\begin{document}

\title{Deeply virtual $\phi$-meson production near threshold  }

\author{Yoshitaka Hatta}
\affiliation{Physics Department, Brookhaven National Laboratory, Upton, NY 11973, USA}
\affiliation{RIKEN BNL Research Center, Brookhaven National Laboratory, Upton, NY 11973, USA}

\author{Henry T. Klest}
\affiliation{Physics Division, Argonne National Laboratory,
9700 S. Cass Ave., Lemont, IL, USA}

\author{Kornelija Passek-K.}
\affiliation{
Division of Theoretical Physics, Rudjer Bo\v{s}kovi\'{c} Institute, HR-10000 Zagreb, Croatia}

\author{Jakob Schoenleber}
\affiliation{RIKEN BNL Research Center, Brookhaven National Laboratory, Upton, NY 11973, USA}

\begin{abstract}

We discuss exclusive $\phi$-meson electroproduction off the proton near threshold  
within the GPD factorization framework. 
We propose the `threshold approximation' in which only the leading term of the 
conformal partial wave expansion of the meson production amplitudes
is kept in both the quark and gluon exchange channels. 
We test the validity of this approximation to next-to-leading order in QCD 
and demonstrate the strong sensitivity of the cross section to the gluon 
and strangeness gravitational form factors. 
We also perform realistic event generator simulations both for Jefferson Lab 
and EIC kinematics and demonstrate the capabilities of future facilities 
for measuring near-threshold $\phi$ electroproduction.     

\end{abstract}

\maketitle

\section{Introduction}

The exclusive production of the $\phi$-meson ($\phi(1020)$) in electron-proton scattering 
serves as a fascinating laboratory for studying the nucleon structure due to the unique status of the $\phi$ in QCD. In electroproduction mediated by a longitudinally polarized photon with high virtuality $Q^2\gg \Lambda_{\rm QCD}^2$, commonly called Deeply Virtual Meson Production (DVMP), a QCD factorization theorem \cite{Collins:1996fb} allows for the first-principle treatment of the process in terms of the Generalized Parton Distributions (GPDs) \cite{Goloskokov:2006hr,Cuic:2023mki}. In contrast to the production of lighter mesons, such as $\pi$, $\rho$, and $\omega$, $\phi$-production gives access to the gluon GPD even at low energy due to its flavor content $s\bar{s}$. Heavier quarkonium states  such as $J/\psi$ (a $c\bar{c}$ bound state) and $\Upsilon$ (a $b\bar{b}$ bound state) are also sensitive to the gluon GPD, but their production typically requires  additional theoretical tools  such as non-relativistic QCD (NRQCD). $\phi$-production can be described by the usual light-cone distribution amplitude (DA), enjoys a higher production rate than $J/\psi$-production, and can be well-identified in experiments  through the $KK$ decay channels. Measurements at different energies have been conducted at Cornell \cite{Dixon:1978vy,Cassel:1981sx}, LEPS \cite{LEPS:2005hax}, HERA \cite{H1:2000hps,H1:2009cml,ZEUS:2005bhf}, and Jefferson laboratory (JLab) \cite{CLAS:2001zwd,CLAS:2008cms,CLAS:2013jlg,Dey:2014tfa}.

One might argue that $\phi$-production is not ideally suited for probing the gluonic content of the proton  because the process can be  `contaminated' by the intrinsic strangeness content. To what extent the latter contributes depends on experimental parameters as well as theoretical interpretations and is presently not well understood. However, one can reverse the argument and convert this feature into an additional advantage. That is, if one has enough knowledge about the gluon GPD from other processes such as $J/\psi$ production, $\phi$-production offers a unique opportunity to access the  strangeness GPDs, which are otherwise very difficult to constrain.   

In this paper, we study $\phi$-meson electroproduction  {\it near  threshold} and at {\it  high photon virtualities} \cite{Hatta:2021can} in the GPD factorization framework.\footnote{Near-threshold $\phi$-meson photoproduction and electroproduction at low-$Q^2$ have been  theoretically studied in  various  nonperturbative frameworks \cite{Laget:2000gj,Ryu:2012tw,Strakovsky:2020uqs,Kou:2021bez,Wang:2022uch,Kim:2024lis}.}    Our goal is to provide a  benchmark calculation with next-to-leading order (NLO)  perturbative QCD ingredients that can be readily used for the ongoing and proposed experimental projects at JLab and, in the long run, to motivate measurements of this process at the future Electron-Ion Collider (EIC) \cite{AbdulKhalek:2021gbh} at Brookhaven National Laboratory.   
Our study is complementary to the recent theoretical \cite{Hatta:2018ina,Hatta:2019lxo,Mamo:2019mka,Wang:2019mza,Du:2020bqj,Boussarie:2020vmu,Mamo:2021krl,Guo:2021ibg,Kharzeev:2021qkd,Sun:2021gmi,Sun:2021pyw,JointPhysicsAnalysisCenter:2023qgg,Guo:2023qgu} and experimental \cite{GlueX:2019mkq,Duran:2022xag,GlueX:2023pev} efforts to understand the near-threshold photoproduction of $J/\psi$. A major driving force of these developments is the recognition that this process is highly sensitive to the proton's gravitational form factors (GFFs)  \cite{Kobzarev:1962wt,Pagels:1966zza,Ji:1996ek}
\beq
\langle p'|T_{q,g}^{\alpha\beta}|p\rangle &=& \bar{u}(p')\left[A_{q,g}(t)\gamma^{(\alpha}P^{\beta)}+B_{q,g}(t)\frac{P^{(\alpha}i\sigma^{\beta)\lambda}\Delta_\lambda}{2M} + D_{q,g}(t)\frac{\Delta^\alpha \Delta^\beta-g^{\alpha\beta}\Delta^2}{4M} + \bar{C}_{q,g} (t)M g^{\alpha\beta}\right]u(p),
\label{gff}
\eeq
where $T^{\alpha\beta}_{q,g}$ are  the quark and gluon parts of the QCD energy momentum tensor ($A^{(\alpha}B^{\beta)}\equiv \frac{A^\alpha B^\beta+A^\beta B^\alpha}{2}$), $M$ is the proton mass, $P=\frac{p+p'}{2}$, and $\Delta=p'-p$. The connection between near-threshold production and the gluon GFFs $A_g,~D_g$, and  $\bar{C}_g$  was originally pointed out using an approach based on holographic QCD at strong coupling \cite{Hatta:2018ina,Hatta:2019lxo} where the energy momentum tensor plays a special role.
Remarkably, the same is true also in GPD-based approaches at weak coupling \cite{Hatta:2021can,Guo:2021ibg} for a quite different reason.  The first attempt to extract the gluon GFFs $A_g(t)$ and $D_g(t)$ has been recently made in \cite{Duran:2022xag} by comparing the latest $J/\psi$ data at JLab with holographic QCD \cite{Mamo:2021krl} and GPD \cite{Guo:2021ibg}   calculations.  However, the result is not without controversy, as the extraction relies on a single observable rather than a combination of multiple observables as is  standard in QCD global analyses. Moreover, there are large theoretical uncertainties in the leading order (LO) perturbative QCD and  holographic QCD  calculations. As a matter of fact, the experimental data can be explained by other theoretical approaches  without any reference to GFFs  \cite{Du:2020bqj,Sun:2021gmi,Sun:2021pyw,JointPhysicsAnalysisCenter:2023qgg}.  

In order to make progress,  one needs more observables with equally good or better  sensitivities to the GFFs. There are a few possibilities. Instead of $J/\psi$, one can consider $\Upsilon$-production \cite{Hatta:2019lxo} where the large $\Upsilon$ mass (9.46 GeV) better justifies the use of peturbative QCD approaches. Alternatively, near-threshold  electroproduction of $J/\psi$  \cite{Boussarie:2020vmu} may be an option, since the photon virtuality $Q^2$ can be adjusted well into the perturbative QCD domain. However, both of these alternatives, although worth pursuing, suffer from low production rates that pose  experimental challenges even with the high luminosities at JLab and the EIC. Furthermore, $\Upsilon$ cannot be produced on proton targets at the current JLab energy.

Under these circumstances,  $\phi$-meson electroproduction near threshold \cite{Hatta:2021can} emerges as a viable and attractive possibility. Due to the relatively low mass of the $\phi$-meson,  the cross section is measurably large even for perturbative values of $Q^2$. 
Compared to $J/\psi$ photoproduction where the hard scale ($J/\psi$ mass) is fixed, a variable hard scale $Q$ provides an additional experimental parameter that can be adjusted for certain purposes.   As we shall demonstrate, the cross section is strongly sensitive to the gluon and strangeness GFFs, in particular the {\it strangeness D-term} $D_s(t=0)$ which was the main focus  of \cite{Hatta:2021can}.   The present work, albeit restricted to the longitudinal cross section, provides a proper QCD-based prediction that should supersede the heuristic discussion in \cite{Hatta:2021can}. 

In addition to enhancing the theoretical framework, we will conduct a feasibility study for experimental measurements at future facilities by integrating the updated theoretical predictions into the \texttt{lAger}~\cite{joosten2021} event generator. The EIC offers the necessary flexibility and luminosity to perform the first measurements of near-threshold meson production at an $e+p$ collider. At Jefferson Lab, the Solenoidal Large Intensity Device (SoLID) provides an excellent platform for utilizing near-threshold $\phi$ and $J/\psi$ electroproduction to study gravitational form factors due to its large acceptance and high instantaneous luminosity.

\section{Kinematics}

\begin{figure}[t]
    \centering

    \begin{minipage}{0.4\textwidth}
        \begin{overpic}[width=\textwidth]{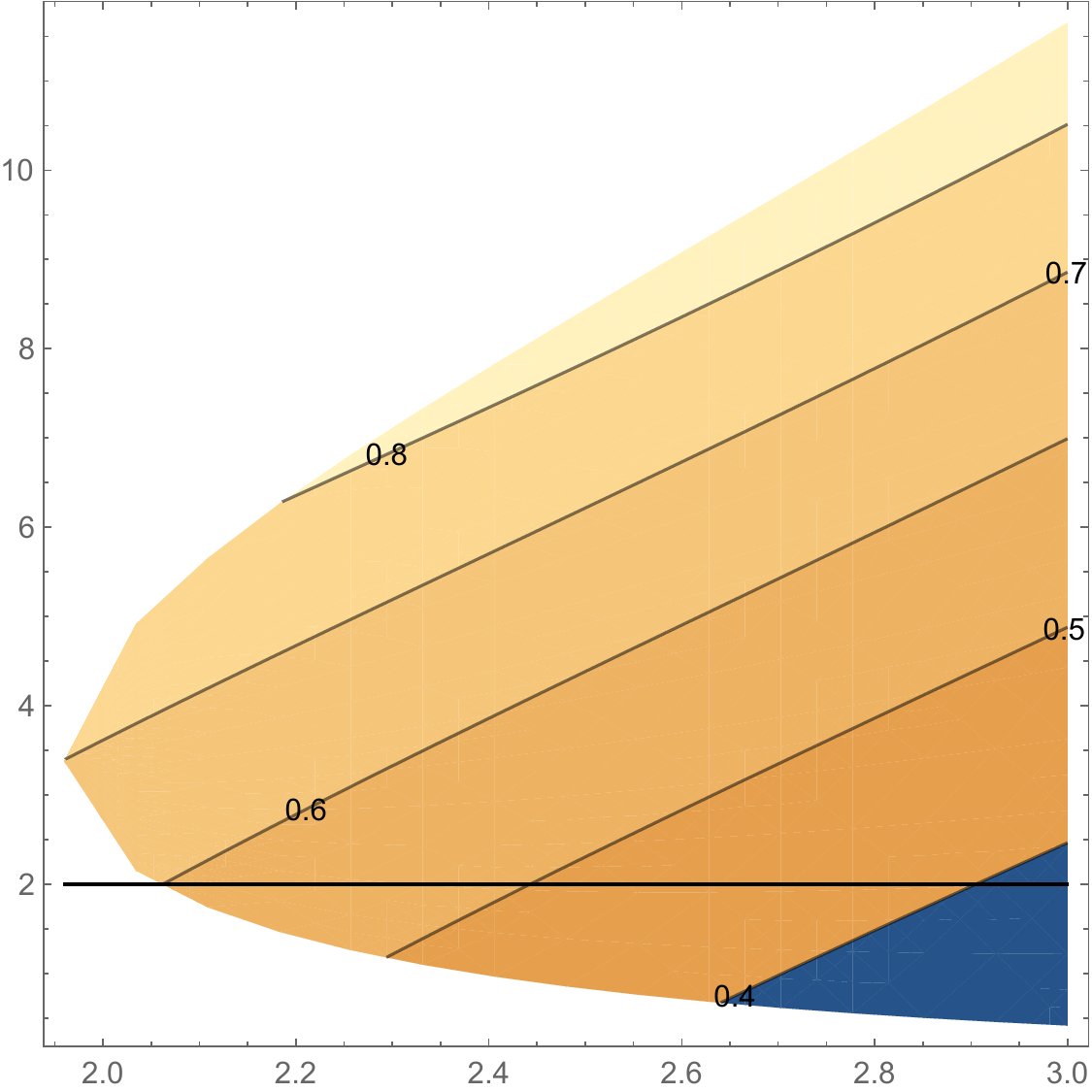}
            \put(-6,50){\rotatebox{90}{\small $|t|\ (\mathrm{GeV}^2)$}}
            
            \put(10,85){{\color{Gray}\raisebox{0.5ex}{\rule{2em}{1pt}}}  Isolines of constant $\xi$}
            
            \put(45,-5){\small $W\ (\mathrm{GeV})$}
        \end{overpic}
    \end{minipage}
    \hspace{0.05\textwidth} 
    \begin{minipage}{0.4\textwidth}
        \begin{overpic}[width=\textwidth]{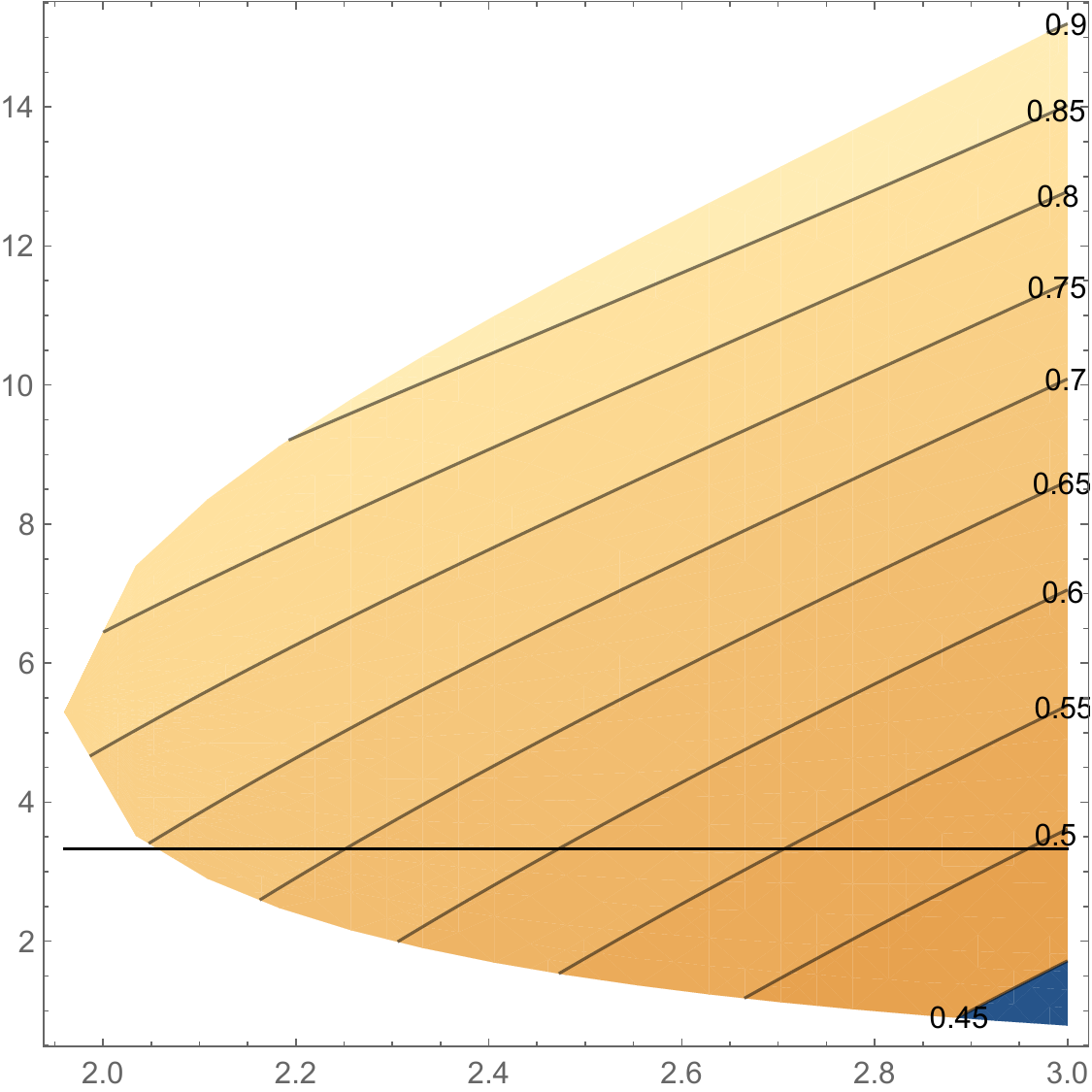}
            \put(-6,50){\rotatebox{90}{\small $|t|\ (\mathrm{GeV}^2)$}}
            \put(10,85){{\color{Gray}\raisebox{0.5ex}{\rule{2em}{1pt}}}  Isolines of constant $\xi$}
            \put(45,-5){\small $W\ (\mathrm{GeV})$}
        \end{overpic}
    \end{minipage}

    \vspace{0.02\textwidth}

    \caption[*]{Contour plots of $\xi$ in the $(W,|t|)$ plane at $Q^2=6$ GeV$^2$ (left) 
    and $Q^2=10$ GeV$^2$ (right). The horizontal line is at $|t|=\frac{Q^2}{3}$.}
    \label{2}
\end{figure}
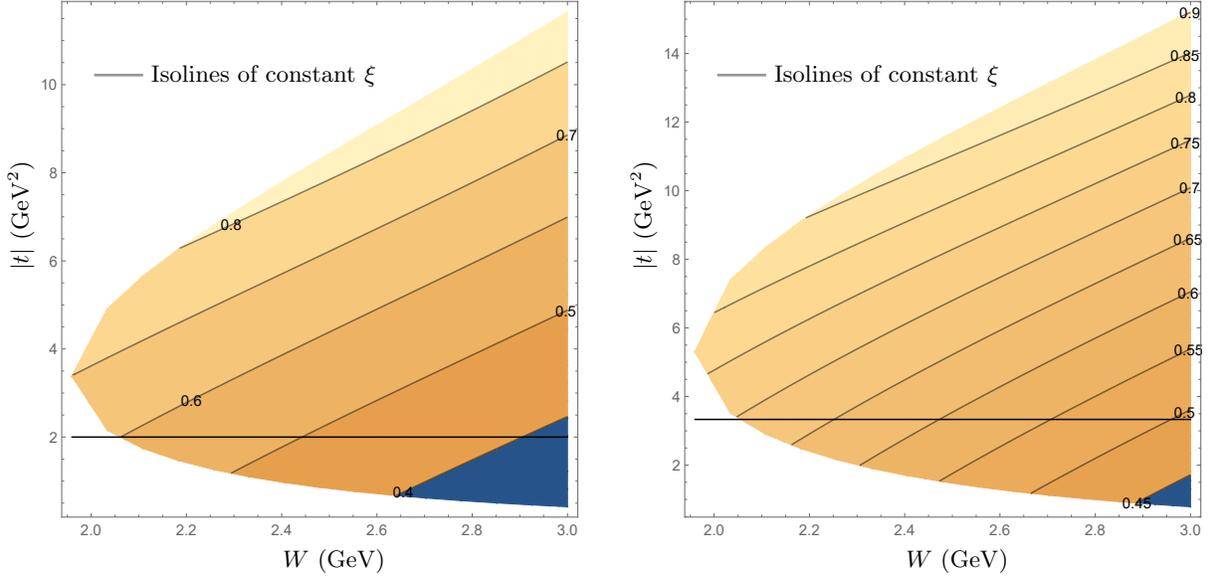

The process of interest is the exclusive production of the $\phi$-meson in unpolarized electron-proton scattering $ep\to e'\gamma^*p \to e'p'\phi$. At fixed $ep$ center-of-mass energy $s_{ep}=(l+p)^2$, the cross section integrated over  azimuthal angles is characterized by the photon virtuality $q^2=-Q^2$, the Bjorken variable $x_B=\frac{Q^2}{2p\cdot q}$ and the proton momentum transfer $t=(p'-p)^2$. The standard inelasticity parameter $y$ in deeply inelastic scattering (DIS)
can be written in terms of these variables:   
\beq
y=\frac{q\cdot p}{l\cdot p} = \frac{Q^2}{s_{ep}x_B}.
\eeq
For the present purpose, it is convenient to introduce the $\gamma^*p$ center-of mass-energy 
\beq 
W^2=(p+q)^2=M^2-Q^2+2p\cdot q=M^2+Q^2\frac{1-x_B}{x_B}, \label{w}
\eeq
to be used interchangeably with $x_B$ and work in the 
 $\gamma^*p$ center-of-mass frame. In this frame,  one can parameterize the momenta in the reaction $\gamma^*(q)+p\to \phi(k)+p'$ as  
\beq
&&p^\mu =(E_{cm},0,0,p_{cm})=\left(\frac{W^2+Q^2+M^2}{2W},0,0,p_{cm}\right),\nn
&& q^\mu = (\sqrt{p_{cm}^2-Q^2},0,0,-p_{cm})= \left(\frac{W^2-M^2-Q^2}{2W},0,0,-p_{cm}\right), \nn
&& p'^\mu = (E'_{cm},\vec{k}_{cm}) = \left(\frac{W^2-M_\phi^2+M^2}{2W},\vec{k}_{cm}\right), \nn
&& k^\mu =(\sqrt{M_\phi^2+k_{cm}^2},-\vec{k}_{cm})=\left(\frac{W^2+M_\phi^2-M^2}{2W},-\vec{k}_{cm}\right), \label{com}
\eeq
where $E_{cm}=\sqrt{M^2+p_{cm}^2}$, $E'_{cm}=\sqrt{M^2+k_{cm}^2}$, and 
\beq
p_{cm}^2&=&\frac{W^4-2W^2(M^2-Q^2)+(M^2+Q^2)^2}{4W^2}, \nn
k_{cm}^2&=& \frac{(W^2-(M_\phi+M)^2)(W^2-(M_\phi-M)^2)}{4W^2}.
\eeq
For a fixed value of $W$, the momentum transfer 
\beq
-t&=&-\left(\sqrt{p_{cm}^2+M^2}-\sqrt{k_{cm}^2+M^2}\right)^2+(\vec{p}_{cm}-\vec{k}_{cm})^2 \nn 
&=& -\left(\frac{Q^2+M_{\phi}^2}{2W}\right)^2+p_{cm}^2+k_{cm}^2-2p_{cm}k_{cm}\cos\theta ,
\eeq
varies in the window 
\beq
-t_{min}<-t<-t_{max}, \label{tminmax}
\eeq
where 
\beq
 -t_{min}  =-\left(\frac{Q^2+M_{\phi}^2}{2W}\right)^2+(p_{cm}-k_{cm})^2,\nn
  -t_{max}  =-\left(\frac{Q^2+M_{\phi}^2}{2W}\right)^2+(p_{cm}+k_{cm})^2.
\eeq
The range (\ref{tminmax}) is plotted in Fig.~\ref{2} for $Q^2=6$ GeV$^2$ (left) and $Q^2=10$ GeV$^2$ (right)  above the threshold energy
\beq
W\ge  W_{th}\equiv M_\phi+M=1.02\, {\rm GeV}+0.94\, {\rm GeV}=1.96 \, {\rm GeV}.
\eeq
Fig.~\ref{2} also shows contours  
 of the skewness variable ($p^+\equiv \frac{1}{\sqrt{2}}(p^0+p^3)$)
\beq
\xi=\frac{p^+-p'^+}{p^++p'^+} &=& \frac{\sqrt{M^2+p_{cm}^2}+p_{cm}-\sqrt{M^2+k_{cm}^2}-k_{cm}\cos\theta}{\sqrt{M^2+p_{cm}^2}+p_{cm}+\sqrt{M^2+k_{cm}^2}+k_{cm}\cos\theta} \nn 
 &=& \frac{2(p_{cm}+E_{cm})(E_{cm}-E'_{cm})-t}{2(p_{cm}+E_{cm})(E_{cm}+E'_{cm})-4M^2+t},
\eeq
which will be an important parameter in this work. There exist various definitions of the skewness variable, but these differences vanish in the limit $Q\rightarrow \infty$. 
Our definition is tied to the $\gamma^* p$ center-of-mass frame. With this choice, $t_{min}$ is related to $\xi$ as 
\beq
t_{min} = -\frac{4\xi^2M^2}{1-\xi^2}, \label{tminxi}
\eeq
which represents the intersection point between a curve of constant $\xi$ and the $t_{min}$ curve in Fig.~\ref{2}. 

The differential cross section takes the form  
\beq
\frac{d\sigma}{dWdQ^2dt} &=& \frac{\alpha_{em}}{4\pi}\frac{W(W^2-M^2)}{(p\cdot l)^2Q^2(1-\varepsilon)}  \frac{d\sigma^{\gamma^*p}}{dt},
\eeq
where 
\beq
\varepsilon=\frac{1-y-\frac{y^2\gamma^2}{4}}{1-y+\frac{y^2}{2}+\frac{y^2\gamma^2}{4}}
\eeq
 is the ratio of the longitudinal and transverse photon fluxes (see below) and $\gamma=\frac{2x_BM}{Q}$ is a parameter characterizing target mass corrections  which is actually not negligible in the present kinematics. 
$\frac{d\sigma^{\gamma^*p}}{dt}$ is the differential cross section in the $\gamma^*p$ subsystem. It consists of the transverse and longitudinal parts
\beq 
\frac{d\sigma^{\gamma^*p}}{dt} &=& \frac{\alpha_{em}}{8(W^2-M^2)Wp_{cm}} 
\left(\frac{1}{2}g_\perp^{\mu\nu}+\varepsilon \epsilon_L^\mu \epsilon_L^\nu\right)  \frac{1}{2}\sum_{spin}\langle p|J^{em}_\mu(-q)|p'\phi\rangle\langle p'\phi|J^{em}_\nu(q)|p\rangle \nn 
&=& \frac{d\sigma_T}{dt}+\varepsilon \frac{d\sigma_L}{dt}, \label{dsigmadt}
\eeq
where $g_\perp^{ij}=-\delta^{ij}$ ($i,j=1,2$) and $\epsilon_L^\mu$ is the longitudinal photon polarization vector. 
The following formula  may be useful to connect seemingly different  expressions in the literature: 
\beq
\frac{x_B^2}{Q^4(1-x_B)\sqrt{1+\gamma^2}} = \frac{1}{2(W^2-M^2)Wp_{cm}}.
\eeq 
Experimentally, it is straightforward to measure the differential (\ref{dsigmadt}) and integrated cross sections 
\beq
\sigma^{\gamma^*p} = \int^{t_{max}}_{t_{min}}dt \frac{d\sigma^{\gamma^*p}}{dt}  =\sigma_T(W,Q)+\varepsilon \sigma_L(W,Q) = \left(\frac{1}{R(W,Q)}+\varepsilon\right)\sigma_L(W,Q), \label{totalcross}
\eeq
where the ratio 
\beq
R(W,Q)= \frac{\sigma_L}{\sigma_T} 
\eeq
is commonly introduced in the literature. 
In principle, it is possible to experimentally measure $R(W,Q)$ by monitoring the $\varepsilon$-dependence (equivalently, the energy $s_{ep}$ dependence) or measuring the decay products of $\phi$. We shall comment more on this in a later section.

At HERA, exclusive $\phi$-production has been measured at high energy, 
up to $W\sim 130$ GeV and $Q^2\lesssim 40$ GeV$^2$ \cite{H1:2000hps,ZEUS:2005bhf,H1:2009cml}, 
primarily motivated by the physics of diffraction at $x_B\ll 1$. 
Such vector meson electroproduction data have been utilized for modeling GPDs 
at LO \cite{Goloskokov:2006hr, Meskauskas:2011aa} and, more recently, 
as inputs for global analyses of GPDs at NLO in deeply virtual Compton scattering (DVCS)  
and DVMP \cite{Lautenschlager:2013uya,Cuic:2023mki}.
Turning to the low energy side, the LEPS collaboration \cite{LEPS:2005hax} has measured the photoproduction ($Q=0$) cross section near the threshold  $W_{th}\le W\le 2.31$ GeV ($1.57\le E_\gamma \le 2.37\, {\rm GeV}$ in terms of the photon energy in the target (proton) rest frame). 
Subsequently, the CLAS collaboration \cite{CLAS:2008cms} has measured the electroproduction cross section in the range $W_{th}<W<3$ GeV and $1.4<Q^2<3.8$ GeV$^2$.  Currently, there are several experimental proposals at JLab to measure electroproduction near the threshold at higher $Q^2$, and the effort can be continued in the future with 
the EIC.

\section{GPD factorization}
\label{Sec:GPD}

\subsection{ General consideration}

The goal of this paper is to provide   predictions for the longitudinal differential $\frac{d\sigma_L}{dt}$ and total $\sigma_{L}$  cross sections  near the threshold and at high values of $Q^2$. By near-threshold, we mean the energy region explored by the CLAS collaboration
\beq
W_{th}<W<3\, {\rm GeV}. \label{wind}
\eeq
Practical reasons for this choice will be explained later. 
On the other hand, since our approach will be based on 
the  QCD factorization framework treating $Q^2$ as a hard scale, we consider $Q^2$ values larger than those explored previously by the CLAS collaboration. Roughly, we have in mind $Q\gtrsim 2$ GeV. The exact value of $Q$ above which the perturbative description becomes valid cannot be sharply determined, but we will make a phenomenological estimate in the numerical section.    

Let us first discuss practical and conceptual issues with the use of QCD factorization in the threshold region.  
First, as is clear from (\ref{w}), when $Q^2$ is large the condition $W\gtrsim W_{th}$ is realized by a large cancellation between $Q^2$ and $2p\cdot q$. This means that the Bjorken variable is close to unity
\beq
 x_B =\frac{Q^2}{2p\cdot q}\approx 1,
\eeq
so the skewness variable is also close to unity
\beq
\xi \approx \frac{x_B}{2-x_B}  \approx 1. 
\eeq
As a matter of fact, GPDs at large $\xi \sim {\cal O}(1)$ are poorly constrained. 
Existing global 
analyses heavily rely on HERA data which were mostly 
taken at $x_B,\xi\ll 1$ 
(see \cite{Kumericki:2016ehc,Favart:2015umi,Cuic:2023mki} and references therein). 
Therefore, GPD models constrained at $\xi \ll 1$ need to be extrapolated to the region $\xi \sim {\cal O}(1)$, and there is no reliable way to do so.  
Moreover, when $\xi\ll 1$ where $t$ is also small, the helicity-flip $E$-type GPDs 
are often neglected since they typically enter cross sections with prefactors 
$\xi^2$  and $t/M^2$ (see for example (\ref{cross}) below). 
Near the threshold where $\xi \sim {\cal O}(1)$, their contributions are no longer negligible,  
but they are less well known compared to the $H$-type GPDs \cite{Kroll:2012sm}. 
Remarkably, however, in this regime, DVMP  scattering amplitudes are  largely insensitive to the $x$-dependence of GPDs. Instead they are dominated by the    gravitational form factors (\ref{gff}) which are much easier to parameterize and evolve. This is actually the main point of this paper, and we will explain it in detail in the next subsections. 

Another issue is that the momentum transfer $|t|$ tends to be large in the threshold region at high-$Q^2$, often much larger than $\Lambda_{\rm QCD}^2$, see Fig.~\ref{2}. At the threshold, 
\beq
|t_{th}|= \frac{M(M_\phi^2 + Q^2)}{M+M_\phi} \approx \frac{Q^2}{2}.
\eeq 
Above the threshold, $|t|$ varies in the window (\ref{tminmax}). It is easy to check that, in the kinematical region we consider ($Q>2$ GeV and $W<3$ GeV),  $|t_{min}|$ grows linearly with  $Q^2$. 
This is worrisome because higher twist corrections in GPD factorization formulas are  of order $t/Q^2$. We can alleviate this problem by avoiding regions too close to the threshold and staying around $t\approx t_{min}$, see Fig.~\ref{2}. However, experiments can in principle measure $\frac{d\sigma}{dt}$ in the entire range of $t$, and  larger $t$-values are interesting since $\xi$ is also larger. As a compromise, we consider 
\beq
|t_{min}|<|t| \lesssim \frac{Q^2}{3\, {\rm or}\, 4}  \label{upper}
\eeq  
to be the limit of applicability for our calculation. 
This limit is indicated by a horizontal line in Fig.~\ref{2}. Above this line, higher twist corrections could be very large.

Finally, near the threshold, the relative velocity between the produced $\phi$ and the recoiling proton is small and can be nonrelativistic (zero at the threshold by definition). This means that there could be final state interactions (FSIs) which potentially break factorization \cite{Hatta:2021can}. 
From a perturbative QCD perspective, 
the standard collinear factorization theorem \cite{Collins:1996fb} assumes that the outgoing proton  is collinear to the incoming proton 
\begin{align}
p'^+ \gg p_{\perp}' \gg  p'^-.
\label{collinear condition}
\end{align} 
This condition becomes increasingly difficult to satisfy as $\xi \to 1$ since $p'^+\propto 1-\xi$. 
Again one can circumvent this problem by avoiding the region too close to the threshold (see also a  discussion in the $J/\psi$ case in \cite{Guo:2021ibg}). In the $\gamma^*p$ center-of-mass frame (\ref{com}), the relative momentum in the final state is 
\beq
(2k_{cm})^2 \approx W^2-W^2_{th} \approx  \left(\frac{1}{x_B}-1\right)Q^2 \approx \frac{1-\xi}{2\xi}Q^2 \equiv \mu_{th}^2.
\eeq
The requirement of a large relative velocity can be expressed in a Lorentz-invariant manner by the condition that  the `threshold scale' $\mu_{th}^2$ should be much larger than the small scales 
\beq
\frac{1-\xi}{2\xi}Q^2 \gg |t|,M^2. \label{scale}
\eeq
As soon as $\xi \gtrsim 1/3$, this becomes a  stronger constraint than $Q^2\gg |t|$ and further shrinks the region of applicability toward $t\sim  t_{min}$ and $\xi \lesssim 0.6$. 
However, whether this effect is numerically important in practice is largely unknown. Ultimately, what matters is the interaction strength between the produced meson and the recoiling proton, which depends on the meson species.
In this regard, we refer to a recent dedicated study \cite{Kim:2021adl,Kim:2024lis} where it has been concluded that, near the threshold, the $\phi+p$ FSI cross section is orders of magnitude smaller than the production cross section. While Ref.~\cite{Kim:2024lis} concerns photoproduction, we expect the same conclusion holds in electroproduction since the final state is the same. We therefore neglect final-state interactions and stick to  (\ref{upper}), although more investigations are certainly desirable.  \\

\subsection{Conformal partial wave expansion}

For light vector meson production, QCD factorization has been proven only for the longitudinally polarized virtual photon \cite{Collins:1996fb}. We thus focus on the longitudinal cross section $\sigma_L$ in this section. The transverse cross section $\sigma_T$ will be phenomenologically included in a later section. Up to higher twist corrections of order $t/Q^2$, the differential cross section in (\ref{dsigmadt}) can be written as 
\beq
\frac{d\sigma_L}{dt}
= \frac{2\pi^2\alpha_{em}}{(W^2-M^2)Wp_{cm}}\left((1-\xi^2)|{\cal H}|^2 -\left(\frac{t}{4M^2}+\xi^2\right)|{\cal E}|^2-2\xi^2{\rm Re}({\cal H}{\cal E}^*)\right), \label{cross}
\eeq
where ${\cal H}$ and ${\cal E}$ are the
DVMP amplitudes which contain the GPDs $H^{q,g}$ and $E^{q,g}$, respectively.  $q$ is the label for quark flavors. Throughout this paper, we consider four active flavors $n_f=4$, or  $q=u,d,s,c$.  ${\cal H}$ and ${\cal E}$ have the following  factorized structure  
\beq
\begin{pmatrix} {\cal H}(\xi,t,Q^2) \\ {\cal E}(\xi,t,Q^2)\end{pmatrix} = e_s\frac{C_Ff_\phi}{N_cQ}\sum_{a=q,g} \begin{pmatrix} H^a(x,\xi,t,\mu^2)  \\ E^a(x,\xi,t,\mu^2)\end{pmatrix} \otimes T^a\left(x,\xi,u,\frac{Q^2}{\mu^2}\right) \otimes \varphi(u,\mu^2), \label{fact}
\eeq
where $e_s=-\frac{1}{3}$ is the electric charge (in units of $|e|$) of the $s$-quark The notations $N_c=3$, $C_F=\frac{N_c^2-1}{2N_c}=\frac{4}{3}$ are standard.  $\varphi(u)$ and $f_\phi$ are the $\phi$-meson distribution amplitude (DA) and the decay constant, respectively. $T^a$ are the hard scattering amplitudes,  also referred to as coefficient functions. The symbols $\otimes$ represent convolutions in $-1<x<1$ and $0<u<1$.  
For simplicity, we have chosen the factorization scales of GPD and DA to be 
the same and equal to the renormalization scale $\mu^2$.

A particularly powerful framework to analyze the amplitude (\ref{fact}) is 
the conformal partial wave (CPaW) formalism developed in \cite{Mueller:2005ed} 
and systematized to NLO  
in \cite{Kumericki:2007sa} for DVCS and in \cite{Muller:2013jur,Duplancic:2016bge} for DVMP.  
It 
combines the conformal partial wave expansion in terms of the Gegenbauer polynomials with 
the Mellin-Barnes  integral technique.  
Compared to the traditional momentum fraction  
representation, the CPaW formalism simplifies GPD evolution to NLO and beyond,  
enables innovative GPD modeling, 
and facilitates efficient code  development for handling GPD evolution 
and data fitting.
While the formalism is applicable to all kinematical regions, so far 
it has mostly been used in the small-$x_B$ region due to data availability  
(\cite{Cuic:2023mki} and references therein). 
We argue here that, 
when applied to the threshold region of DVMP (large-$x_B$ and large-$\xi$), 
the CPaW formalism can be reduced to  a very simple effective model which captures 
the physics of the QCD energy momentum tensor and GFFs 
\cite{Boussarie:2020vmu,Hatta:2021can,Guo:2021ibg,Guo:2023qgu},  
while retaining the connection to QCD factorization.

We first introduce the conformal moments for  the meson DA
\beq
\varphi_k(\mu^2)= \frac{2(2k+3)}{3(k+1)(k+2)} \int_0^1 du C_k^{3/2}(2u-1)\varphi(u,\mu^2), \qquad (k:{\rm even})
\eeq
where $C_k^{3/2}$ is the Gegenbauer polynomial. The inverse transform is 
\beq
\varphi(u)= 6u(1-u)\sum_{k=0}^{\rm even} \varphi_k C_k^{3/2}(2u-1).
\eeq
As for the quark and gluon GPDs, we define 
\beq
H^q_j(\xi,t)=\frac{\Gamma(3/2)\Gamma(j+1)}{2^j\Gamma (j+3/2)} \frac{1}{2}\int_{-1}^1dx \xi^j C_j^{3/2}(x/\xi) H^{q(+)}(x,\xi,t), \qquad (j:{\rm odd}) \label{cpw}
\eeq
\beq
H^g_j(\xi,t)=\frac{\Gamma(3/2)\Gamma(j+1)}{2^j\Gamma (j+3/2)} \frac{1}{2}\int_{-1}^1dx \frac{3}{j}\xi^{j-1} C_{j-1}^{5/2}(x/\xi) H^g(x,\xi,t), \qquad (j:{\rm odd}) \label{cpw2}
\eeq
and similarly for $E_j^{q,g}(\xi,t)$. The notation  $H^{q(+)}(x,\xi,t)=H^q(x,\xi,t)-H^q(-x,\xi,t)$ denotes the C-even part. 
The gluon GPD is normalized such that, in the forward limit, $H^g(x,0,0)=xG(x)$, where $G(x)$ is the unpolarized gluon PDF. 
These moments reduce to the standard Mellin moments of PDFs in the $\xi\to 0$ limit.\footnote { Note that the Gegenbauer polyanomials in (\ref{cpw}), (\ref{cpw2}) are defined for any $-\infty < x/\xi < \infty$  which is outside of the Efremov-Radyushkin-Brodsky-Lepage (ERBL) region $-1 < x/\xi < 1$ where they form a set of orthogonal polynomials. Compare with (\ref{Hq exp}) where the latter condition is enforced by  $\theta(\xi -|x|)$.  } The first moments $j=1$ are the quark and gluon GFFs (\ref{gff}) 
 \beq
&& H_1^{q,g}(\xi,t,\mu^2)= A_{q,g}(t,\mu^2)+\xi^2 D_{q,g}(t,\mu^2),  \nn 
&& E_1^{q,g}(\xi,t,\mu^2)= B_{q,g}(t,\mu^2)-\xi^2 D_{q,g}(t,\mu^2). \label{gff2} 
\eeq 
In the moment space, the amplitudes (\ref{fact}) take the form 
\beq
\begin{pmatrix} {\cal H}(\xi,t,Q^2) \\ {\cal E}(\xi,t,Q^2)\end{pmatrix} =\kappa  \sum_{j=1}^{\rm odd}\sum_{k=0}^{\rm even}\sum_a \frac{2}{\xi^{j+1}} \begin{pmatrix} H_j^a(\xi,t,\mu^2) \\ E_j^a(\xi,t,\mu^2)\end{pmatrix} T^a_{jk}(Q^2/\mu^2)\varphi_k(\mu^2), \qquad  \kappa\equiv e_s\frac{C_Ff_\phi}{N_cQ} . \label{me} 
\eeq
Throughout this paper, we keep only the $k=0$ term for which $\varphi_0=1$. This is tantamount to assuming the asymptotic $\mu^2\to \infty$ formula for the DA $\varphi(u)\approx 6u(1-u)$. While the validity of this approximation is not fully understood, it is usually considered permissible and widely used in practice. If necessary, higher-$k$ terms \cite{Hua:2020gnw,Hu:2024tmc} can be included.

\subsection{Threshold approximation}

We now come to the crucial step of this work. We implement the `threshold approximation' by keeping only the $j=1$  term in the sum (\ref{me})  
\beq
\begin{pmatrix} {\cal H}(\xi,t,Q^2) \\ {\cal E}(\xi,t,Q^2)\end{pmatrix}  \approx  \frac{2\kappa}{\xi^2} \sum_a \begin{pmatrix} H_1^a(\xi,t,\mu^2) \\ E_1^a(\xi,t,\mu^2) \end{pmatrix} T^a_{10}(Q^2/\mu^2). \label{j=1}
\eeq
In contrast to the $k=0$ truncation, the $j=1$ truncation (\ref{j=1}) 
is unusual  and certainly does not hold in general.   
DVMP amplitudes 
are complex, and the imaginary part can only be recovered by performing the analytic continuation  of the infinite sum (\ref{me}) in  the Mellin-Barnes representation \cite{Mueller:2005ed,Muller:2014wxa}.  By keeping  only the $j=1$ moment, one neglects the imaginary part from the beginning. However, scattering amplitudes near  threshold are indeed  dominantly real since there is no extra particle  production by definition. In fact, (\ref{j=1}) is known to be a good approximation when $\xi\approx 1$  in the gluon sector $a=g$ at least to leading order (LO) in perturbation theory. In other words, 
DVMP amplitudes 
in the gluon channel can be characterized by the gluon GFFs (\ref{gff2}). 
This phenomenon has been first pointed out in  \cite{Hatta:2021can,Guo:2021ibg} in the context of near-threshold $J/\psi$ production using the Mellin moment expansion of the scattering amplitude.  Subsequently, it has been noticed in   \cite{Guo:2023qgu} that the approximation is even better if one uses  the conformal moment  expansion.

On the other hand, in the quark channel $a=q$  relevant to the production of light mesons (including $\phi$),  keeping only the first term of the Mellin moment expansion is not as good of an approximation as in the gluon channel   
\cite{Hatta:2021can}. In the next subsection we demonstrate that, if one switches to the conformal moment expansion, the threshold approximation (\ref{j=1}) is actually quite decent even in the quark sector. 
Furthermore, for the first time, we test the validity of the threshold approximation to next-to-leading order (NLO)  in  both the quark and gluon channels. 

For this purpose, let us first write down (\ref{j=1}) explicitly. 
The NLO  hard coefficients $T_{jk}$ for DVMP are available in the literature   
\cite{Muller:2013jur,Duplancic:2016bge,Cuic:2023mki}. 
Keeping only the $T_{10}$ component and adjusting the flavor content to $\phi$-production, we find 
\beq
{\cal H}(\xi,t,Q^2)&\approx &\frac{2\kappa}{\xi^2} \frac{15}{2}\Biggl[\left\{\alpha_s(\mu)+\frac{\alpha_s^2(\mu)}{2\pi}\left(25.7309-2n_f +\left(-\frac{131}{18}+\frac{n_f}{3}\right)\ln \frac{Q^2}{\mu^2}\right)\right\}(A_s(t,\mu)+\xi^2D_s(t,\mu)) \label{hxi} \\  
&& \qquad +\frac{\alpha_s^2}{2\pi}\left(-2.3889+\frac{2}{3}\ln \frac{Q^2}{\mu^2}\right) \sum_q (A_{q}+\xi^2 D_{q})  + \frac{3}{8}\left\{\alpha_s+\frac{\alpha_s^2}{2\pi}\left(13.8682-\frac{83}{18}\ln \frac{Q^2}{\mu^2}\right)\right\} (A_g+\xi^2D_g) \Biggr],\nonumber 
\eeq
\beq
{\cal E}(\xi,t,Q^2)&\approx&\frac{2\kappa}{\xi^2} \frac{15}{2}\Biggl[\left\{\alpha_s(\mu)+\frac{\alpha_s^2(\mu)}{2\pi}\left(25.7309-2n_f+\left(-\frac{131}{18}+\frac{n_f}{3}\right)\ln \frac{Q^2}{\mu^2}\right)\right\}(B_s(t,\mu)-\xi^2 D_s(t,\mu)) \label{exi}\\
&& \qquad +\frac{\alpha_s^2}{2\pi}\left(-2.3889+\frac{2}{3}\ln \frac{Q^2}{\mu^2}\right)  \sum_q(B_q-\xi^2 D_{q})  + \frac{3}{8}\left\{\alpha_s+\frac{\alpha_s^2}{2\pi}\left(13.8682-\frac{83}{18}\ln \frac{Q^2}{\mu^2}\right)\right\} (B_g-\xi^2D_g) \Biggr] .\nonumber
\eeq 
\begin{figure}[t]
    \centering
    \includegraphics[width=0.3\linewidth]{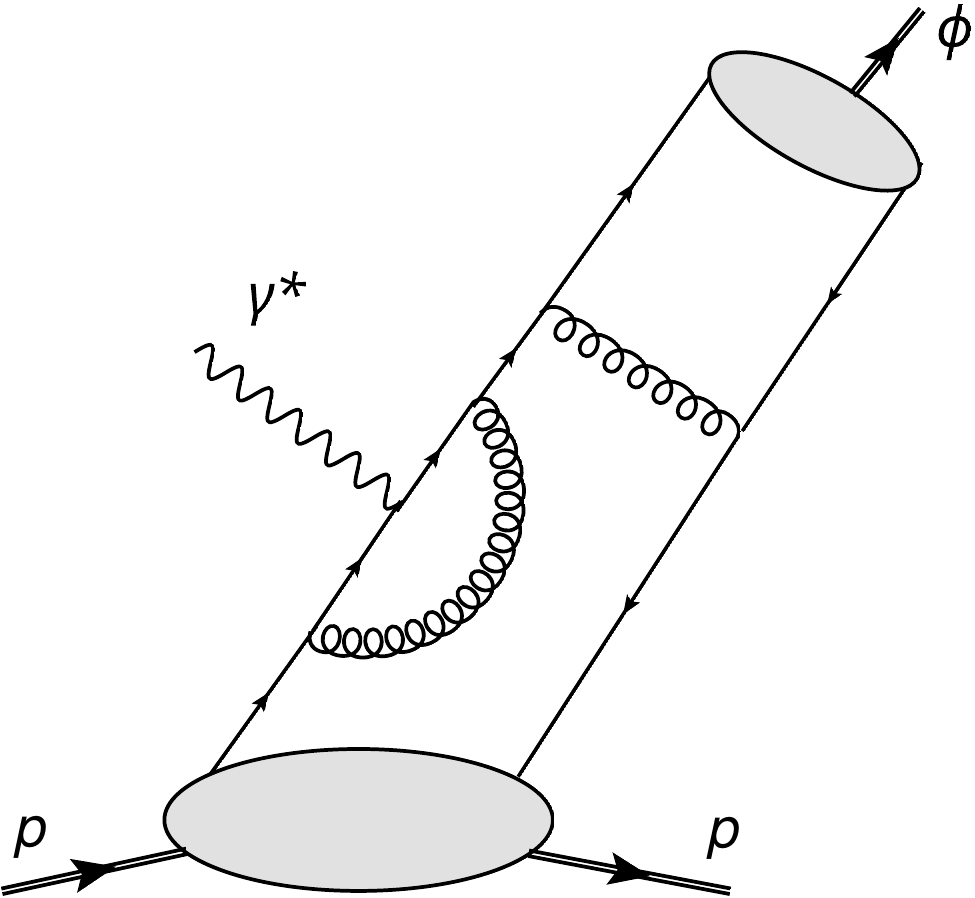}
    \includegraphics[width=0.3\linewidth]{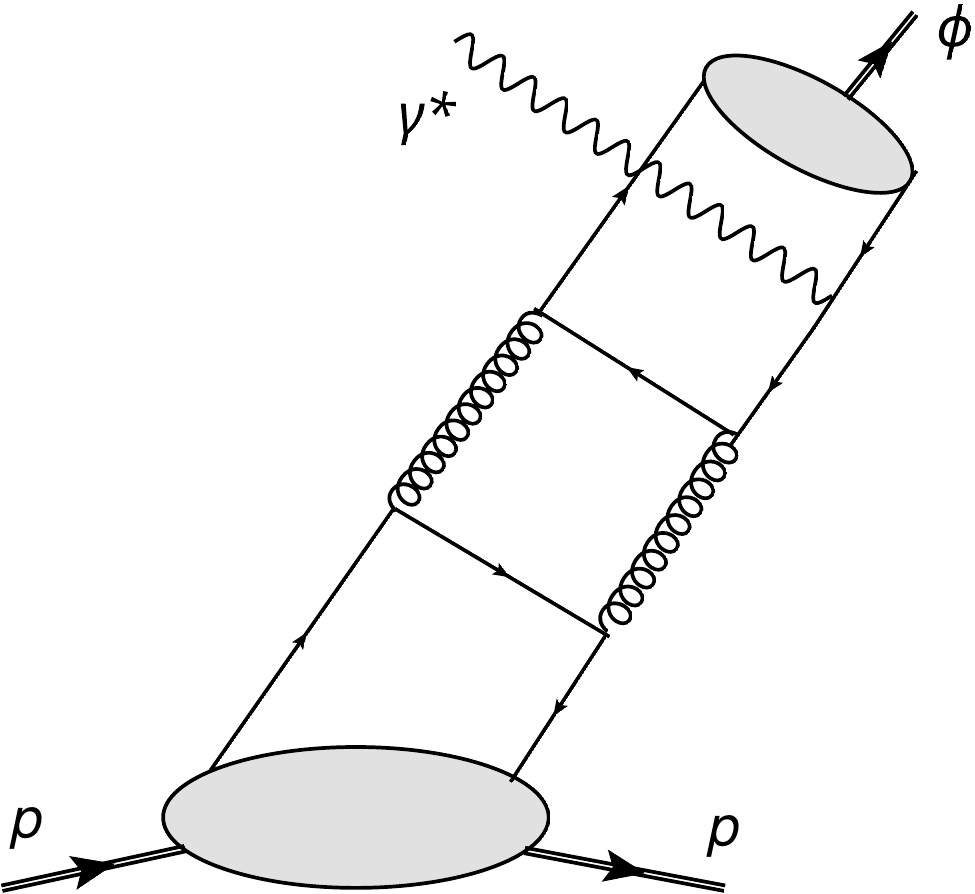}
    \includegraphics[width=0.3\linewidth]{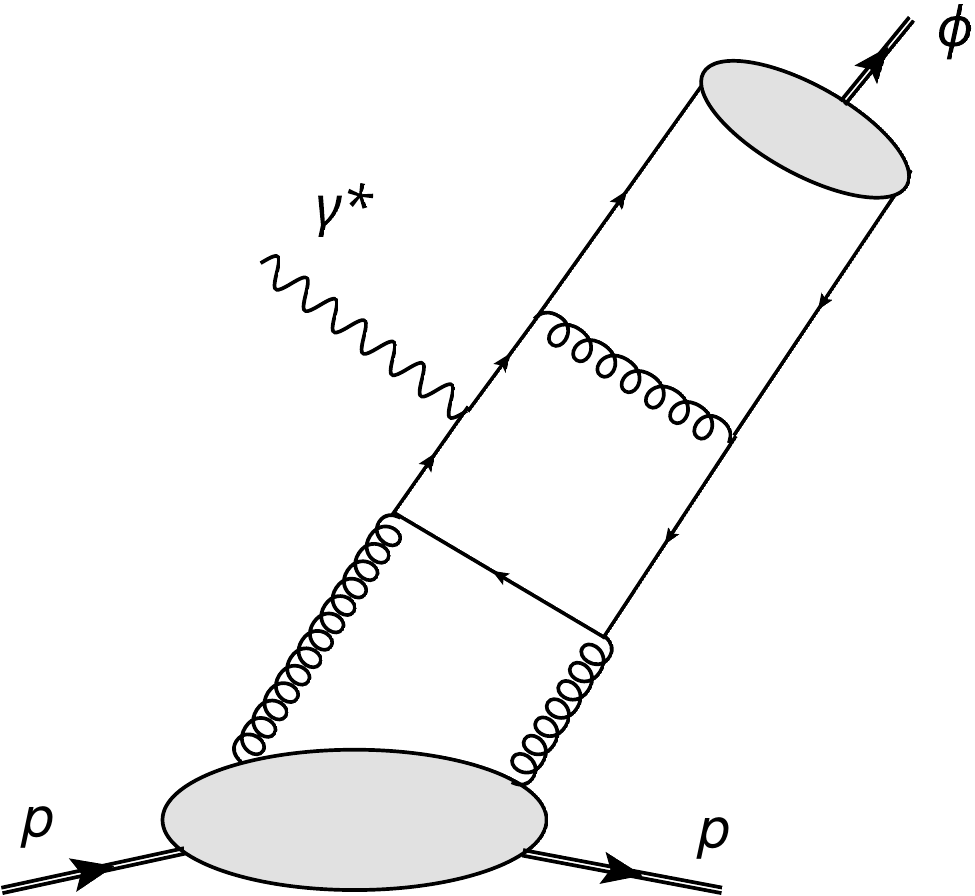}
    \caption{Representative NLO diagrams for the three terms in (\ref{hxi}) and (\ref{exi}). Left: $s$-quark exchange. Middle: `pure singlet' exchange. Right: gluon exchange.  }
    \label{3diagrams}
\end{figure}
The three contributions are depicted in Fig.~\ref{3diagrams}. 
One can roughly estimate the size of each contribution using the following rule of thumb: 
\beq
 A_{u+d}\sim A_g\sim 10A_s, \qquad \frac{\alpha_s}{\pi}\sim 0.1, \label{rough}
\eeq 
deferring the detailed numerical analysis to the next section. 
If we compare only the $A_{q,g}$ terms and assume a similar $t$-dependence for the quark and gluon GFFs, clearly the gluon exchange  dominates in ${\cal H}$. The NLO correction for this term ${\cal O}(\alpha_s^2)$ is sizable, about 60-70\% of the leading order result. We also notice that the contribution from the $s$-quark exchange is comparable to that from  $u,d$-quarks despite $A_s\ll A_{u+d}$ because the latter only enters at NLO. The relative minus sign suggests a significant cancellation between the two contributions.  However, the inclusion of the $D_{q,g}$ terms may drastically change this expectation. Since  $\xi^2\sim {\cal O}(1)$ and $D_{q,g}$ are supposedly   negative, there can be a significant cancellation in $A_{q,g}+\xi^2 D_{q,g}$ if $D_{q,g}$ are order unity. However, this cancellation may not be effective in the $s$-quark sector because $A_s\ll 1$. Therefore, if $D_s\sim {\cal O}(1)$, the $s$-quark contribution could become dominant \cite{Hatta:2021can}.   $D_{q,g}$ play an even more important role in ${\cal E}$ since there are indications  that $B_{q,g}$ are very small  \cite{Teryaev:1999su,LHPC:2007blg}. 
 As is clear from (\ref{cross}), when $\xi\sim {\cal O}(1)$, the ${\cal E}$ term is no longer negligible relative to the ${\cal H}$ term.

\subsection{Accuracy of the threshold approximation}
\label{sec: accuracy}

Let us investigate for which values of $\xi$ the threshold approximation (\ref{hxi}), (\ref{exi}) can be  a decent approximation to the full DVMP  amplitude.  We shall focus on the GPD $H$ and the associated amplitude ${\cal H}$, but the discussion for $E$ and ${\cal E}$ is entirely analogous. In the  asymptotic limit $\mu^2\to \infty$ (corresponding to $Q^2\to \infty$) where $H_g(x,\xi)\propto \theta(\xi-|x|)(1-\frac{x^2}{\xi^2})^2$, the $j=1$ conformal moment by definition accounts for 100\% of the LO amplitude, irrespective of the value of $\xi$. (In the Mellin moment expansion, the  $j=1$  moment accounts for 80\% of the amplitude in the limit $\xi\to 1$ \cite{Hatta:2021can,Guo:2021ibg}.) If $\mu^2$ is finite this is not the case, however.
Indeed, the inverse transforms of \eqref{cpw} and \eqref{cpw2} are given by the formal series
\begin{align}
H^{q(+)}(x,\xi,t) &``=" \sum_{j=1}^{\infty} \frac{2^{j+1} \Gamma(5/2+j)}{\xi^{j+1}\Gamma(3/2) \Gamma(3+j)} \left ( 1- \frac{x^2}{\xi^2} \right ) H_j^q(\xi,t) C_j^{3/2}(x/\xi) \Theta(\xi - |x|),
\label{Hq exp}
\\
H^g(x,\xi,t) &``=" \sum_{j=1}^{\infty} \frac{3}{\xi^j} \frac{2^{j+1}\Gamma(5/2+ j)}{\Gamma(3/2)\Gamma(4+j)} \left ( 1- \frac{x^2}{\xi^2} \right )^2 H_j^g(\xi,t) C_{j-1}^{5/2}(x/\xi) \Theta(\xi - |x|),
\end{align}
where the symbol $``="$ indicates that the summation is generally divergent, unless the GPDs have support only in the ERBL region $-\xi <x<\xi$, which is not the case for finite $\mu^2$. 
However, one may as well  expand the GPDs  in terms of the basis $C_j^{3/2}(x)$, $C_j^{5/2}(x)$ defined over their  correct support region $-1<x<1$ \cite{Belitsky:2005qn}. 
The modified moments are given by
\begin{align}
\mathbb{H}^q_j(\xi,t)=\frac{\Gamma(3/2)\Gamma(j+1)}{2^j\Gamma (j+3/2)} \frac{1}{2}\int_{-1}^1dx \, C_j^{3/2}(x) H^{q(+)}(x,\xi,t), \label{bb}
\end{align} 
and similarly for the gluon. 
The inverse of (\ref{bb}) is a convergent series
\begin{align}
H^{q(+)}(x,\xi,t) &= \sum_{j=1}^{\infty} \frac{2^{j+1} \Gamma(5/2+j)}{\Gamma(3/2) \Gamma(3+j)} \left ( 1- x^2 \right ) \mathbb H_j^q(\xi,t) C_j^{(3/2)}(x).
\label{Hq exp convergent}
\end{align}
It is clear that, by truncating the divergent series in \eqref{Hq exp}, we obtain 
the corresponding truncation of the converging series \eqref{Hq exp convergent} as $\xi \rightarrow 1$. In particular the first moment agrees exactly $\mathbb{H}_1^q = H_1^q$, so that \eqref{j=1} essentially corresponds to truncating the convergent series \eqref{Hq exp convergent} after the first term. 

\begin{figure}
    \centering
    \includegraphics[width=0.6\linewidth]{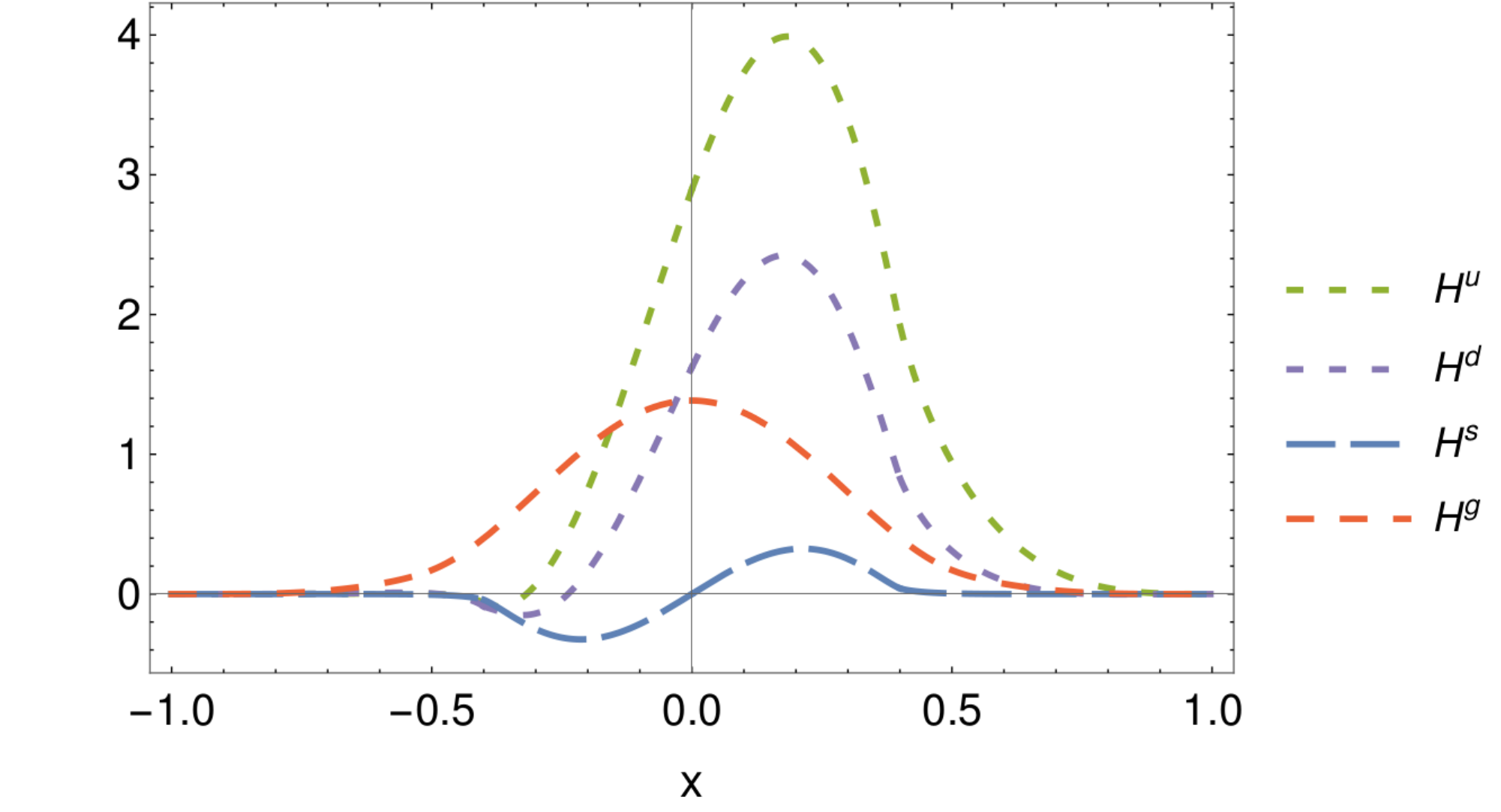}
    \caption{GK model for $H^{u,d,s,g}(x,\xi,t=0,\mu^2)$ at $\xi = 0.5$ and $\mu=2$ GeV with PDF parameters fitted to the PDF sets \cite{Alekhin:2017kpj,Alekhin:2018pai}. }
    \label{fig: GK model}
\end{figure}

On general grounds, one expects that convergent orthogonal series are dominated by low moments because higher moments involve stronger oscillating functions  which destructively interfere with more slowly varying functions like GPDs. 
It then remains to be seen whether the convolution with the coefficient functions $T^a(x,\xi)$  preserves this expectation.  In order to quantitatively test this,   
we use the Goloskokov-Kroll (GK) model \cite{Goloskokov:2006hr,Goloskokov:2007nt}  for $H^{q,g}$ with PDF parameters fitted to the 
PDF sets \cite{Alekhin:2017kpj,Alekhin:2018pai}.\footnote{This model has been used for the numerical analysis in \cite{Braun:2022bpn}.} The GK model was originally designed for the small-$x_B$, small-$t$ region. Assuming an exponential $t$-dependence in GPDs, the model  successfully describes the HERA data on meson production cross sections. However, it  has not been tested in the region of our interest, namely, $\xi\sim {\cal O}(1)$ where $|t| \geq |t_{min}|=\frac{4\xi^2 M^2}{1-\xi^2}$ becomes large and one starts to see the power-law behavior in $t$ \cite{Kroll:2012sm}.  
Given the potentially strong model dependence at large-$t$, in the following we use the GK model evaluated at an unphysical point $H^{q,g}(x,\xi,t=0)$ (see also \cite{Guo:2023qgu}) which can be constructed relatively unambiguously using the double distribution technique \cite{Radyushkin:1998bz}. We will comment on the finite-$t$ effect later.  
Fig.\,\ref{fig: GK model} shows $H^{u,d,s,g}(x,\xi,t=0,\mu^2)$ at a representative point  $\xi=0.5$ and $\mu=2$ GeV.

Within this model, we numerically evaluate the convolution integral (\ref{fact})  using the asymptotic DA $\varphi(u)=6u(1-u)$ and the NLO coefficient functions  
\beq
&&T^q\left(x,\xi,u,Q^2/\mu^2\right) = \alpha_s\left(\frac{1}{1-u}\frac{1}{\xi-x-i\epsilon} -\frac{1}{u}\frac{1}{\xi+x-i\epsilon}\right)  +{\cal O}(\alpha_s^2) ,\nn
&& T^g\left(x,\xi,u,Q^2/\mu^2\right) = \frac{\alpha_s}{4C_F}\frac{1}{u(1-u)}\frac{1}{x}\left(\frac{1}{\xi-x-i\epsilon} -\frac{1}{\xi+x-i\epsilon}\right)  +{\cal O}(\alpha_s^2),
\eeq
where the NLO ${\cal O}(\alpha_s^2)$ terms can be found in 
\cite{Muller:2013jur} with a correction in \cite{Duplancic:2016bge}. 
The direct integral in $x$-space is numerically challenging due to the poles of the coefficient functions at $x =\pm\xi$ which must be circumvented according to the $\xi \rightarrow \xi - i\epsilon$ prescription. Straightforward methods work fine at LO but become increasingly numerically unfeasible for the NLO kernel and beyond.
Given that one has a piecewise analytic (up to branch cuts) GPD model, a good way to compute the convolution numerically is by integrating on a set of contours in the complex plane as follows \cite{Braun:2020yib}. Let $H(x,\xi) = \theta(x-\xi) H_1(x,\xi) + \theta(\xi - x) H_2(x,\xi)$. Then 
\begin{align} 
\int_0^1 dx\, T(x,\xi - i\epsilon) H(x,\xi) = \int_{\gamma_1} dx\, T(x,\xi) H_2(x,\xi) + \int_{\gamma_2} dx\, T(x,\xi) (H_1(x,\xi) - H_2(x,\xi) )  + \int_{\gamma_3} dx\, T(x,\xi) H_1(x,\xi). 
\end{align}
The contours $\gamma_j$ can be  defined as straight lines with respect to an arbitrary point $z \in \mathbb C$ with $\text{Im}\,z > 0$ in the following way. $\gamma_1$ goes from $0$ to $z$, $\gamma_2$ goes from $\xi$ to $z$ and $\gamma_3$ goes from $z$ to $1$.  In practice, $z$ should be chosen for optimal numerical stability. 

\begin{figure}
    \centering
    \qquad 
    \includegraphics[width=.8\linewidth]{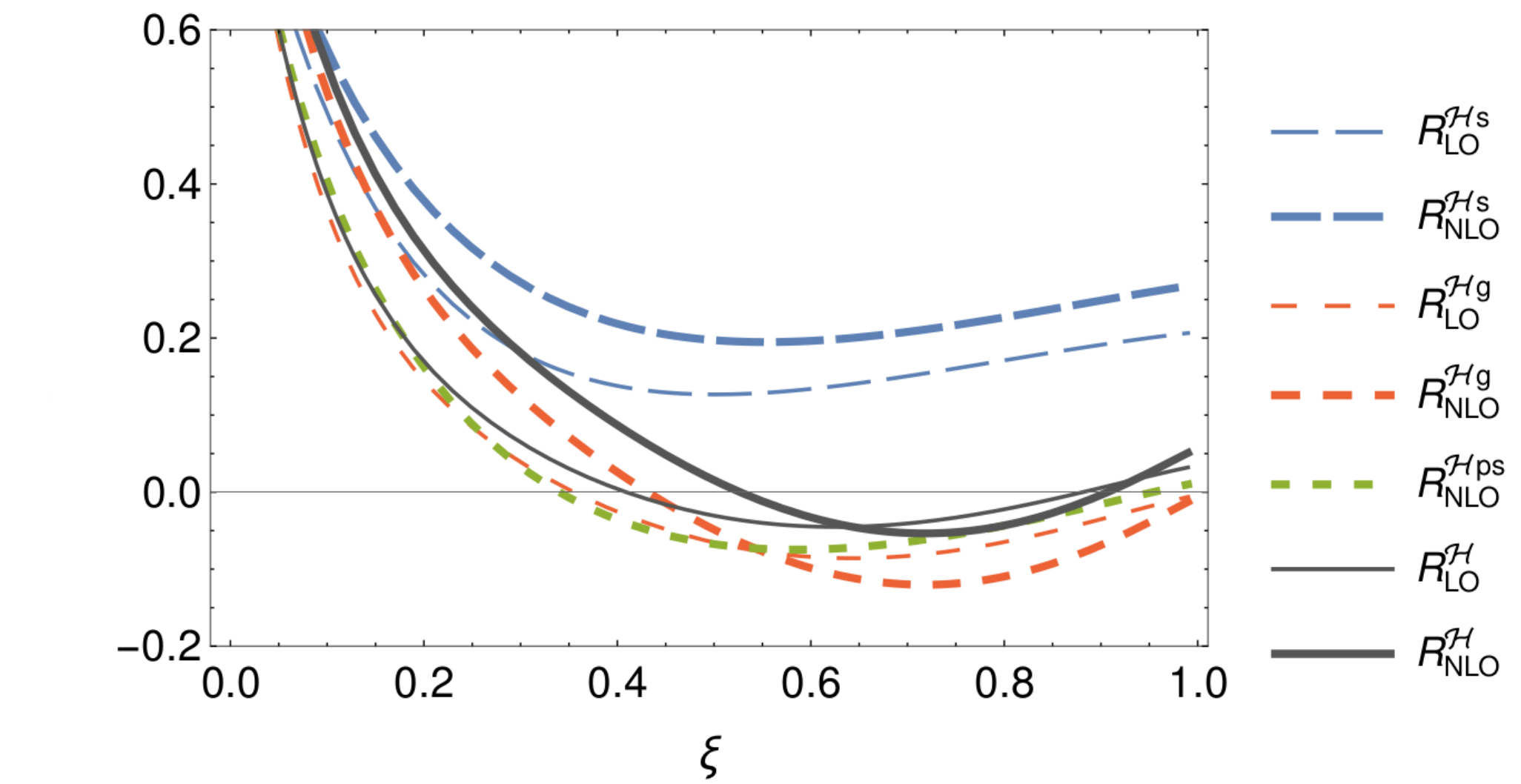}
    \caption{Relative error for the amplitude $\mathcal H$ from truncating the conformal partial wave expansion after the first term. We plot the $\xi$ dependence at $t = 0$. Plotted quantities are  defined in \eqref{eq: R defs}. The subscript denotes whether the leading order (LO) or next-to-leading order (NLO) coefficient function has been used. We have set $\kappa=1$ and $Q = \mu = 2\,\text{GeV}$ corresponding to $\alpha_s = 0.34$ (from (\ref{alphas})). 
}
    \label{fig: trunc}
\end{figure}

We then compare the resulting LO and NLO amplitudes ${\cal H}$ with  
their truncated versions  
${\cal H}_{\rm trunc}$ (\ref{hxi}) separately in the three channels ${\cal H}={\cal H}^s+{\cal H}^{ps}+{\cal H}^g$ corresponding to the three terms in (\ref{hxi}). (The superscript $ps$ stands for `pure singlet' which enters only at NLO, see the middle diagram of Fig.~\ref{3diagrams}.) This is plotted in Fig.~\ref{fig: trunc}  
in terms of  the relative errors
\begin{align}
R^s = 1 - \frac{|\mathcal H^{s}|}{|\mathcal H^{s}_{\rm trunc}|}, \quad R^{ps} = 1 - \frac{|\mathcal H^{ps}|}{|\mathcal H^{ps}_{\rm trunc}|}, \quad R^g = 1 - \frac{|\mathcal H^g|}{|\mathcal H^g_{\rm trunc}|}, \quad R^{\rm tot} = 1 - \frac{|\mathcal H|}{|\mathcal H_{\rm trunc}|},
\label{eq: R defs}
\end{align}
where $|{\cal H}^a|=\sqrt{({\rm Re}\,{\cal H}^a)^2+({\rm Im}\,{\cal H}^a)^2}$. (Note that ${\cal H}^a_{\rm trunc}$ are real.)  
\begin{figure}
    \centering
    \includegraphics[width=0.47\linewidth]{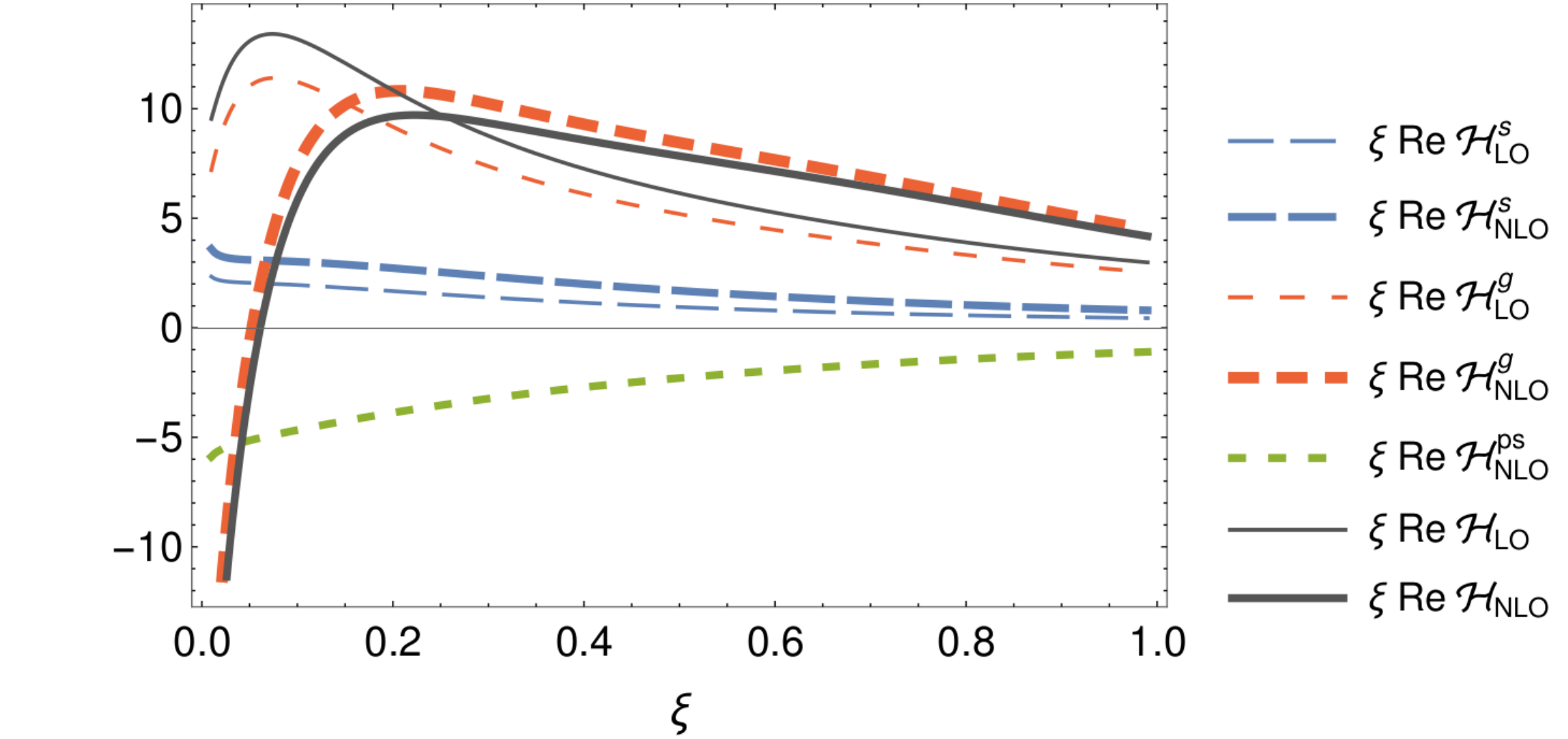}
    \qquad 
    \includegraphics[width=0.47\linewidth]{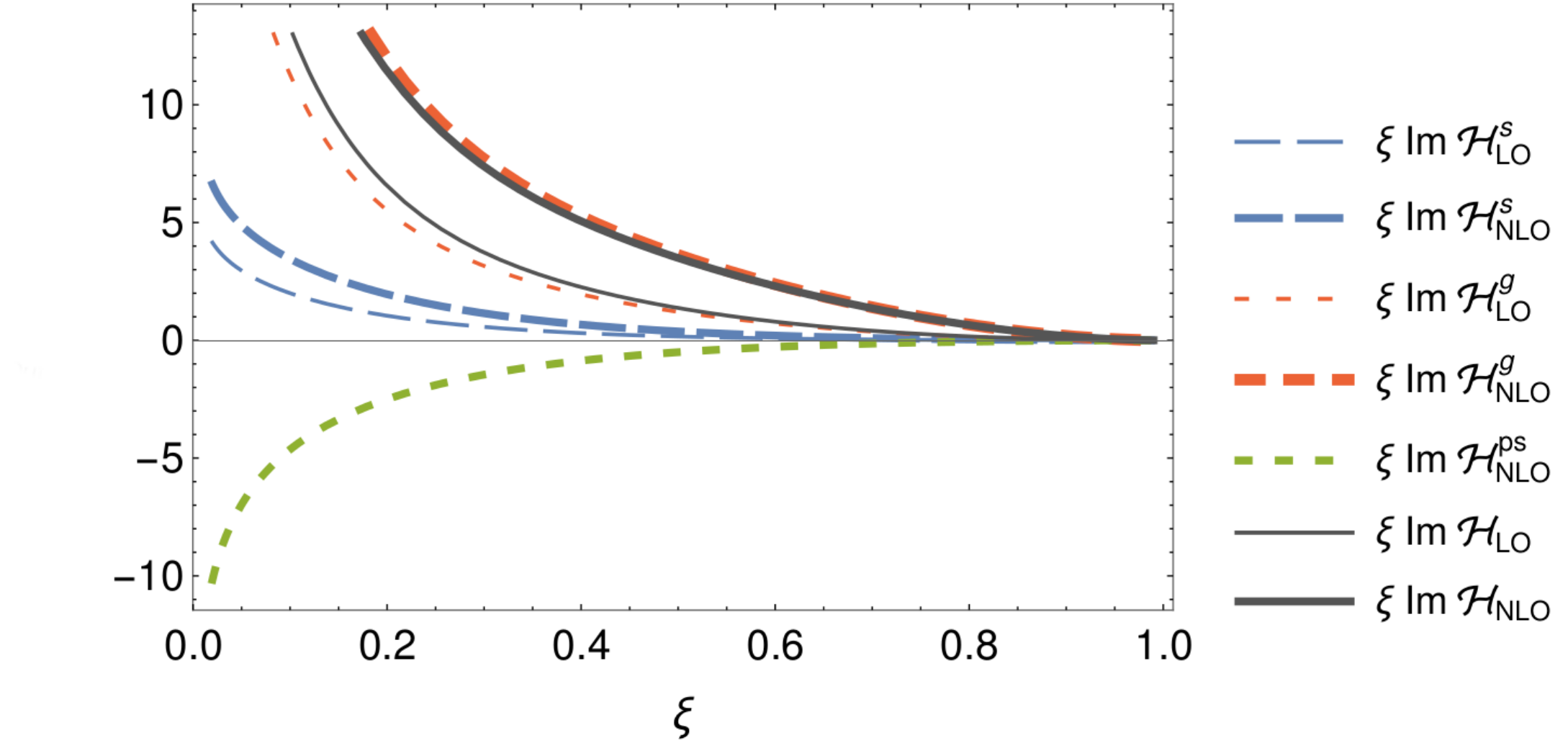}
    \caption{The real (left) and imaginary (right) parts of $\mathcal H$ at $\mu=2$ GeV separated into the $s$-quark, pure-singlet and gluon contributions. Results for both LO and NLO predictions are also shown. 
    }
    \label{fig: LO and NLO}
\end{figure}
It can be seen that the truncation error in the $s$-quark contribution is the largest, at best 20\% around $\xi \approx 0.5$. The error does not decrease by increasing $\xi$, indicating that higher moments are important. Still, the situation is much better than in the Mellin moment expansion where the truncation error can be much larger \cite{Hatta:2021can}.   For the gluon and pure-singlet  contributions, the threshold approximation is very good, with errors being below  10\% starting already at $\xi \approx 0.35$. Interestingly, the approximation is the best for the total amplitude ${\cal H}$ for $\xi>0.5$, which may be a special feature of $\phi$-production. This is due to a significant cancellation between the $s$-quark and pure singlet contributions both in the truncated  and full amplitudes, see the discussion below (\ref{rough}) and  
Fig.~\ref{fig: LO and NLO} where $\mathcal H^{s}_{\rm NLO} \approx -\mathcal H^{ps}_{\rm NLO}$ in the entire $\xi$ range. As a result, $\mathcal H$ is almost entirely dominated by the gluon contribution $\mathcal H \approx \mathcal H^g$. We expect that the threshold approximation gets even better as $\mu^2\sim Q^2$ is increased, since the GPDs approach their asymptotic limits.

 An important question is whether the above conclusion, obtained at an unphysical point $t=0$, remains qualitatively valid in the phenomenologically relevant region $t\approx t_{min}$.  Empirically, it is known that the gluon GPD and form factors fall faster with increasing  $|t|$  than the quark counterparts \cite{CLAS:2001zwd,ZEUS:2005bhf,H1:2007vrx}. One can then imagine a scenario  where the $\sum_q A_q$  term in (\ref{hxi}) cancels the $A_g$ term (instead of the $A_s$ term as mentioned above)  at large-$t$. If this rather accidental  cancellation occurs, $R^{\rm tot}$ can become very large $|R^{\rm tot}|\gg 1$ due to a small denominator  $|{\cal H}_{\rm trunc}|$ even though the threshold approximation works well for individual contributions $|R^{s,ps,g}|\ll 1$. In fact, this is what happens if one evaluates ${\cal H}$ using  $H(x,\xi,t=t_{min}(\xi))$ and the default parameter set of the GK model \cite{Goloskokov:2007nt}. We however think that this is an artifact of the GK model which predicts a too rapid falloff of the gluon GPD and GFFs  at large-$t$ as compared to the quark ones. A more reasonable estimate of the $t$-dependence may come from lattice QCD \cite{Hackett:2023rif}, where it was found that $\sum_q A_q(t)$ and $A_g(t)$ have quite similar $t$-dependencies. A simple numerical check using (\ref{rough}) and the parametrization of $A_{q,g}(t)$ obtained in \cite{Hackett:2023rif} shows that the $A_g$ term safely dominates over the $\sum_q A_q$ term in (\ref{hxi})  in the phenomenologically relevant region $|t|<2$ GeV$^2$. Therefore, for the moment we exclude the above scenario, although further investigations are certainly needed.     

Another caveat is  that the GK model that has been used here does not have a D-term, and hence no information about the $D$-type contributions in \eqref{hxi} can be gained from our analysis. However, the above discussion indicates that the dominance of the $j=1$ term is a generic feature at large skewness that arises from the  combined effects of  the suppression of the  DGLAP region $\xi<|x|<1$,  the smooth behavior of GPDs, and  the property of the coefficient functions.  Recalling that the $D$-term has support only in the ERBL region $|x|<\xi$, we can expect  a similar level of accuracy after including the $D$-type  contributions.  The relative signs of the contributions might differ. For instance, we might have $D_s > 0$ (whereas $D_{u+d+s} < 0$), which reverses the aforementioned cancellation between the $s$-quark and pure-singlet contributions. 

Overall, we conclude that errors due to the threshold approximation are at worst $20\%$  and can be as good as  $10\%$ or less for moderately large $\xi \gtrsim 0.4$. Considering also the discussion around (\ref{scale}), it appears that the range $0.4<\xi< 0.6$ is the best region to focus on.

\section{Numerical analysis}

In this section we present our numerical results on the differential cross section $\frac{d\sigma_L}{dt}$ (\ref{cross}), where ${\cal H},{\cal E}$ are given by (\ref{hxi}), (\ref{exi}). The inputs for the NLO computation are as follows. The one-loop running coupling is 
\beq
\alpha_s(\mu)= \frac{4\pi}{\beta_0\ln (\mu^2/\Lambda_{\rm QCD}^2)},
\qquad \beta_0= \frac{11N_c}{3}-\frac{2n_f}{3}, \label{alphas}
\eeq
where $n_f=4$. Following \cite{Cuic:2023mki}, we  set $\frac{\alpha_s}{2\pi}=0.0606$ at $Q^2=2.5$ GeV$^2$, which gives $\Lambda_{\rm QCD}^2=0.0476$ GeV$^2$. The $\phi$-meson decay constant is 
\beq
f_\phi = 0.221\, {\rm GeV},
\eeq 
as in \cite{Cuic:2023mki}. 
As for the GFFs, we use the dipole and tripole parameterizations for the $A$-type and $D$-type form factors, respectively \cite{Tanaka:2018wea,Tong:2021ctu}
\beq
&&A_{q,g}(t,\mu) = \frac{A_{q,g}(\mu)}{(1-t/m_A^2)^2}, \label{ad} \\
&& D_{q,g}(t,\mu)= \frac{D_{q,g}(\mu)}{(1-t/m_D^2)^3}. \nonumber
\eeq
In principle, the mass parameters $m_A$ and $m_{D}$ depend on parton species, but for simplicity we assume common values \cite{Duran:2022xag} 
\beq
m_A=1.6\, {\rm GeV}, \qquad m_D=1.1\, {\rm GeV}.  \label{lama}
\eeq
The forward values $A_a(\mu)$ represent the momentum fraction of the proton carried by partons  $\sum_{a=q,g}A_a(\mu)=1$. We consider their one-loop QCD evolution using  
\beq
A_s(\mu_0)=0.03, \qquad A_g(\mu_0)=0.42, \qquad A_{u+d+c}(\mu_0)=1-A_s(\mu_0)-A_g(\mu_0),
\eeq
at the reference scale  $\mu_0=2$ GeV \cite{Hou:2019efy}.  
The one-loop evolution of the D-terms $D_{q,g}(\mu)$  is the same as that for $A_{q,g}(\mu)$ and is explicitly given by
\beq
D_q(\mu)&=& \frac{D}{4C_F+n_f} + \frac{1}{n_f\left(4C_F+n_f\right)}\left(\frac{\alpha_s(\mu_0)}{\alpha_s(\mu)}\right)^{-\frac{2}{3\beta_0}\left(4C_F+n_f\right)}\bigl(4C_F D-\left(4C_F+n_f\right)D_g(\mu_0)\bigr)\nn 
&& +\left(\frac{\alpha_s(\mu_0)}{\alpha_s(\mu)}\right)^{-\frac{8C_F}{3\beta_0}}\left(D_q(\mu_0)-\frac{1}{n_f}\sum_{q'}D_{q'}(\mu_0)\right),
\eeq
\beq
D_g(\mu)
&=& \frac{4C_F}{4C_F+n_f}D
- \frac{1}{4C_F+n_f}\left(\frac{\alpha_s(\mu_0)}{\alpha_s(\mu)}\right)^{-\frac{2}{3\beta_0}
\left(4C_F+n_f\right)}\left(4C_F D -(4 C_F+n_f) D_g(\mu_0)\right),
\eeq
where the total D-term $D=\sum_q D_q(\mu)+D_g(\mu)$ 
is $\mu$-independent. 
Our main interests in this paper are the strangeness and gluon D-terms $D_s(\mu)$ and $D_g(\mu)$. 
We therefore assume that, 
at the reference scale  $\mu_0=2$ GeV 
\cite{Hackett:2023rif},
\beq
D_{u+d+c}(\mu_0)=-1.2 
\eeq
and treat $D_{s}(\mu_0)$ and $D_{g}(\mu_0)$  as  free parameters.  
In the following, we use the simpler notations $D_s(\mu_0)\to D_s$ and $D_g(\mu_0)\to D_g$. 
Finally, we neglect the $B$-type GFFs $B_{q,g}(t)$ altogether \cite{Teryaev:1999su,LHPC:2007blg}. 
\begin{figure}[t]
    \centering
    \begin{overpic}[
       width=0.6\textwidth,
       trim={0.0cm 0.0cm 0.6cm 1.5cm}, 
       clip
    ]{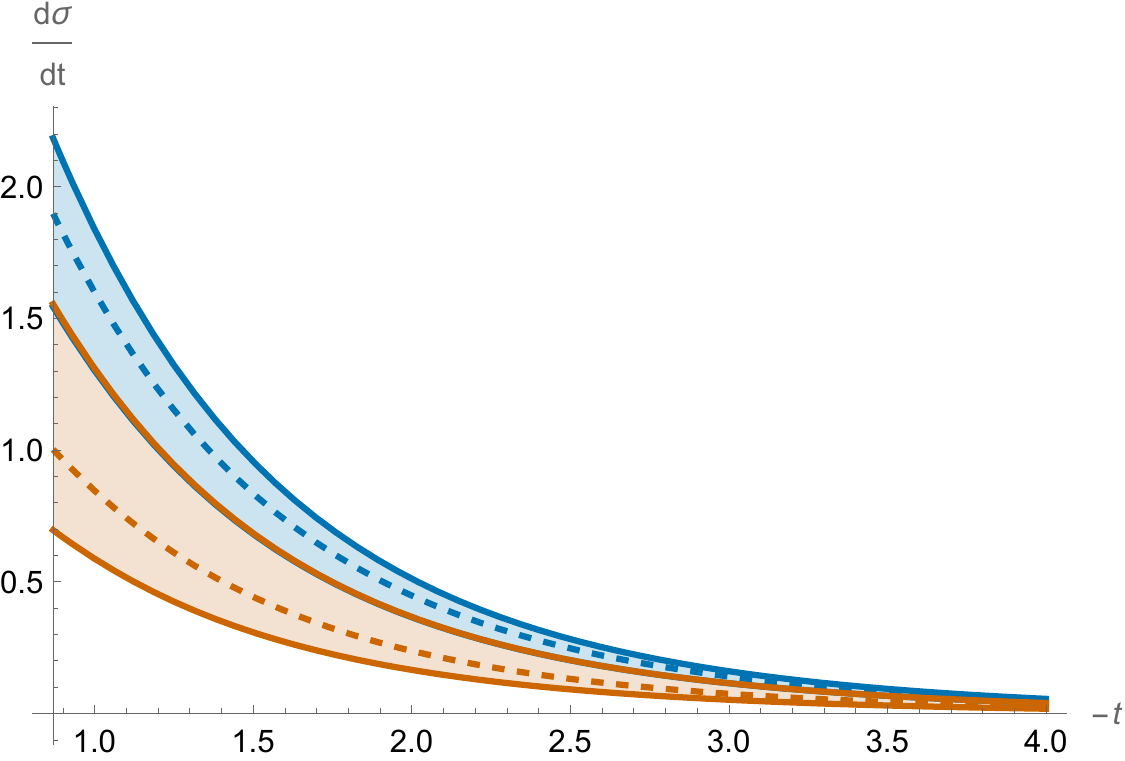}

      \put(-6,20){\rotatebox{90}{\small $d\sigma_L/d|t|\;(\mathrm{nb}/\mathrm{GeV}^2)$}}

      \put(45,-4){\small $|t|\;(\mathrm{GeV}^2)$}

        \put(35,45){
           {\color{Bittersweet}\raisebox{0.75ex}{\rule{1.5em}{2pt}}}%
           \,\large{LO}}
        \put(35,50){
           {\color{MidnightBlue}\raisebox{0.75ex}{\rule{1.5em}{2pt}}}%
           \,\large{NLO}}

    \end{overpic}
    \vspace{0.5cm}
    \caption[*]{Differential $\gamma^*p$ cross section in units of nb/GeV$^2$ 
      at $W=Q=2.5$ GeV, $D_g=-1$, and $D_s=0$ as a function of $|t|$. 
      The orange and blue bands represent 
      the LO and NLO cross sections, respectively, with the renormalization scale 
      varied in the range $Q/2 < \mu < 2Q$.}
    \label{band}
\end{figure}
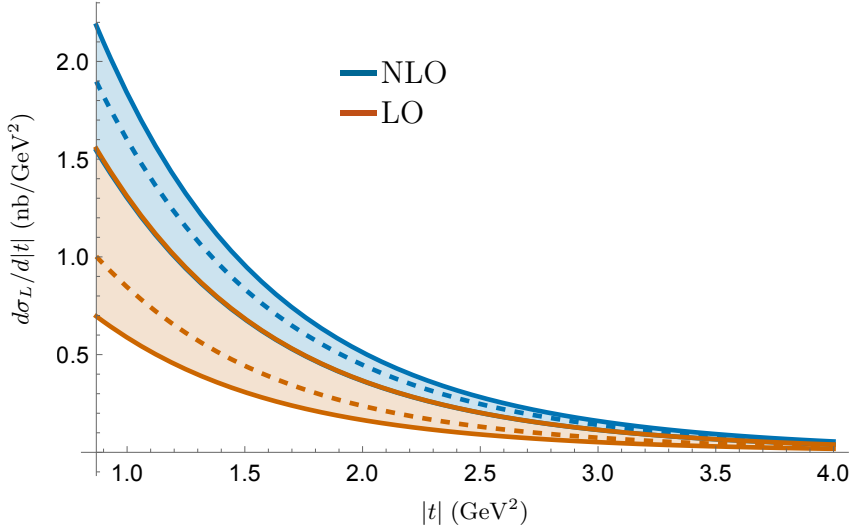

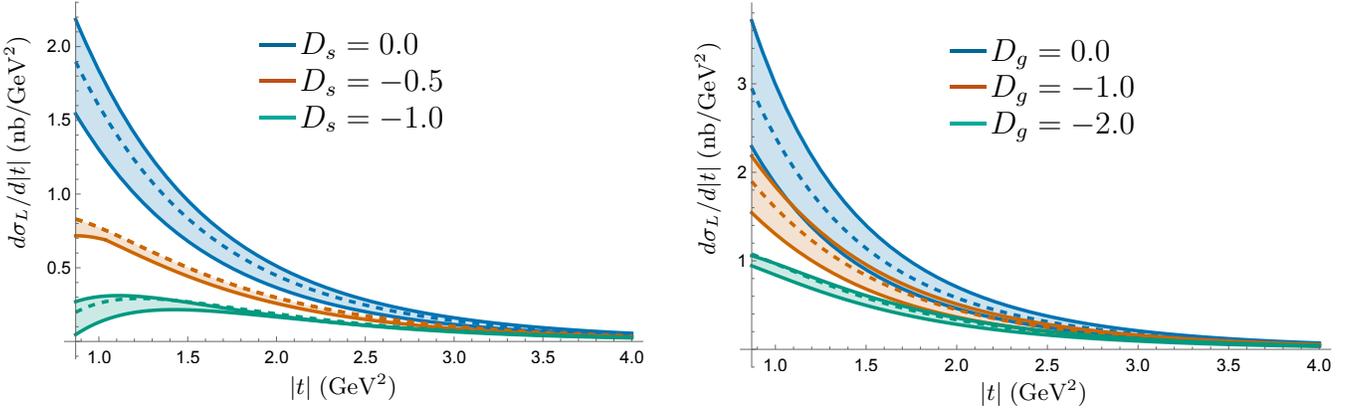
\begin{figure}[t]
    \centering
    \begin{minipage}{0.45\textwidth}
      \begin{overpic}[
         width=\textwidth,
         trim={0 0 0.6cm 1.5cm}, 
         clip
       ]{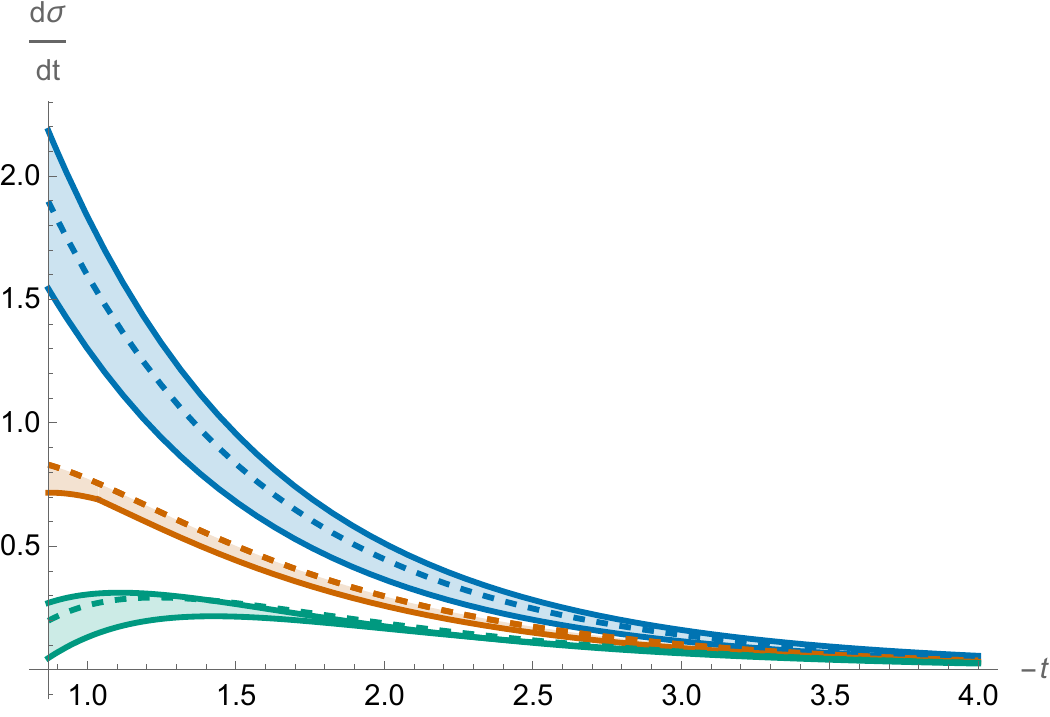}

        \put(-7,20){\rotatebox{90}{\small $d\sigma_L/d|t|\;(\text{nb}/\text{GeV}^2)$}}
        \put(40,-4){\small $|t|\;(\text{GeV}^2)$}

        \put(35,40){
           {\color{JungleGreen}\raisebox{0.75ex}{\rule{1.5em}{1.5pt}}}%
           \,\large $D_s=-1.0$
        }
        \put(35,46){
           {\color{Bittersweet}\raisebox{0.75ex}{\rule{1.5em}{1.5pt}}}%
           \,\large $D_s=-0.5$
        }
        \put(35,52){
           {\color{MidnightBlue}\raisebox{0.75ex}{\rule{1.5em}{1.5pt}}}%
           \,\large $D_s=0.0$
        }
      \end{overpic}
    \end{minipage}
    \hspace{0.05\textwidth}
    \begin{minipage}{0.45\textwidth}
      \begin{overpic}[
         width=\textwidth,
         trim={0 0 0.6cm 1.5cm}, 
         clip
       ]{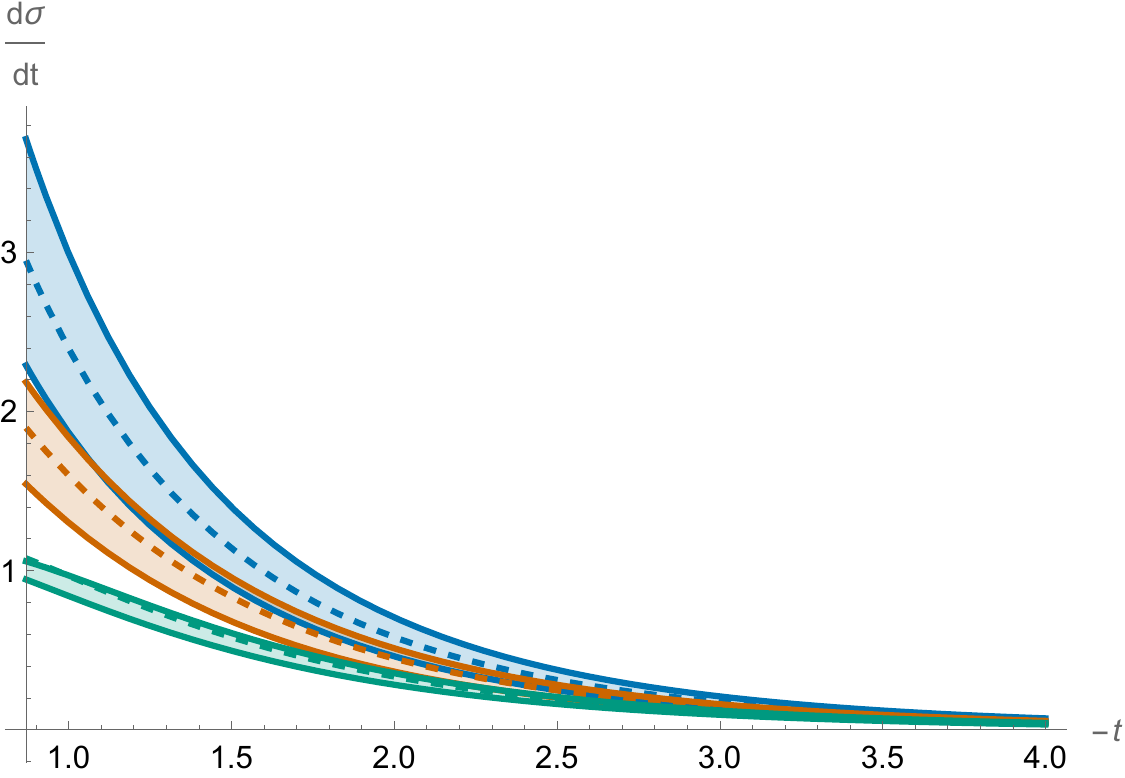}

        \put(-7,20){\rotatebox{90}{\small $d\sigma_L/d|t|\;(\text{nb}/\text{GeV}^2)$}}
        \put(40,-4){\small $|t|\;(\text{GeV}^2)$}

        \put(35,40){
           {\color{JungleGreen}\raisebox{0.75ex}{\rule{1.5em}{1.5pt}}}%
           \,\large $D_g=-2.0$
        }
        \put(35,46){
           {\color{Bittersweet}\raisebox{0.75ex}{\rule{1.5em}{1.5pt}}}%
           \,\large $D_g=-1.0$
        }
        \put(35,52){
           {\color{MidnightBlue}\raisebox{0.75ex}{\rule{1.5em}{1.5pt}}}%
           \,\large $D_g=0.0$
        }
    
      \end{overpic}
    \end{minipage}
    \vspace{0.5cm}
    \caption[*]{NLO longitudinal cross section at $W=Q=2.5$ GeV 
      as a function of $|t|$. 
      Left: $D_s=0,-0.5,-1$ from top to bottom at fixed $D_g=-1$. 
      Right: $D_g=0,-1,-2$ from top to bottom at fixed $D_s=0$.  }
    \label{dsigma}
\end{figure}

We are now ready to present the numerical results. First, the differential cross section (\ref{cross}) at $W=Q=2.5$ GeV, $D_g=-1$, $D_s=0$ is shown in Fig.~\ref{band}.  The purple and yellow lines represent the LO and NLO cross sections evaluated at $\mu=Q$, respectively. The orange and green bands are the respective uncertainty bands obtained by varying $\mu$ in the range $Q/2<\mu<2Q$. As expected, the scale uncertainty is reduced by going to NLO.   The result in the large-$|t|$ region should be taken with a grain of salt because  higher twist corrections might be very large there. In the present kinematics, the condition (\ref{upper}) reads $|t|\lesssim \frac{Q^2}{3}\approx 2.1$ GeV. 

Next, in 
Fig.\ref{dsigma} we show $\frac{d\sigma_L}{dt}$ at NLO for three  different values of $D_s$ at fixed $D_g$ (left) and for three different values of $D_g$ at fixed $D_s$ (right). Again, the band attached to each curve represents the scale uncertainty $Q/2<\mu< 2Q$. (The same comment applies to the plots below and will not be repeated.) It is reassuring to see that the sensitivities to $D_{s,g}$ are larger than the theory uncertainty bands, making this observable ideally suited for constraining  $D_{s,g}$. As we increase $W$, the three curves tend to approach one another. But they are still well separated when $W\lesssim 3$ GeV, which is the limit of applicability of our calculation, see below.  The bending of $d\sigma/dt$ in the small-$|t|$ region for large negative values of $D_s$ was previously observed in   \cite{Hatta:2021can}. We note that a recent model calculation found  $D_s\approx -0.5$  \cite{Won:2023ial}.

\begin{figure}[t]
\centering
 \begin{minipage}{0.45\textwidth}
   \begin{overpic}[
         width=\textwidth
         ]{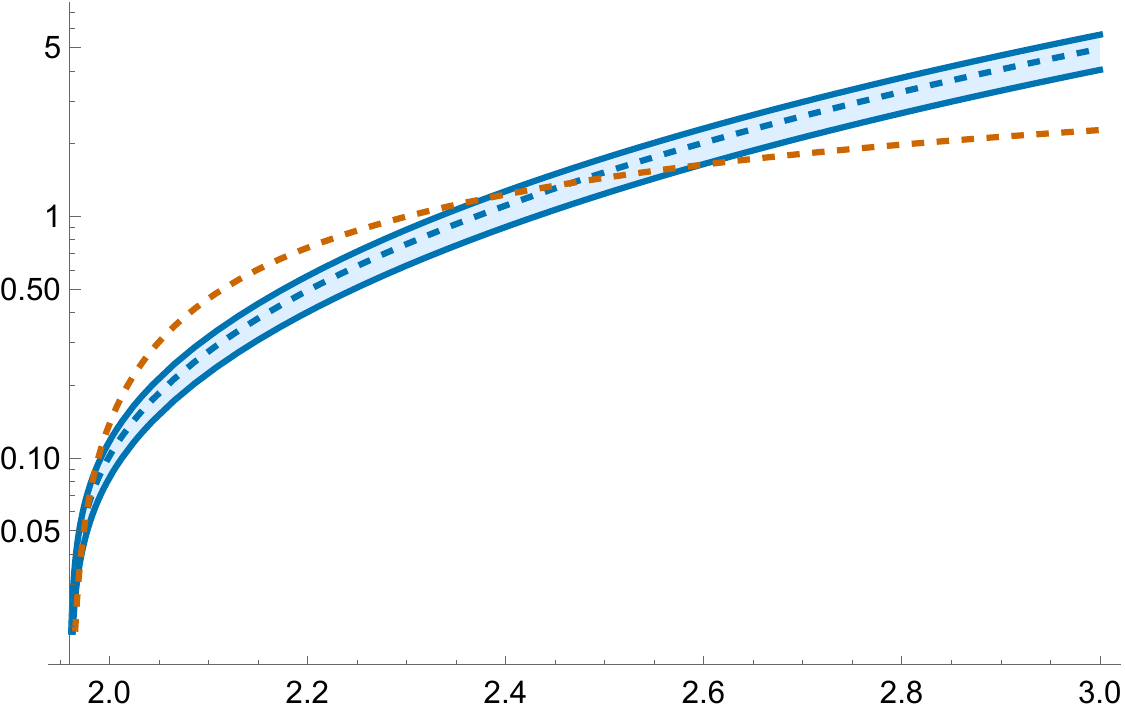}
          \put(-7,30){\rotatebox{90}{\small $\sigma_L\; (\text{nb})$}}
              \put(40,-4){\small $W\; (\text{GeV})$}
         \end{overpic}
\end{minipage}
\hspace{0.05\textwidth}
 \begin{minipage}{0.45\textwidth}
\begin{overpic}[width=\textwidth]{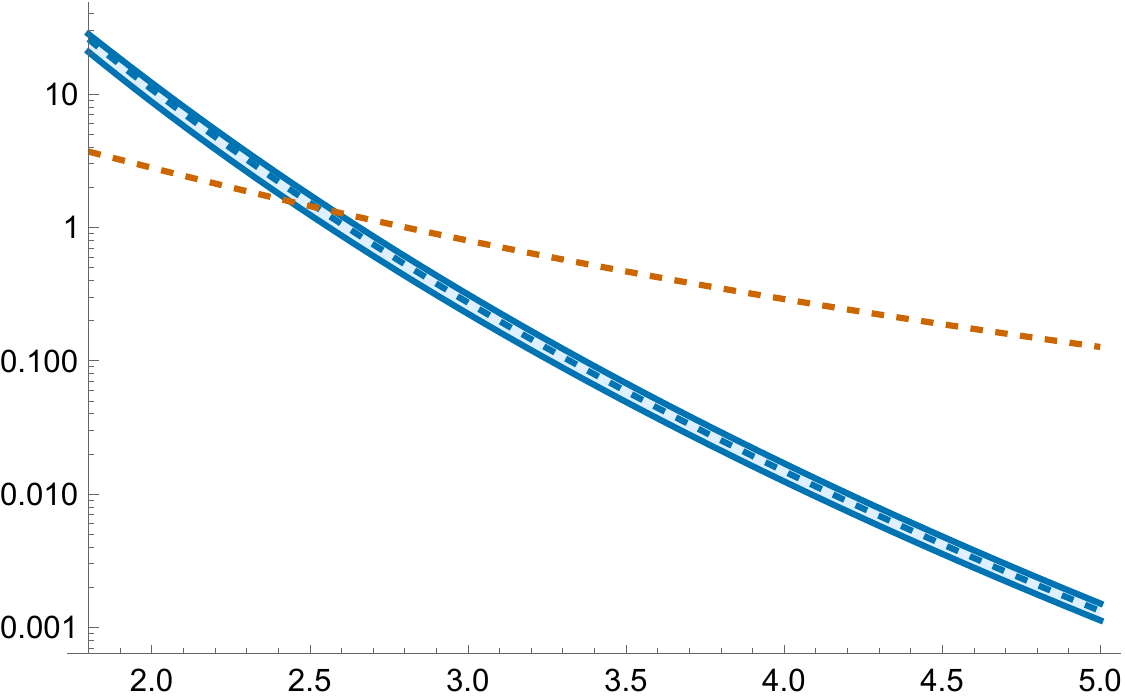}
   \put(-7,30){\rotatebox{90}{\small $\sigma_L\;(\text{nb})$}}
        \put(40,-4){\small $Q\;(\text{GeV})$}
\end{overpic}
\end{minipage} 
\vspace{0.5cm}
\caption[*]{NLO longitudinal cross section integrated over $t$ with $D_s=0$ and $D_g=-1$ as a function of $W$ at fixed $Q=2.5$ GeV (left) and as a function of $Q$ at fixed $W=2.5$ GeV (right). The orange dashed curve is the CLAS parametrization (\ref{clas}). }
\label{wdep}
\end{figure}

In Fig.~\ref{wdep} (left), we  plot the integrated  cross section 
\beq
\sigma_L(W,Q)=\int_{t_{min}}^{t_{max}} dt \frac{d\sigma}{dt}, \label{tota}
\eeq
at fixed $Q=2.5$ GeV. 
Again, the integrand in the large-$t$ region is not reliable due to potentially large higher twist corrections. Fortunately, however, the integral is dominated by the small-$t$ region $t\sim t_{min}$. Due to  the spin-2 nature of the energy momentum tensor, the cross section rises rather fast with increasing $W$. While this is a definite prediction of theory near the threshold, the strong dependence on $W$ cannot be extrapolated to high-$W$ values. A comparison with the experimental data taken at high energy (such as at HERA, see below) suggests that already $W=3$ GeV is too high to be considered within the threshold region. We feel comfortable applying our result to, say, $W\lesssim 2.8$ GeV.

Finally, in Fig.~\ref{wdep} (right), we plot the $Q$-dependence of the NLO integrated cross section (\ref{tota}) at fixed $W=2.5$ GeV. We find that the cross section falls very rapidly with increasing $Q$  as
\beq
\sigma_L\propto \frac{1}{Q^{a}} ,\qquad 9<a<10. \label{power}
\eeq
 At first sight, this is surprising because superficially the formula (\ref{cross}) suggests that, at fixed $W$,\footnote{  Note that, in the usual Bjorken limit $Q^2\to \infty$ at fixed $x_B$, $\frac{d\sigma_L}{dt}\sim Q^{-6}$ up to logarithms in $Q$, although this has tension with the HERA data at $x_B\ll 1$, see the discussion in \cite{Cuic:2023mki}. } 
\beq
\frac{d\sigma_L}{dt} \propto \frac{1}{Q^4} \qquad (Q^2\to \infty). \label{usu}
\eeq
 Two powers of $1/Q$ come from the hard coefficients (\ref{me}) squared, and the remaining factor $1/Q^2$ comes from $p_{cm}\sim Q^2$ at high-$Q^2$. The extra  suppression (\ref{power}) mainly comes from the gravitational form factors. As mentioned earlier, $t_{min}$ grows quadratically with $Q$
\beq
|t_{min}| =a(W) Q^2+\cdots, \label{quad}
\eeq
 already when $Q\gtrsim 2$ GeV. Although the coefficient $a$ is small in the sense of  (\ref{upper}), (\ref{quad}) eventually leads to the behavior  
\beq
A_g(t) \propto \frac{1}{(t-m_A^2)^2} < \frac{1}{(t_{min}-m_A^2)^2}  \sim \frac{1}{Q^4},
\eeq
which partly explains (\ref{power}). Typically in high energy  experiments, one measures the region $|t|\lesssim 1$ GeV$^2$ and does not expect any correlation between $t$ and $Q^2$. While the faster-than-expected falloff  (\ref{power}) is a unique feature of near-threshold production, it also makes  measurements at high-$Q^2$ challenging.      

\section{Experimental Projections for the EIC and JLab}

We turn now to determining whether the restrictive kinematic requirements laid out in the previous sections can be feasibly satisfied by an experiment. To summarize these requirements, the value of $Q^2$ should be a factor of 3 or more greater than $|t|$, $\xi$ should be greater than 0.4, and $W$ should be less than 3 GeV.  To explore the sensitivity of various experiments, we utilize the  \texttt{lAger} Monte Carlo event generator, which is capable of simulating a variety of exclusive reactions in both photoproduction and electroproduction~\cite{joosten2021}.

In Monte Carlo simulations, events with all possible values of $W,Q^2$ are  generated. Also, the virtual photon can be both transversely and longitudinally polarized. In contrast, our perturbative calculation  is valid only near the threshold (low-$W$) and at high-$Q^2$, and only for the longitudinally  polarized photons. Increasing $W$ is roughly equivalent to decreasing skewness $\xi$. If we require $\xi>0.4\sim 0.5$, Fig.~\ref{2} already suggests that we cannot go much higher than $W\sim 3$ GeV. To estimate at which value of $Q$ our result becomes unreliable, in Fig.~\ref{wdep} we have also plotted (dashed curve) the following parametrization of world data  by the CLAS collaboration \cite{CLAS12Phi}: 
\beq
\sigma^{\rm CLAS}_L(W,Q)= R(Q)\sigma^{\rm CLAS}_T(W,Q), \qquad \sigma^{\rm CLAS}_T(W,Q)=400\left(1-\frac{W_{th}^2}{W^2}\right)\frac{W^{0.32}}{\left(1+\frac{Q^2}{M_\phi^2}\right)^3}, \label{clas}
\eeq   
where $W,Q$ are in units of GeV and the $L/T$ ratio is assumed to be $W$-independent 
\beq
R(Q)= \frac{0.4Q^2}{M_\phi^2}. \label{ltratio}
\eeq
A more recent and more elaborate parametrization of $R$ from the HERA data can be 
found in \cite{Cuic:2023mki}.  
(\ref{clas}) features the usual scaling relation $d\sigma_L\sim 1/Q^4$  (\ref{usu}) at high-$Q^2$ as well as a Pomeron-like behavior (gradual increase with $W$) at high-$W$. These constraints mainly come from the HERA data at high-$W$. On the other hand, at low-$W$, the fit is poorly constrained due to the scarcity of  electroproduction data.  In particular, the CLAS data \cite{CLAS:2008cms}  are limited to $Q^2<3.8$ GeV$^2$.  

The right plot of Fig.~\ref{wdep}  suggests that our formula can be used  at $Q\gtrsim 2.4$ GeV. Below this scale, it appears necessary to tame the singular behavior $1/Q^2$ in the perturbative kernel. 
The left plot provides another argument that indeed the threshold approximation is limited to $W\lesssim 3$ GeV.  In order to implement this transition in our  simulation, we evaluate the cross section in a given point in phase space by the smaller of the two  
\beq
{\rm min}\{ \sigma^{\rm CLAS}, \sigma^{\rm pQCD}\}. \label{min}
\eeq
More precisely, to facilitate the implementation we have parametrized the NLO cross section in the form\footnote{We set $D_g=-1$, $D_s=0$, $Q>2.5$ GeV and $\mu=Q$ in this parametrization.  One may think of more complicated parametrizations in which $W$- and $Q$-dependencies are not factorized.  However, we think the simple form (\ref{fit}) is enough for the present demonstrative purpose. }
\beq
\sigma^{\rm pQCD}_L(W,Q)=\frac{140}{Q^{9.4}}(W^2-1.96^2)^{0.71}W^{3.8} .\label{fit}
\eeq
Notice the very different $W$- and $Q$-dependencies from (\ref{clas}) as already discussed. 
We then apply the same $R$-factor (\ref{ltratio}) to (\ref{fit}) to obtain a model for $\sigma_T^{\rm pQCD}$ and hence also the sum (\ref{totalcross}). We further  assume the following $t$-dependence motivated by the dipole form factor  (\ref{ad})
\beq
\frac{d\sigma_{T/L}}{dt}= \frac{3(m_g^2-t_{min})^3}{(m^2_g-t)^4}\sigma_{T/L}, 
\label{FormFactor}
\eeq
where $m_g=1$ GeV in the CLAS model \cite{CLAS12Phi} and $m_g=m_A=1.6$ GeV (\ref{lama}) in our model. In this way, we evaluate (\ref{min})  at the level of the differential cross section $\frac{d\sigma}{dt}$ and generate events according to this probability.

In practice, an actual experiment will measure the $e+p$ cross section within a certain kinematic range and then extract the longitudinal cross section using the formula (\ref{totalcross}). While one may use a model for $R$ as in our simulations,  $R$ can be experimentally determined if the spin-density matrix elements (SDMEs) of the $\phi$ are measured.\footnote{$R$ may also be measured via the Rosenbluth separation technique, wherein cross sections are measured at the same kinematics ($x_B,Q^2,t$) but different values of $\varepsilon$. In practice, $\varepsilon$ is typically varied by altering the beam energy. Measurements of this kind are well suited to high-luminosity spectrometer experiments where the kinematics are known precisely and point-to-point uncertainties are small. A measurement of $R$ in exclusive $\phi$ electroproduction using the Rosenbluth method would provide a valuable cross-check of the SDME method. However, no Rosenbluth separation of exclusive $\phi$ production has been performed to date.} Assuming $s$-channel helicity conservation (SCHC), the value of $R$ is given as
\beq
R = \frac{T_{00}^{2}}{T^2_{11}} = \frac{1}{\varepsilon} \frac{r^{04}_{00}}{1-r^{04}_{00}}, \label{h1}
\eeq
in the Schilling and Wolf convention~\cite{Schilling:1973ag}, where $T_{00}$ and $T_{11}$ are the helicity amplitudes for a longitudinally polarized photon to transition to a longitudinally polarized vector meson and a transversely polarized photon to transition to a transversely polarized vector meson, respectively. $r^{04}_{00}$ is the dominant spin-density matrix element describing the longitudinal photon to longitudinal vector meson transition and is measurable from the angular distribution of the $\phi$ decay products. In light of the violation of SCHC observed by some experiments, the H1 experiment \cite{H1:2009cml} employed a more rigorous formula than (\ref{h1})  and measured $R$ in bins of $Q^2$ with a precision of approximately 10\% at values of $Q^2$ similar to those relevant here. In principle, the value of $R$ can be  determined  for each bin in $x_B,Q^2,W,$ and $t$ independently if the statistical and systematic uncertainties in the bin are small enough to permit measurement of SDMEs. This should be possible with SoLID at JLab and the EIC at BNL, the future large-acceptance experiments that will be able to detect the $\phi$ decay products. For this reason, in the following subsections we provide projections for the measurement of the longitudinal cross section at SoLID and the EIC.

\subsection{EIC}
The major challenge for measuring near-threshold meson production at a collider arises from the fact that the produced meson and the scattered proton must be relatively close in momentum and angle. This means that for events with $W\approx W_{th.}$, the meson decay products are generally lost down the beampipe. This limited the H1 and ZEUS measurements of exclusive $\phi$ production to $W>35$ GeV. However, the flexibility of the EIC in terms of beam energies and the large detector acceptances provide the first opportunity to measure near-threshold meson production at an $e+p$ collider. 

The ePIC experiment will be operational at the beginning of EIC running and located at the 6 o'clock interaction point. ePIC offers excellent acceptance and capability for measuring the momentum and species of long-lived particles such as pions, kaons, and protons. To make projections for measurements of exclusive $\phi$ electroproduction we will assume acceptances, resolutions, and particle identification abilities similar to what should be achieved by ePIC, summarized in Table~\ref{tab:detector_eff}~\cite{ePICTDR}. However, it is worth noting that a second EIC detector, coming online after ePIC and the first phase of EIC running,  could offer improved capabilities for these measurements. The configuration of the EIC which collides 5 GeV electrons with 41 GeV protons (5x41)\footnote{In the following, the notation $A$x$B$ will refer to collisions of electrons with momentum $A$ 
with protons of momentum $B$.} offers the best experimental access to the near-threshold region because the charged kaons produced in the decay of the $\phi$ enter the central detector region of $-3.5<\eta<3.5$. These $\phi$ decay kaons predominantly populate the forward region of the detector, where ePIC is equipped with a dual-radiator Ring-Imaging Cherenkov (dRICH) detector capable of efficiently separating pions from kaons up to approximately 50 GeV~\cite{Chatterjee:2024zrn}. For the purposes of this study, we assume that the dRICH and time-of-flight system can provide perfect identification of kaons from $\phi$ decay, which have typical momenta of 5 GeV. 

\begin{figure}[h]
    \centering
    \includegraphics[
        width=0.8\textwidth,
        trim=0 360 0 23,
        clip
    ]{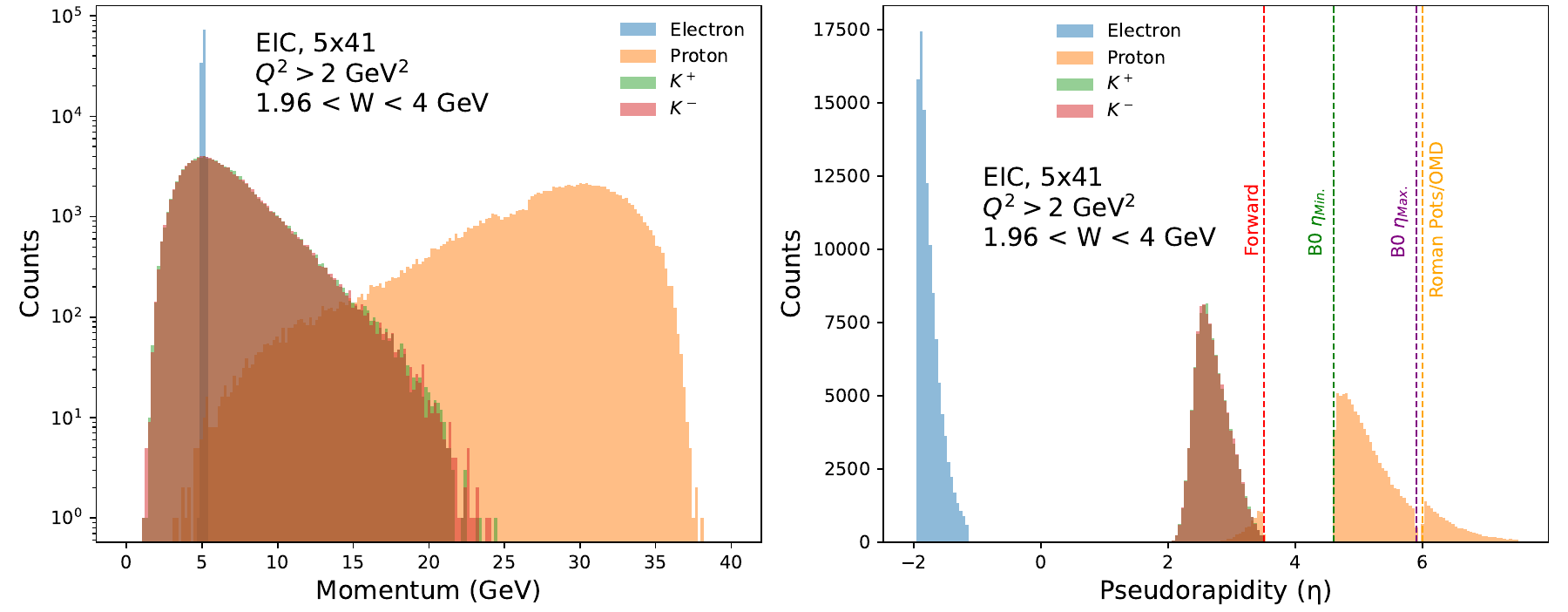}
    \caption{Momentum (Left) and pseudorapidity (Right) distributions for particles in events with $1.96 < W < 4$ GeV and $Q^2>2$ GeV$^2$ where all four final-state particles are reconstructed. }
    \label{EICAcceptance}
\end{figure}

The momentum spectra of all four final-state particles in reconstructed near-threshold events are shown in Fig.~\ref{EICAcceptance}. The scattered electron and the kaons from the $\phi$ decay are required to be within the central detector acceptance $|\eta| < 3.5$. In near-threshold events, the scattered proton is most often measured in the B0 detector system, which spans approximately $4.6<\eta<5.9$. However, the azimuthal angle acceptance of the B0 is not 100\% for all polar angles. To compensate for the missing azimuthal acceptance, the assumed detection efficiency of the B0 detectors is set to 70\%. Finally, at the smallest scattering angles, the proton can be detected in the Roman Pots (RP) or Off-momentum Detectors (OMD), which reside inside the beam vacuum. Since these detectors rely on particles passing through the accelerator magnet optics along certain trajectories, they do not have acceptance for all momenta. To crudely simulate this effect, we include a longitudinal momentum dependent acceptance of $0.65 < p_z/p_{\mathrm{beam}} < 1$ and $0.3 < p_z/E_\mathrm{beam} < 0.6$. There will also be a dependence of the acceptance on $p_T$, which we make no attempt to capture here. We also note that the above numbers hold for the 18x275 magnet lattice, but the acceptance will likely change in the 5x41 magnet lattice. A second detector at the EIC will have a still-different lattice and may make use of a second focus point in the accelerator magnets to significantly enhance the RP/OMD acceptance. The present assumption of 95\% efficiency in the RP/OMD is optimistic. However, for the range of $W$ that we focus on, namely $2.4 < W < 2.8$, less than 10\% of scattered protons are produced at $\eta > 5.9$.  We therefore expect our results to hold regardless of the details of the RP/OMD acceptance. 

The momentum resolutions of the central, B0, and RP/OMD sections of the detector are in line with what can be expected from ePIC. The angular resolutions are somewhat less studied than the momentum resolutions. We posit that the angular resolution of the B0 and RP/OMD detectors, all of which are equipped with high resolution tracking detectors, will be somewhat better than the central detector due to the significantly longer distance that particles travel before interacting with the far-forward detectors. The number of 3 mrad is perhaps conservative for the central detector as a whole, but as can be seen from Fig.~\ref{EICAcceptance}, most of the electrons and kaons do enter the central detector at relatively steep angles where the angular resolution is expected to deteriorate somewhat.

\begin{table}[ht]
\centering
\renewcommand{\arraystretch}{1.15}
\begin{tabular}{lccc}
\hline
\textbf{Detector Region} & \textbf{Efficiency} & \textbf{Momentum Resolution} & \textbf{Angular Resolution} \\
\hline
\textbf{Central $(|\eta|<3.5)$} & 95\% & 
$\displaystyle \frac{\Delta p}{p} =
\begin{cases}
1\%, & |\eta|<2.5\\
2.5\%, & 2.5<|\eta|<3.5\\
\end{cases}$ & 
$\displaystyle \sigma_{\theta,\phi} = 3\,\text{mrad\quad} $\\[1.2em]

\textbf{B0 $(4.6<\eta<5.9)$} & 70\% & 
$\displaystyle \frac{\Delta p_T}{p_T} = 6\%$ & 
$\displaystyle \sigma_{\theta,\phi} = 1\,\text{mrad\quad} $\\[0.8em]

\textbf{RP/OMD $(\eta>6.0)$} & 95\% & 
$\displaystyle \frac{\Delta p_T}{p_T} = 8\%$ & 
$\displaystyle \sigma_{\theta,\phi} = 1\,\text{mrad\quad} $\\[0.8em]
\hline
\end{tabular}
\caption{Efficiencies, momentum resolution, and angular resolution in different pseudorapidity regions. The combined acceptance of the Roman Pots and Off-momentum detectors has an additional requirement that $\theta<5\,\text{mrad}$ and $p_z/p_{\mathrm{beam}} \in [0.65,1.0]\cup[0.3,0.6]$.}
\label{tab:detector_eff}
\end{table}

\begin{figure}[t]
\begin{center}
\includegraphics[width=0.45\textwidth]{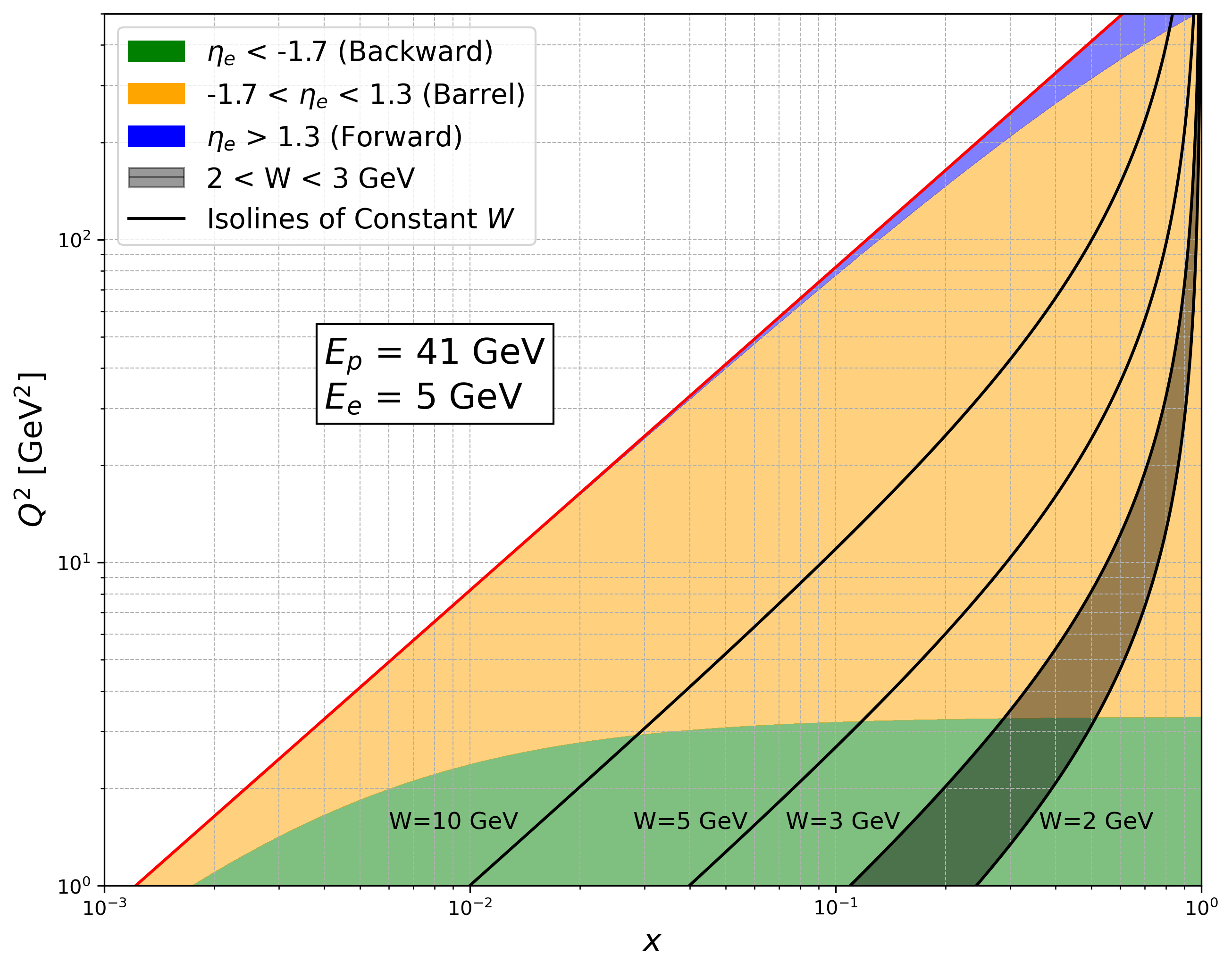}
\includegraphics[width=0.45\textwidth]{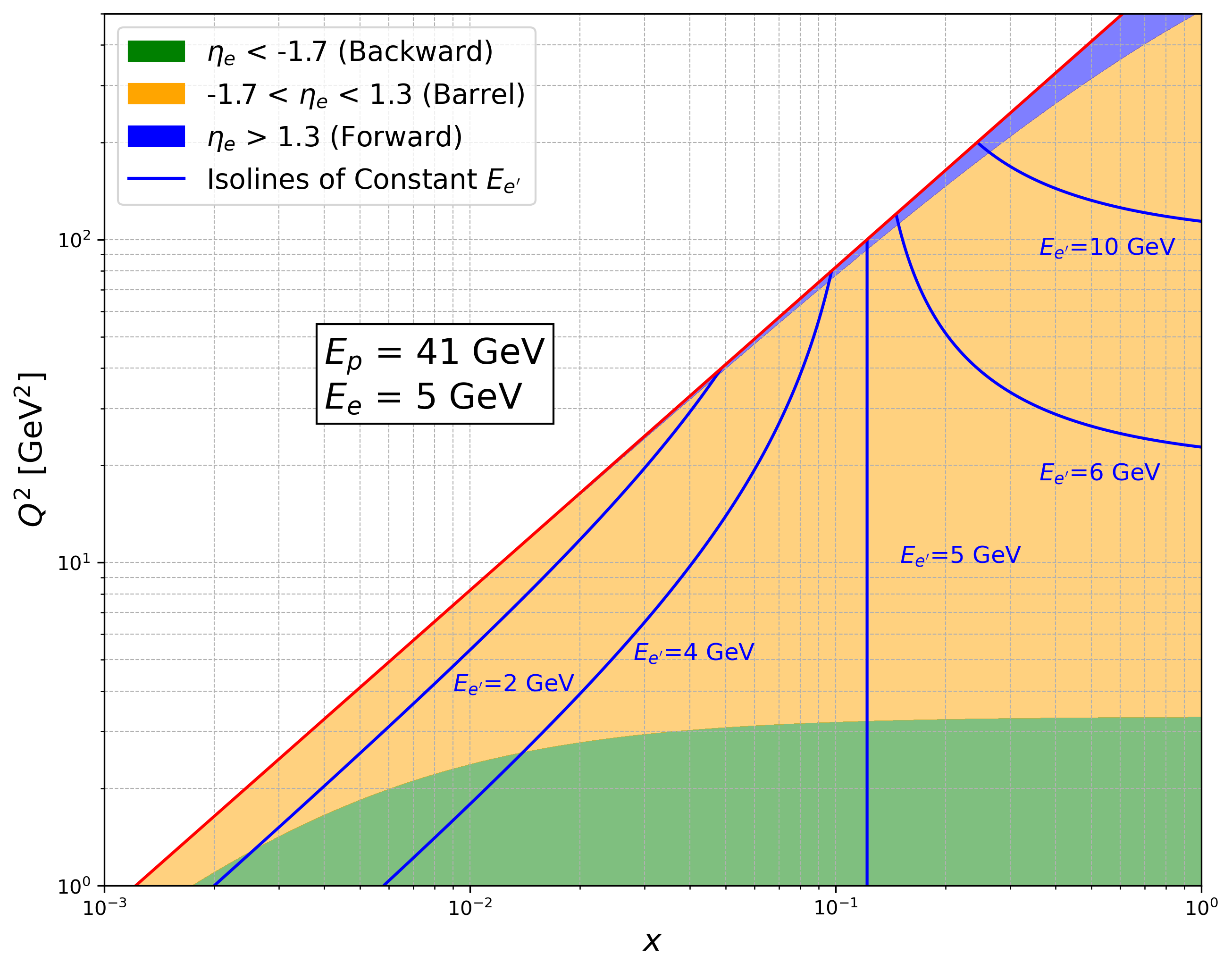}
\end{center} 
\caption[*]{$x_B, Q^2$ plane for the 5x41 beam energy configuration of the EIC. Left: Kinematic plane with iso-lines of constant $W$. The region of interest, $W<3$ GeV, is shaded in gray. Right: Kinematic plane with iso-lines of constant scattered electron energy $E_{e'}$.}
\label{EICKinematics}
\end{figure}

A key challenge in reconstructing near-threshold events at the EIC is accurately measuring the event kinematics. Relying solely on the scattered electron yields poor resolution at low $W$ because small mis-measurements of the electron energy can lead to large uncertainties in the reconstructed kinematic variables, as can be seen from the right panel of Fig.~\ref{EICKinematics}. In contrast, reconstructing the entire hadronic final state—the approach assumed in these projections—significantly improves the $W$ resolution from about 50\% down to a few percent. When all final-state particles are measured, the full suite of inclusive kinematic reconstruction techniques (see Ref.~\cite{Bassler:1994uq} for a succinct review) can be used to reconstruct $W$ via $Q^2$ and $x_B$  to great effect. Beyond these standard methods, $W$ can also be reconstructed as the invariant mass of the hadronic final state, potentially achieving even better resolution. Furthermore, the exclusivity of the events provides additional constraints that can be leveraged through kinematic fitting or ML-based approaches, leading to more precise event reconstruction overall.

The current projection for the EIC is that the integrated luminosity per year will be 5 fb$^{-1}$ for the 5x41 beam energy configuration. We assume in our projections a total integrated luminosity of 10 fb$^{-1}$, corresponding to approximately two years of 5x41 running. The 10x100 beam energy configuration offers a factor of 10 higher instantaneous luminosity, making it an attractive option. However, for the 10x100 beam energy, in the near-threshold kinematics the produced kaons almost exclusively fall in the pseudorapidity region $3.5 < \eta < 4.6$. In ePIC this pseudorapidity range is occupied primarily by the beampipe and is not instrumented with tracking and paricle identification. A second EIC detector with forward tracking and PID acceptance out to $\eta=4$ or greater could recover a significant fraction of these events and perhaps do a more precise measurement at the 10x100 beam energy.

Using the acceptances provided in Table~\ref{tab:detector_eff} for the 5x41 configuration, the acceptance for events with $W<3$ GeV to be reconstructed with all four final-state particles is around 4\%. The overall acceptance is primarily hindered by the acceptance for the scattered proton. We assume in our results a 10\% systematic uncertainty, likely to arise from the understanding of the acceptance. This may be an underestimate, but this uncertainty will necessarily depend on the final implementation and performance of the subdetectors. An enhancement in the event statistics can be gained by not requiring the proton to be detected and instead using only the electron and kaons to reconstruct the event, but this comes at the cost of higher backgrounds and worsened resolution.

\begin{figure}[h]
    \centering
    
    \begin{subfigure}[b]{0.49\textwidth}
        \centering
        \includegraphics[width=\textwidth]{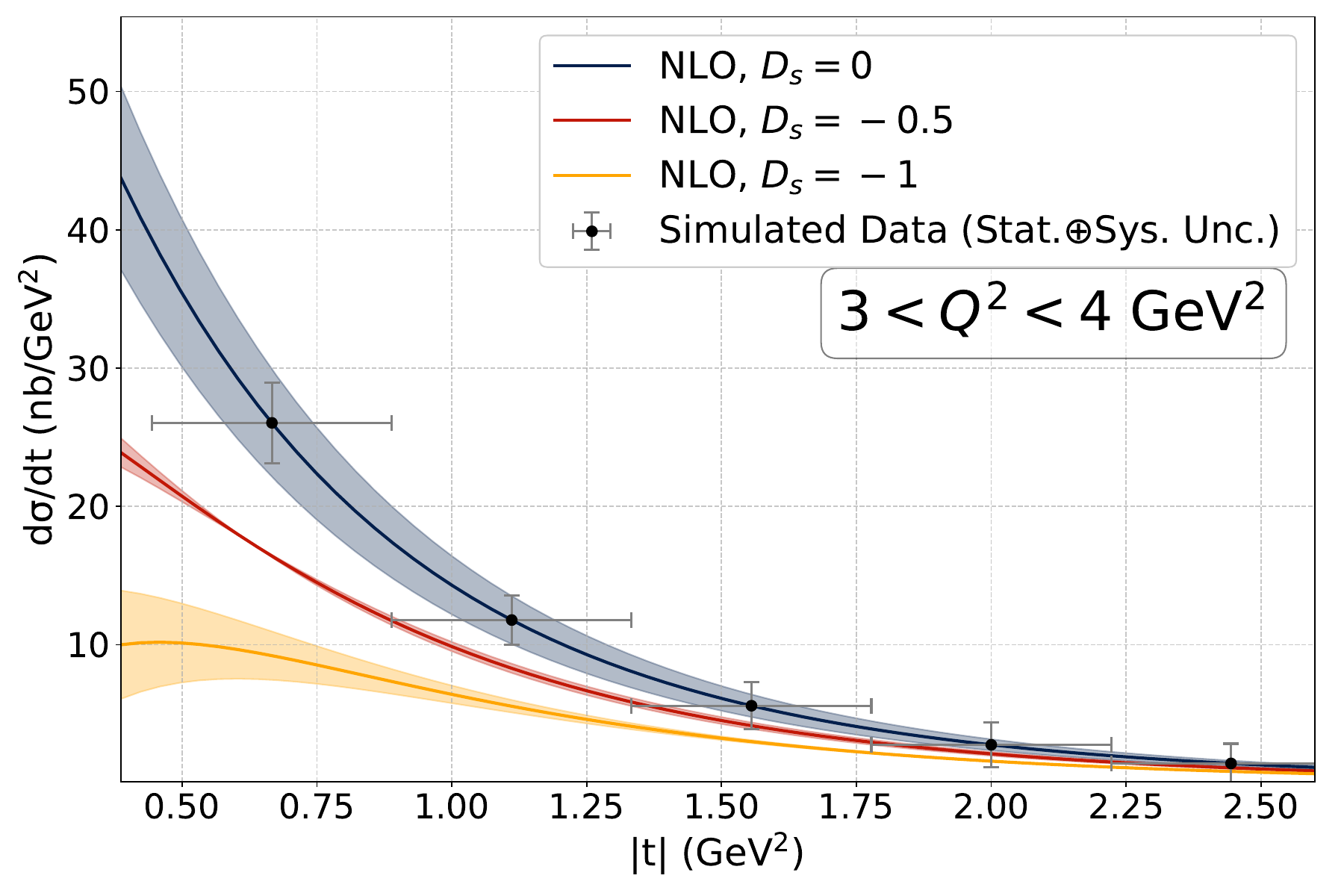}
        \label{fig:plot1}
    \end{subfigure}
    \hspace{0.0\textwidth} 
    \begin{subfigure}[b]{0.49\textwidth}
        \centering
        \includegraphics[width=\textwidth]{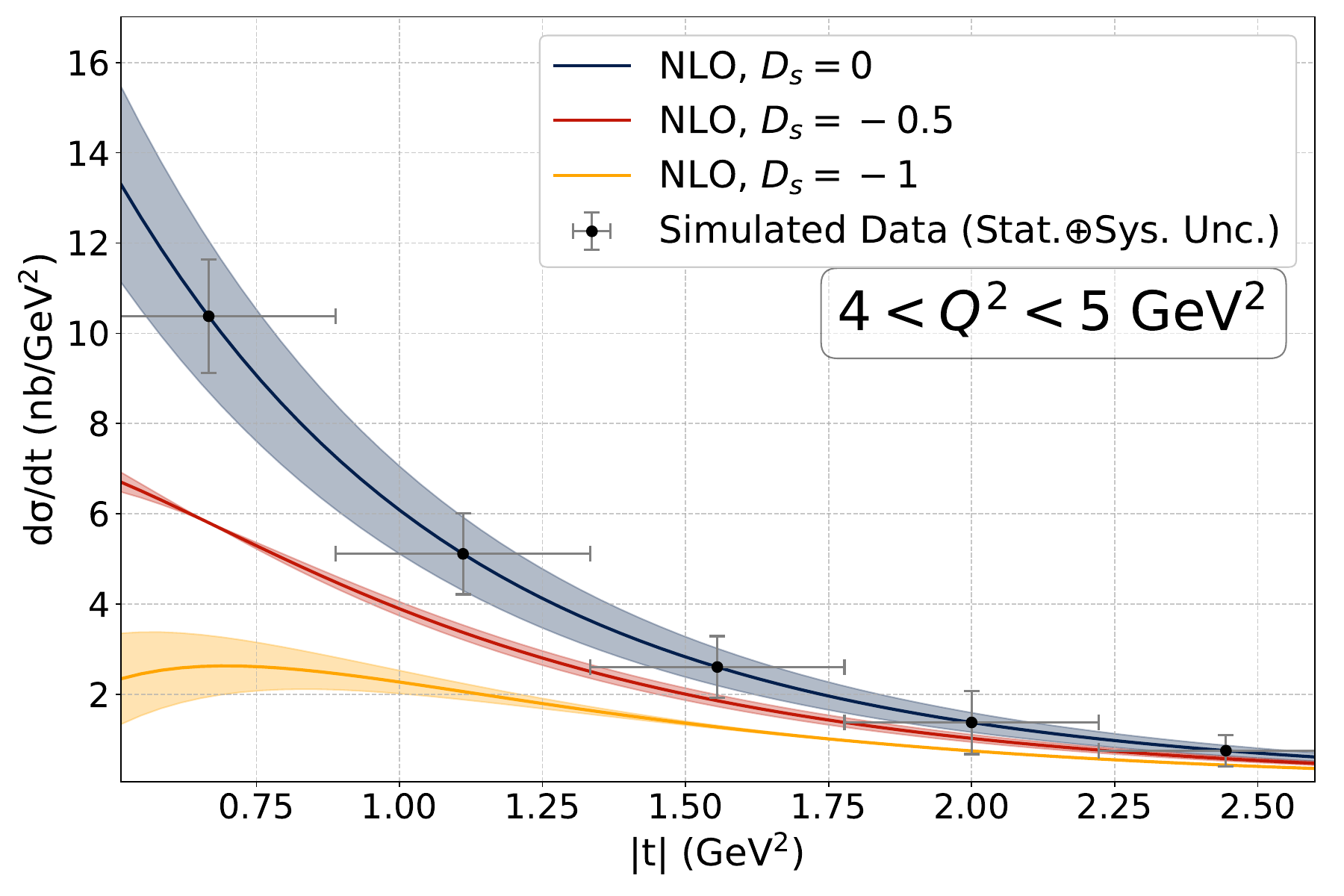}
        \label{fig:plot2}
    \end{subfigure}
    
    \vspace{1em} 
    
    \begin{subfigure}[b]{0.49\textwidth}
        \centering
        \includegraphics[width=\textwidth]{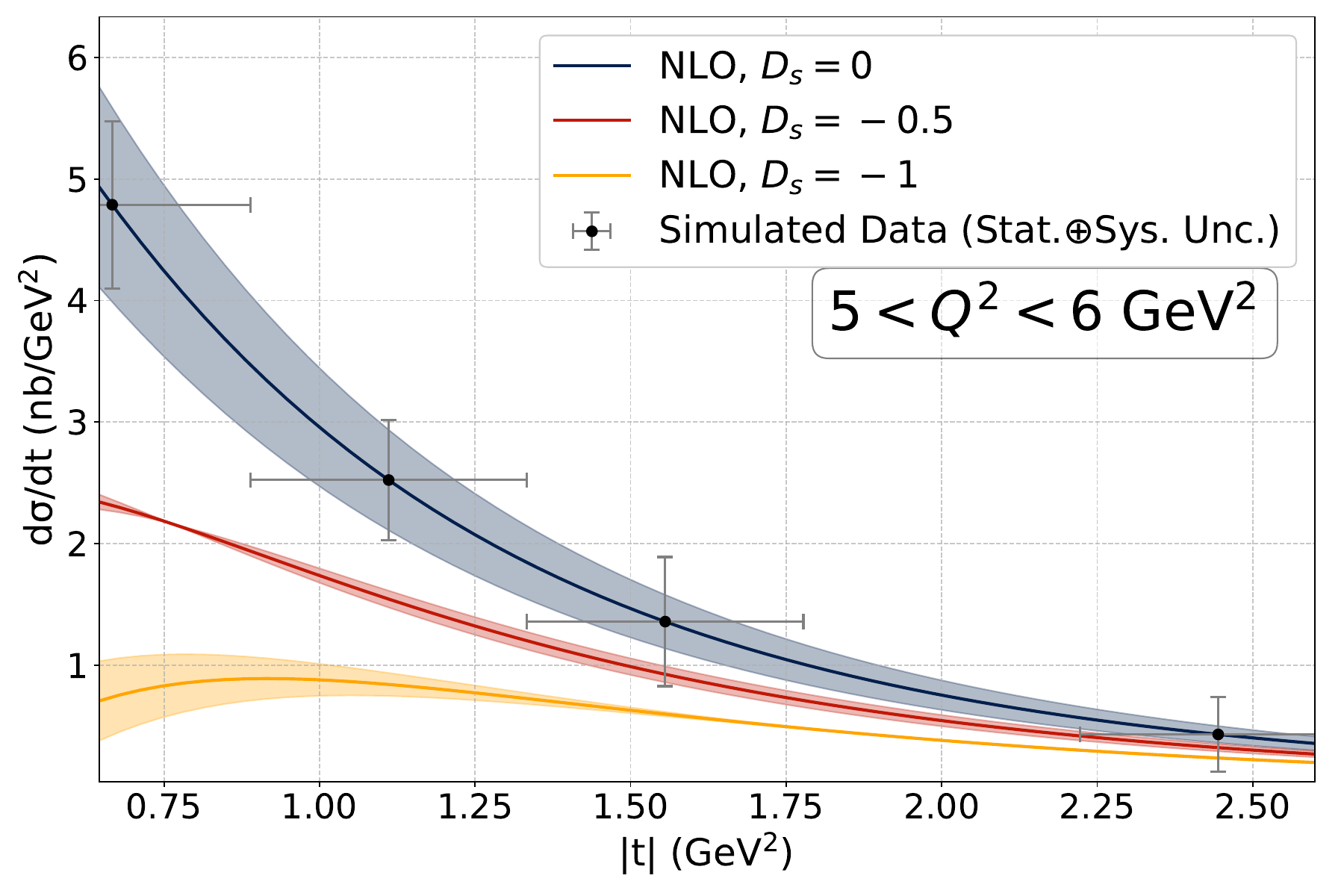}
        \label{fig:plot3}
    \end{subfigure}
    \hspace{0.0\textwidth} 
    \begin{subfigure}[b]{0.49\textwidth}
        \centering
        \includegraphics[width=\textwidth]{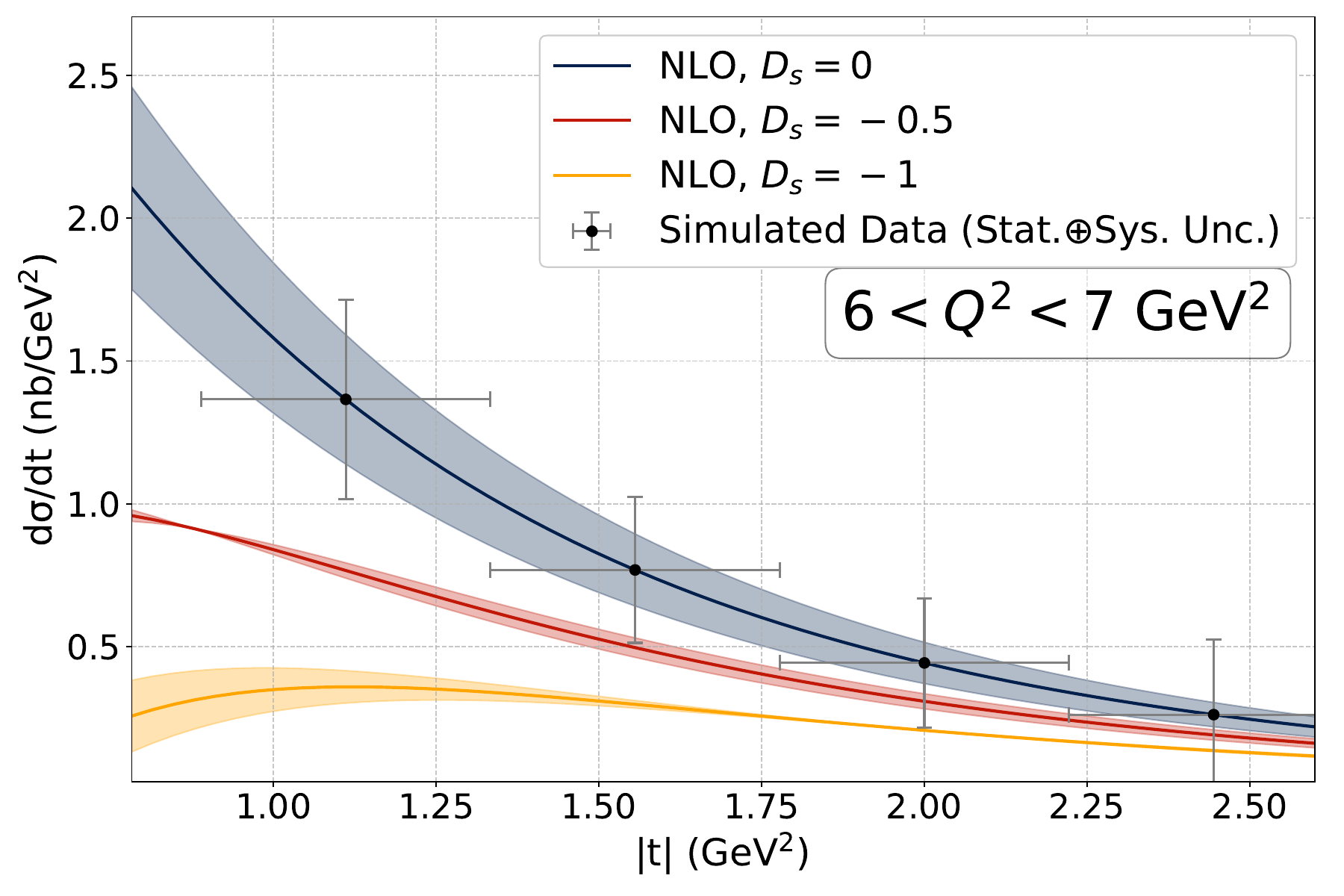}
        \label{fig:plot4}
    \end{subfigure}
    \vspace{1em}
    \hspace{0.0\textwidth} 
    \begin{subfigure}[b]{0.49\textwidth}
        \centering
        \includegraphics[width=\textwidth]{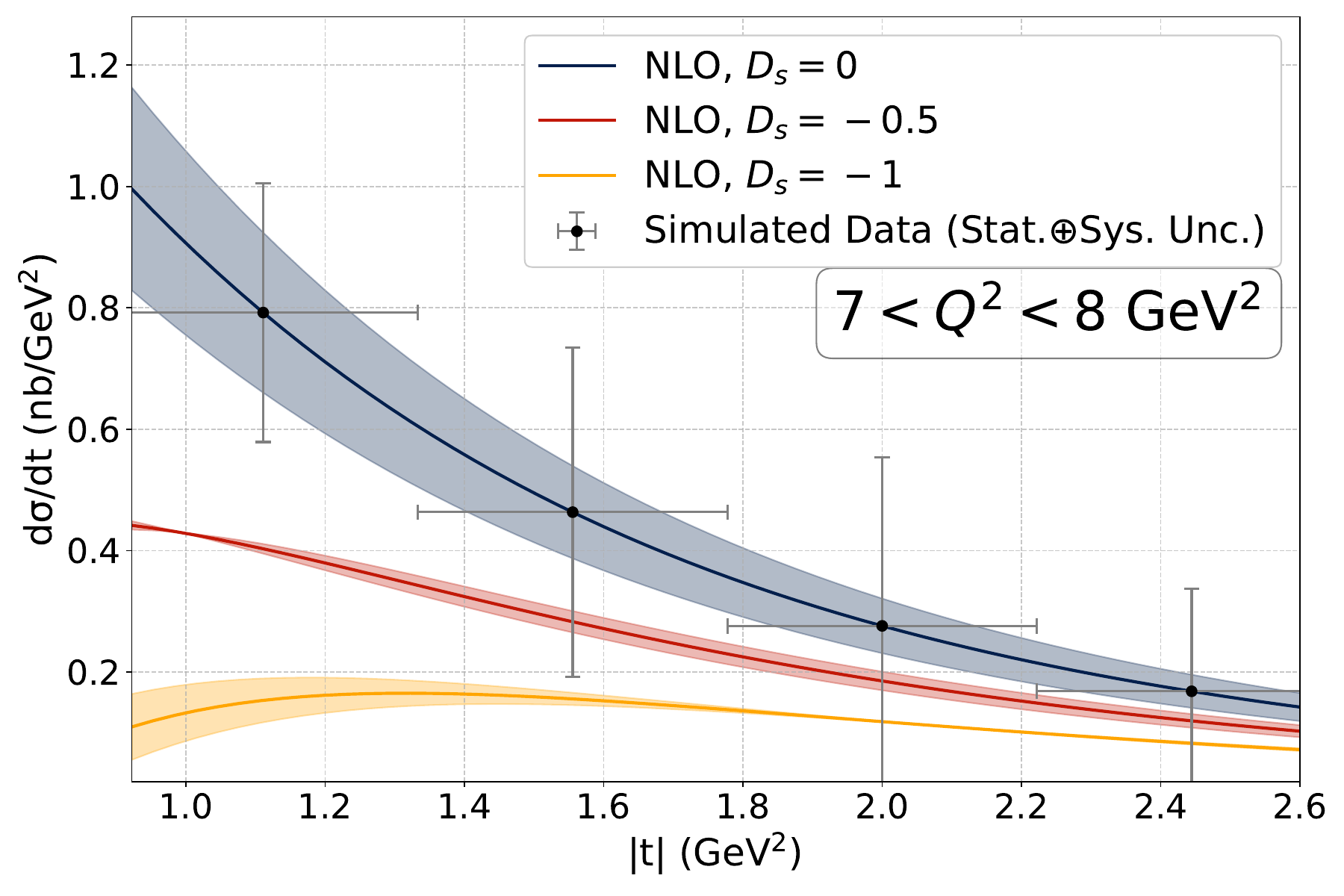}
        \label{fig:plot5}
    \end{subfigure}
    
    \caption{Projected results at the EIC for the cross section $\gamma^*_{L}+p \rightarrow \phi+p$ differential in $|t|$ and $Q^2$ in the near-threshold regime of $2.4 < W < 2.8$. The data points are fixed to the central value of the theoretical curve. Theoretical uncertainties are determined via a scale variation in the same way as Fig.~\ref{band}. The assumed integrated luminosity is 10 fb$^{-1}$ at the 5x41 beam energy. }
    \label{fig:EICResults}
\end{figure}

The projected sensitivity to $d\sigma_L/d|t|$ at the EIC is shown in Fig.~\ref{fig:EICResults}. For simplicity, the NLO theory curves are generated at fixed values of $W$ and $Q^2$ corresponding to the center of the bin, i.e. $W=2.6$ GeV and $Q^2=3.5$ GeV$^2$ for the lowest $Q^2$ bin. The reconstructed fully exclusive event yield is around 1500 in 10 fb$^{-1}$ for $2.4 < W < 2.8$ GeV and $Q^2>3$ GeV$^2$. In all bins, the statistical uncertainty is larger than the assumed 10\% systematic uncertainty.  We present our results only in the range $2.4 < W < 2.8$ GeV because this range provides the best discriminating power for $D_s$. At lower $W$, the event statistics are too small to perform a reliable measurement of the $|t|$-distribution. At larger $W$, despite the somewhat higher measured event statistics, the sensitivity to $D_s$ is diminished.  Besides, as explained before, our calculation is not reliable for $W>3$ GeV.  
Therefore we predict that $2.4 < W < 2.8$ GeV will be an optimal range for this measurement if the B0 detector meets the performance presented in Table~\ref{tab:detector_eff} and the acceptance of the forward particle identification is not significantly reduced compared to the nominal $\eta<3.5$ boundary. 

In summary, we present the first estimates for measuring near-threshold electroproduction of $\phi$ mesons at the EIC. Our preliminary calculations, based on naive assumptions about detector acceptances and resolutions, indicate that such a measurement is feasible. However, because event statistics at high $Q^2$ are low, extracting $D_s$ with high precision is challenging using only the nominal EIC detector acceptances. This measurement can be incrementally improved by collecting more than 10 fb$^{-1}$ of integrated luminosity at the 5x41 GeV beam energy. However, to achieve substantial gains in precision, the forward and far-forward detector acceptances need to exceed what was assumed here. A second EIC detector at IP8, optimized for exclusive measurements and leveraging the IP8 second focus, with forward particle identification out to $\eta=4$ (approximately a $2\degree$ polar angle), could exploit the higher luminosities available at higher beam energies and enable a much more precise measurement.

\subsection{SoLID}
At JLab, the future SoLID program in Hall A provides the luminosity, particle identification, and acceptance needed to perform precise measurements of rare processes. An experiment to study electroproduction of $\phi$ mesons in SoLID could run concurrently with the already approved SoLID $J/\psi$ experiment. The $J/\psi$ experiment plans to take 50 days of physics running with an 11 GeV $e^-$ beam at a current of 3 $\mu$A on a 15cm long LH$_2$ target, producing an instantaneous luminosity of approximately 10$^{37}$ cm$^{-2}$/s and collecting an integrated luminosity of 43.2 ab$^{-1}$. To evaluate SoLID's capabilities for this measurement, we simulated 43.2 ab$^{-1}$ of $Q^2>2$ GeV$^2$ events using  \texttt{lAger} at a beam energy of 10.6 GeV. A comprehensive summary of the SoLID detector is provided in Ref.~\cite{JeffersonLabSoLID:2022iod}. 

\begin{figure}[h]
\begin{center}
\includegraphics[width=0.5\textwidth]{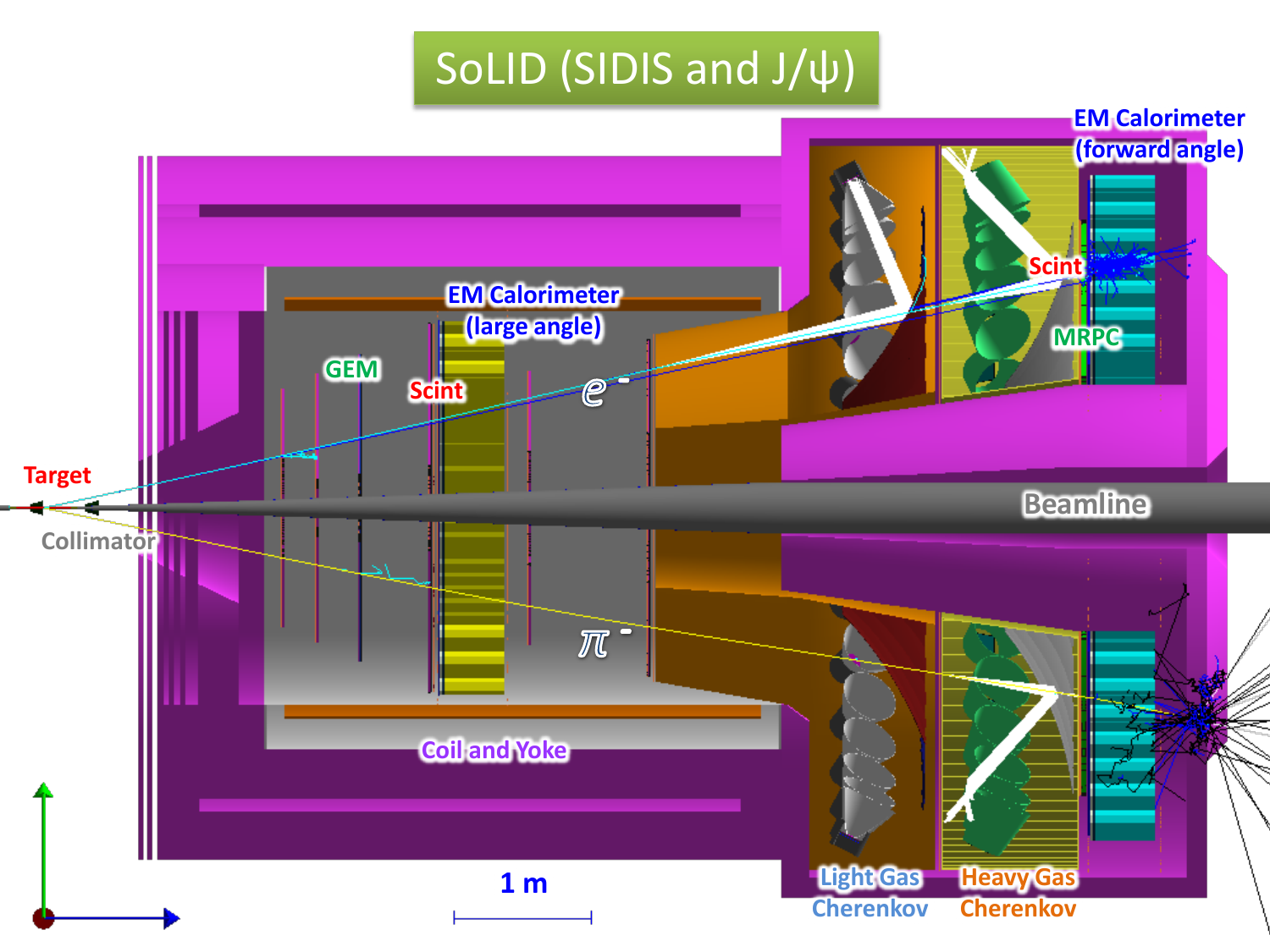}
\end{center} 
\caption[*]{The SoLID detector in the configuration planned for the $J/\psi$ and SIDIS experiments. Image reproduced from Ref.~\cite{JeffersonLabSoLID:2022iod}}
\label{SoLIDDetector}
\end{figure}

SoLID consists of two annular sections covering the whole azimuth, known as the forward-angle detector (FAD) and the large-angle detector (LAD). As shown in Fig.~\ref{SoLIDDetector}, the FAD is equipped with GEM tracking planes, a light gas Cherenkov detector for separation of electrons and pions, a heavy gas Cherenkov detector for separation of pions from kaons, a time-of-flight layer for identification of protons and low momentum hadrons, and an electromagnetic calorimeter. The LAD is equipped with tracking, a scintillator layer for separation of electrons from photons, and electromagnetic calorimetry. The FAD covers the region of polar angle from $8\degree$ to $15\degree$ and the LAD covers $15\degree$ to $24\degree$, where polar angle is defined with respect to the center of the target. 

The SoLID data acquisition system will be designed to handle a trigger rate of up to 100 kHz, motivated by the expected trigger rate for the SIDIS experiments. The trigger planned for the $J/\psi$ experiment is a triple coincidence of the scattered electron and an electron+positron pair from the decay of the $J/\psi$. The trigger rate for this topology is around 800 Hz, leaving ample room for additional triggers to run in parallel. Fully reconstructible near-threshold $\phi$ events occur on the order of 100 Hz, but the selection of these events at trigger-level is complicated by the large rate of random coincidences. Using a quadruple coincidence of an electron, identified via a high energy cluster in the large-angle calorimeter, and three charged hadrons, identified via hits in the scintillators and calorimeter, the trigger rate was roughly estimated to be around 100 kHz. A reduction of the trigger rate to acceptable levels without prescaling could be achieved if even rudimentary tracking information could be included in the trigger. Still further improvements could be gained by moving to a software-based trigger system. A software trigger permits real-time reconstruction, allowing advanced analysis cuts such as two-track invariant mass to be implemented in the trigger. The ability to place analysis cuts at trigger-level enables utilizing the powerful constraints imposed by the exclusivity of the process to completely eliminate random coincidences. This paradigm has been demonstrated at large-scale by the LHCb collaboration~\cite{LHCb:2018zdd} in a complex detector system not unlike SoLID. Such an upgrade to the data acquisition system would be highly beneficial for the study of exclusive $\phi$ production and the SoLID physics program overall, allowing SoLID to make full use of the unparalleled combination of luminosity and acceptance without the otherwise strict limitations imposed by random coincidence trigger rates. This upgrade route furthermore has synergies with the High Performance Data Facility (HPDF) led by Jefferson Lab. In the projections provided in this section, we assume an un-prescaled data acquisition system capable of reading out all $\phi$ events. All of this being said, a trigger prescale of a factor of two would not substantially endanger the measurement, as will be shown in the remainder of this section.

\begin{figure}[h]
\begin{center}
\includegraphics[width=0.45\textwidth]{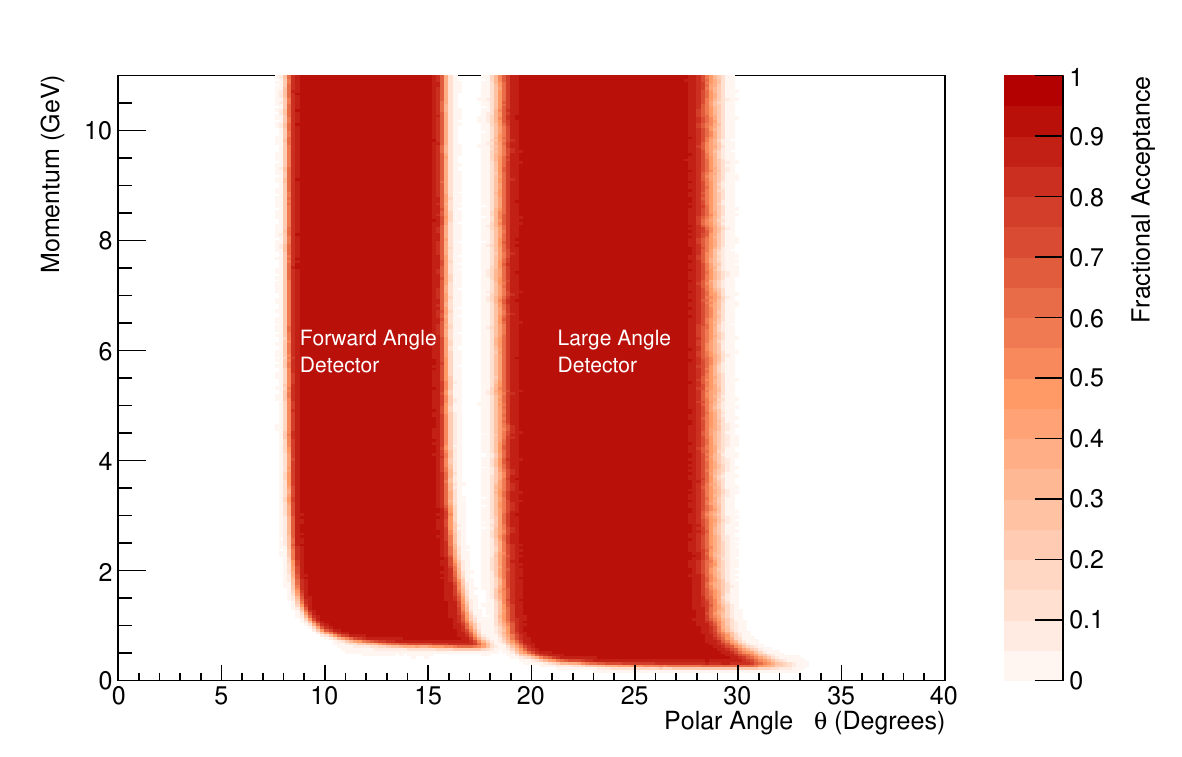}
\includegraphics[width=0.45\textwidth]{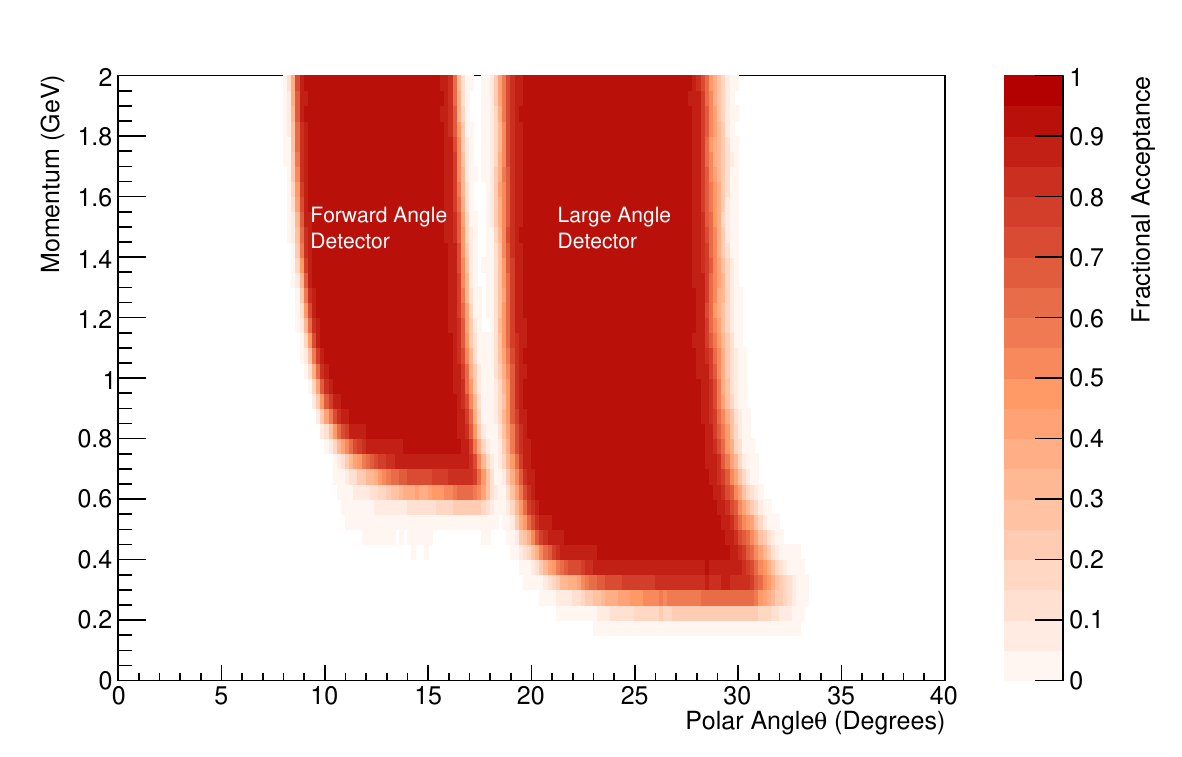}
\end{center} 
\caption[*]{The fractional acceptance of SoLID for charged particles in momentum and polar angle. Left: Full acceptance for particles of momenta up to 11 GeV. Right: Zoomed acceptance for particles in the momentum range particularly relevant for reactions at low values of $|t|$.}
\label{SoLIDAcceptance}
\end{figure}

\begin{figure}[h]
    \centering
    \includegraphics[
        width=0.8\textwidth,trim=0 360 0 23,clip
 ]{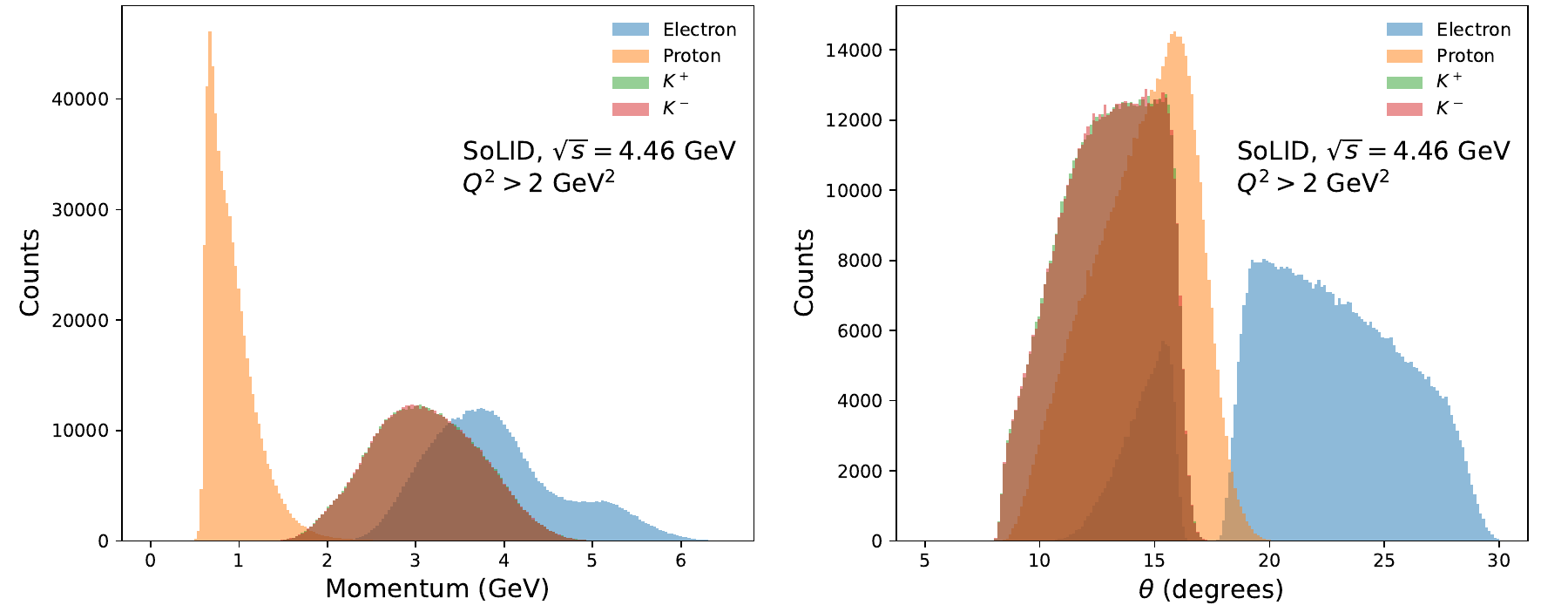}
    \caption{Momentum (Left) and polar angle (Right) distributions for particles in fully reconstructed events with $1.96 < W < 4$ GeV and $Q^2>2$ GeV$^2$. The gap between the forward-angle and large-angle detectors is visible around 16 degrees.}
    \label{SoLIDAcceptedParticles}
\end{figure}

 The resolutions and detector acceptances assumed in this study were determined by the SoLID collaboration with a dedicated GEANT4 simulation in the GEMC framework~\cite{Software:GEMC,GitHub:libsolgem,GitHub:solid-gemc,GitHub:SoLIDTracking,Software:Geant4}. We make use of these results in a fast MC to smear the four-vectors of generated particles in accordance with the expected resolutions. The single-particle acceptance of the SoLID $J/\psi$ configuration in momentum and polar angle $\theta$ is shown in Fig.~\ref{SoLIDAcceptance}. The distributions in momentum and polar angle for electrons, kaons, and protons in events where all four final-state particles are reconstructed are provided in Fig.~\ref{SoLIDAcceptedParticles}. The fast MC queries the acceptance map and randomly discards or accepts particles based on their momentum and polar angle in accordance with the expected acceptance. Compared to the acceptance map determined with the SoLID GEANT4 simulation, for the purposes of making realistic projections we scale the acceptance by a factor of 0.9 to simulate tracking inefficiencies in the high-rate environment at SoLID. We furthermore only consider events in which the proton and kaons enter the FAD, such that they can be efficiently identified by the particle identification detectors. We assume perfect identification of kaons and protons in the FAD, which should be a reasonable approximation since background pions can be rejected by the heavy gas Cherenkov and protons can be identified by the time-of-flight. 
 
 Since statistical uncertainty is less of a limitation at SoLID, the bin sizes in $|t|, W,$ and $Q^2$ were chosen to be a factor of 3 or more larger than the detector resolution on those quantities. This should permit a reasonably stable unfolding and bin-by-bin extraction of $R$. To approximate the effect of systematic uncertainties, we assume a uniform systematic uncertainty of 10\% on each bin of $|t|$.

\begin{figure}[ht]
    \centering

    \begin{minipage}[t]{0.23\textwidth}
    \includegraphics[width=\textwidth]{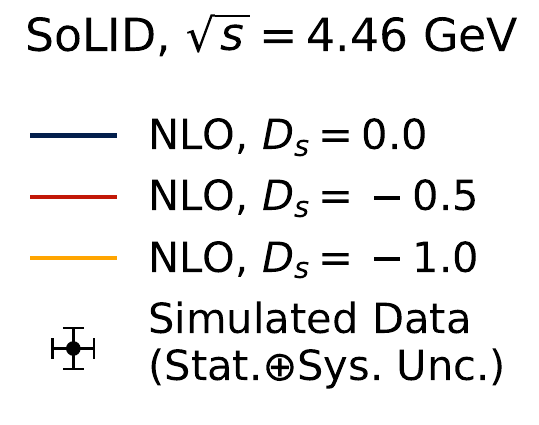}
      \includegraphics[width=\textwidth]{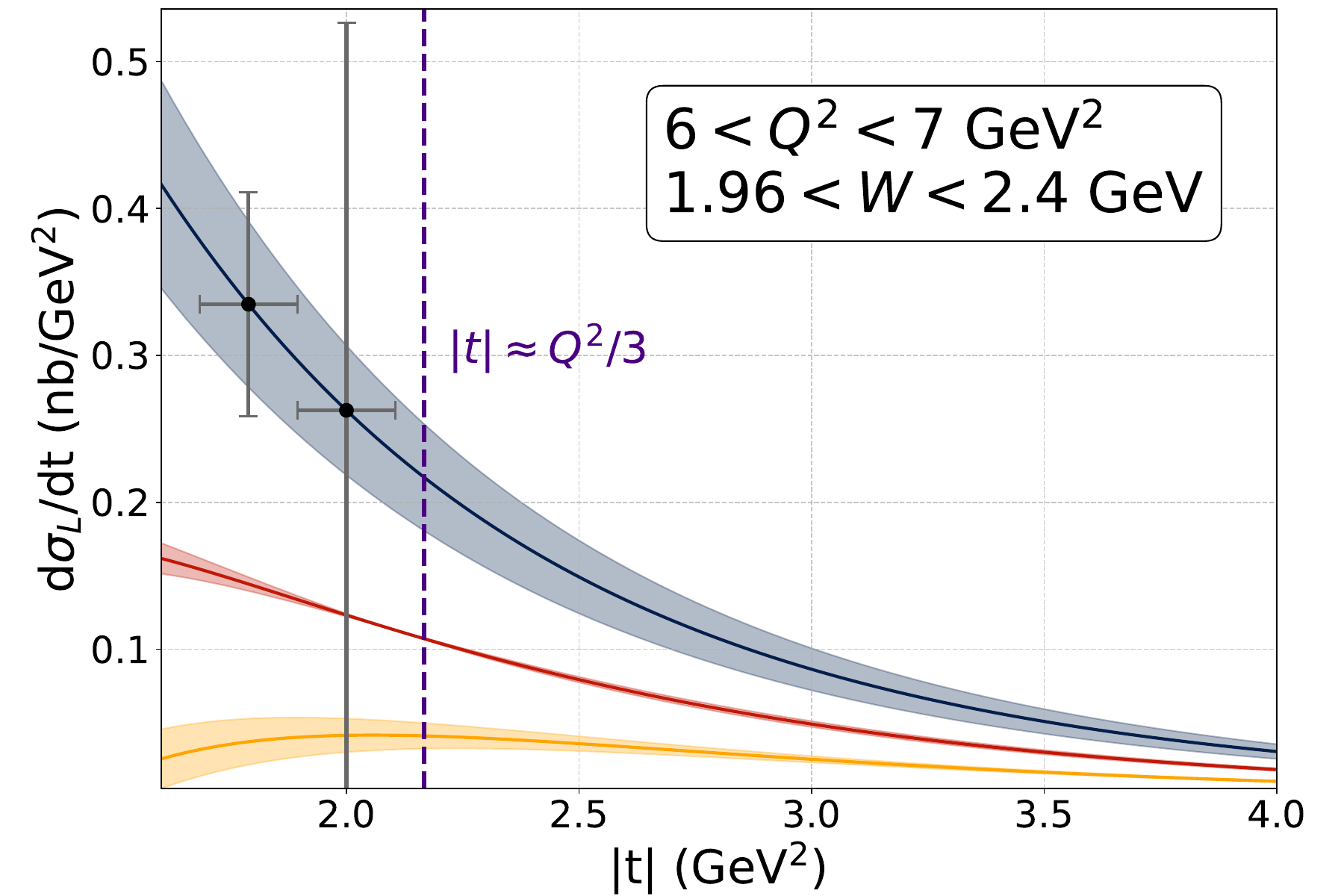}\\
      \includegraphics[width=\textwidth]{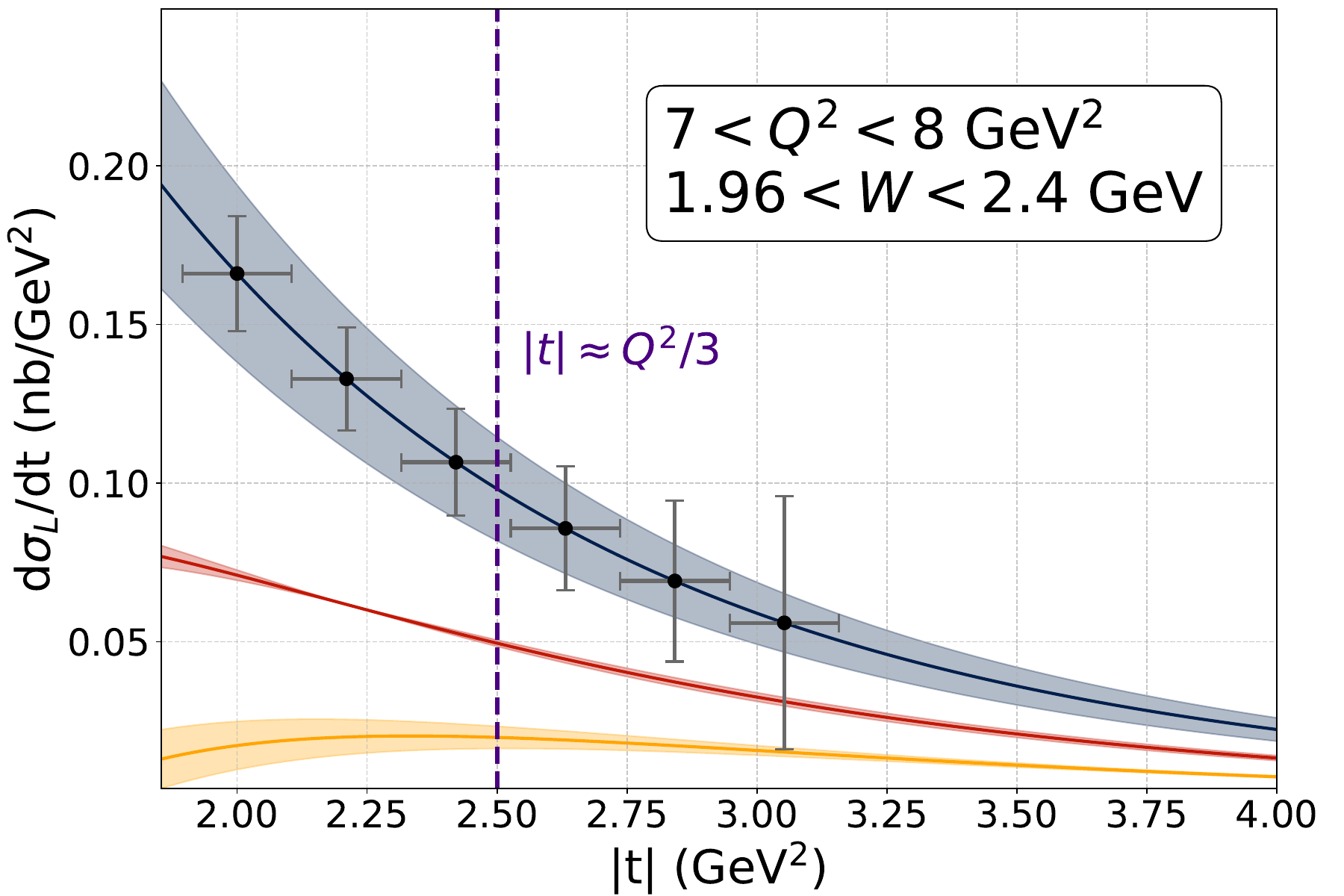}\\
      \includegraphics[width=\textwidth]{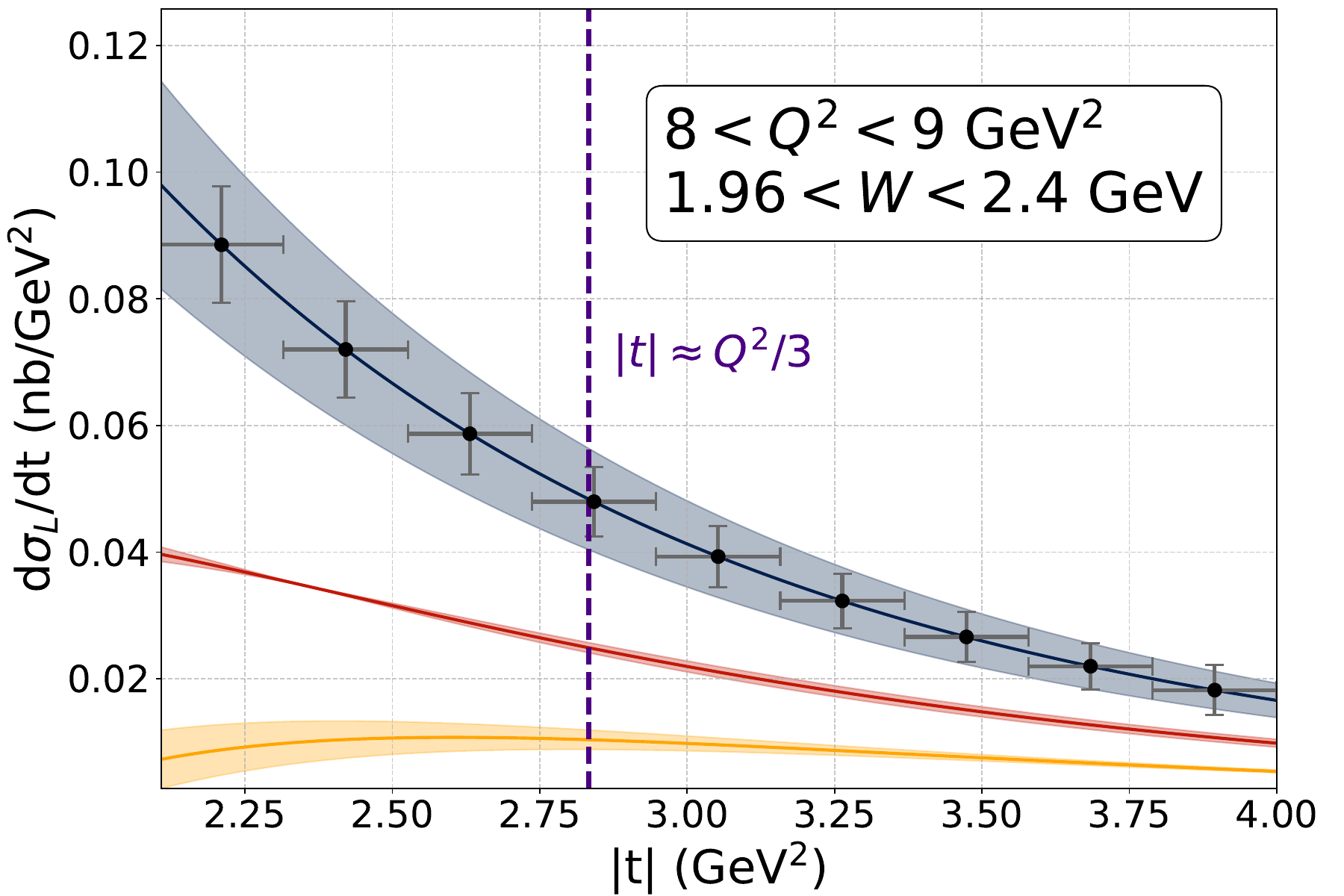}\\
      \includegraphics[width=\textwidth]{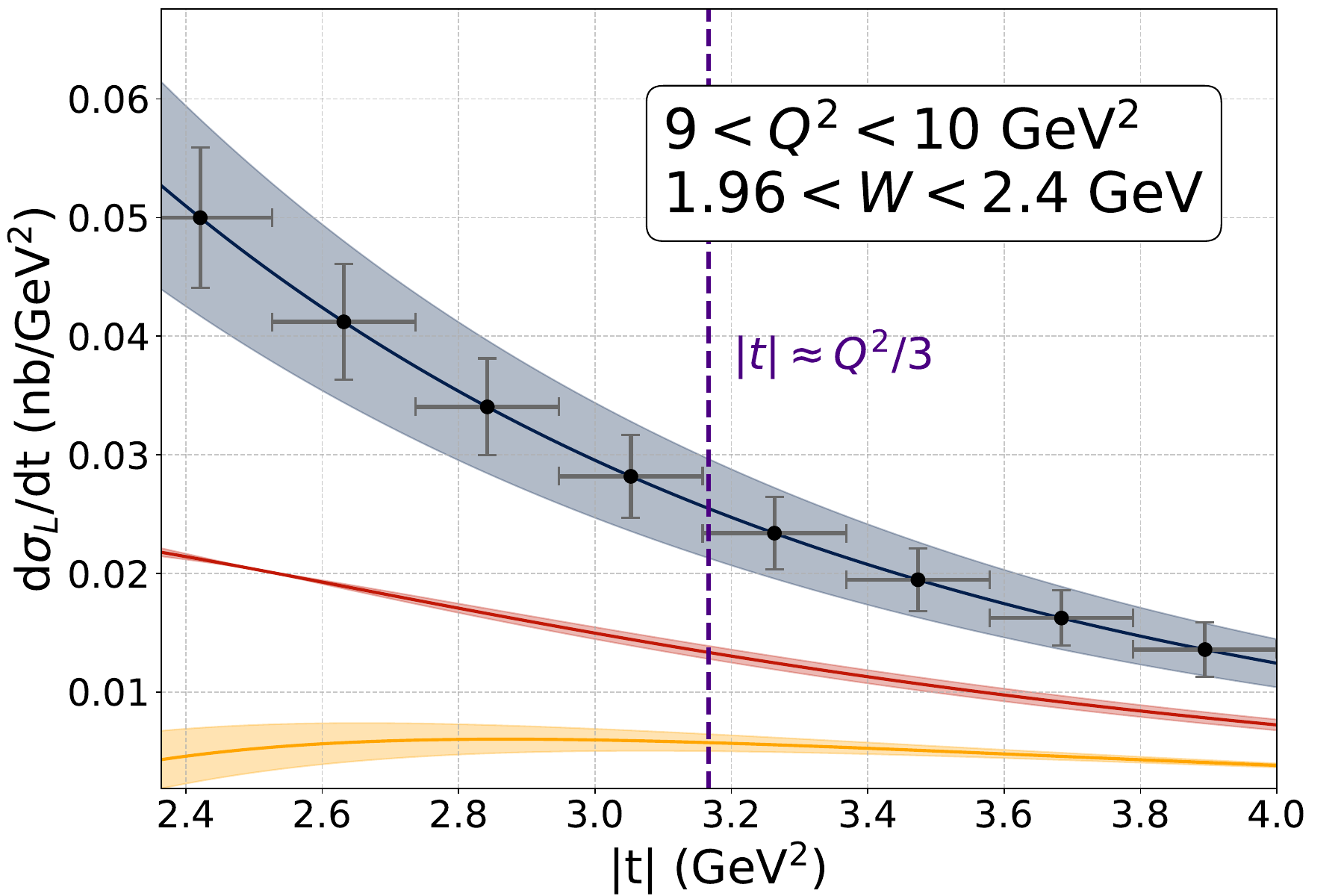}
      \par\phantomsection\label{fig:w218} 
    \end{minipage}
    \hfill
    \begin{minipage}[t]{0.23\textwidth}
      \includegraphics[width=\textwidth]{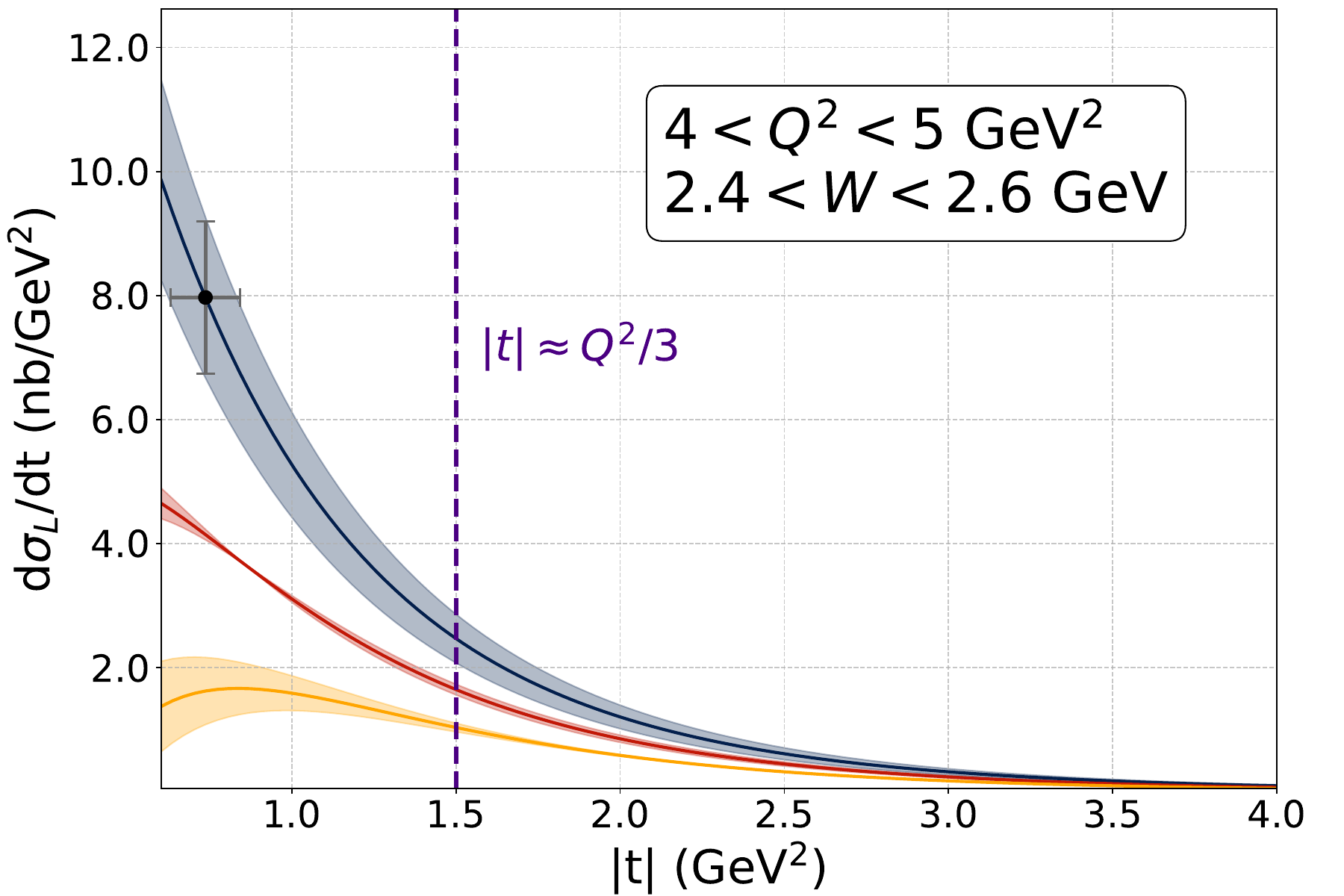}\\
      \includegraphics[width=\textwidth]{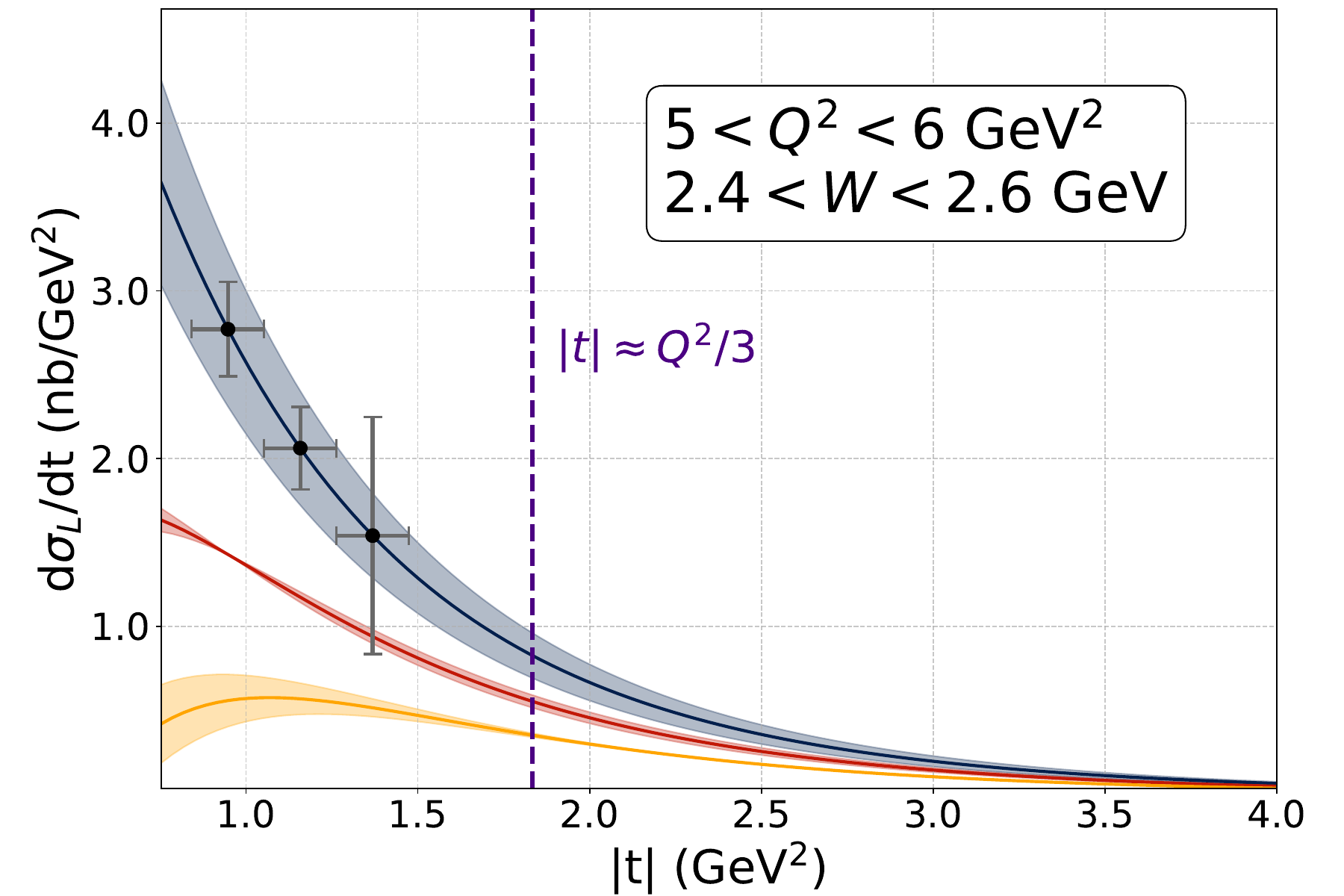}\\
      \includegraphics[width=\textwidth]{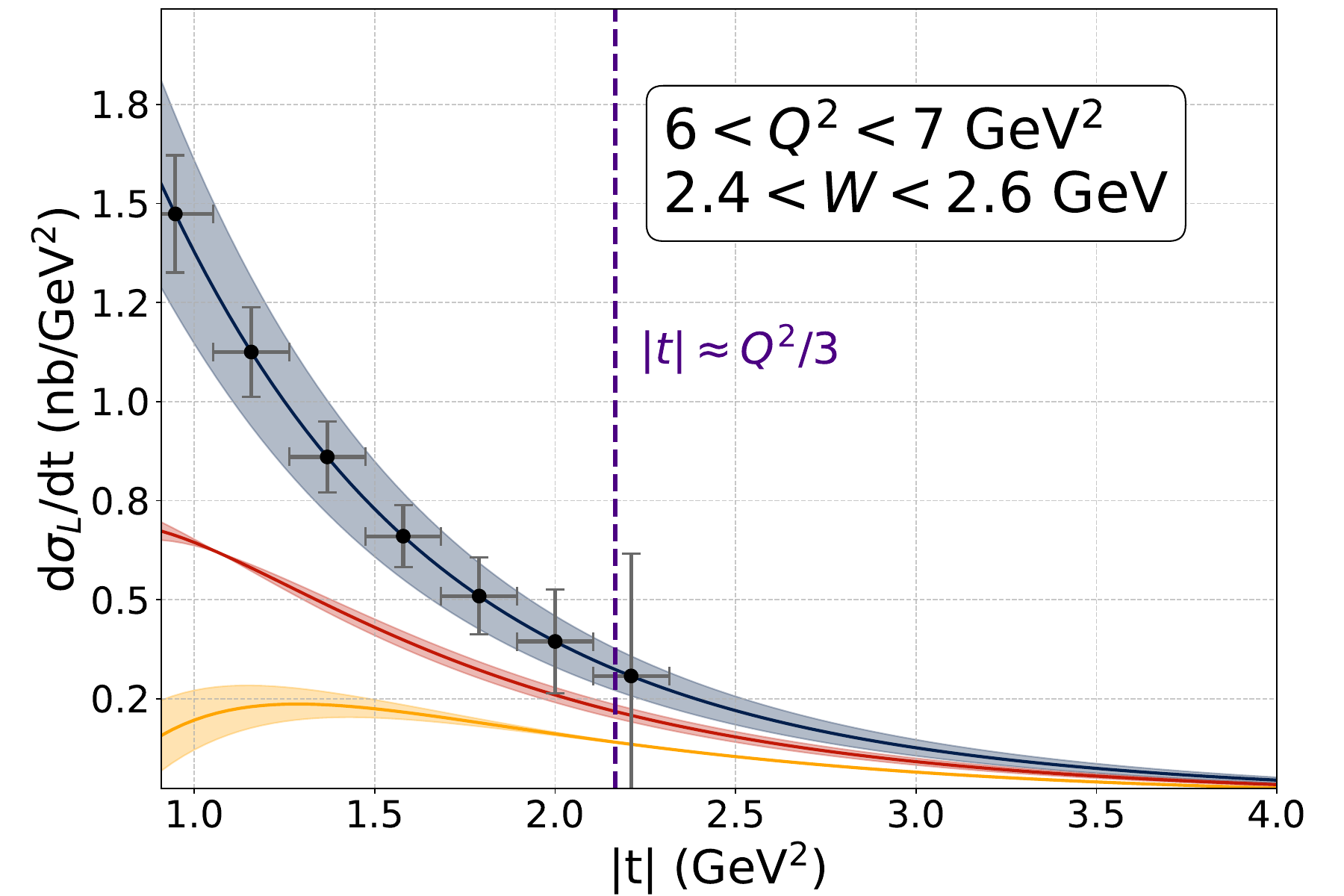}\\
      \includegraphics[width=\textwidth]{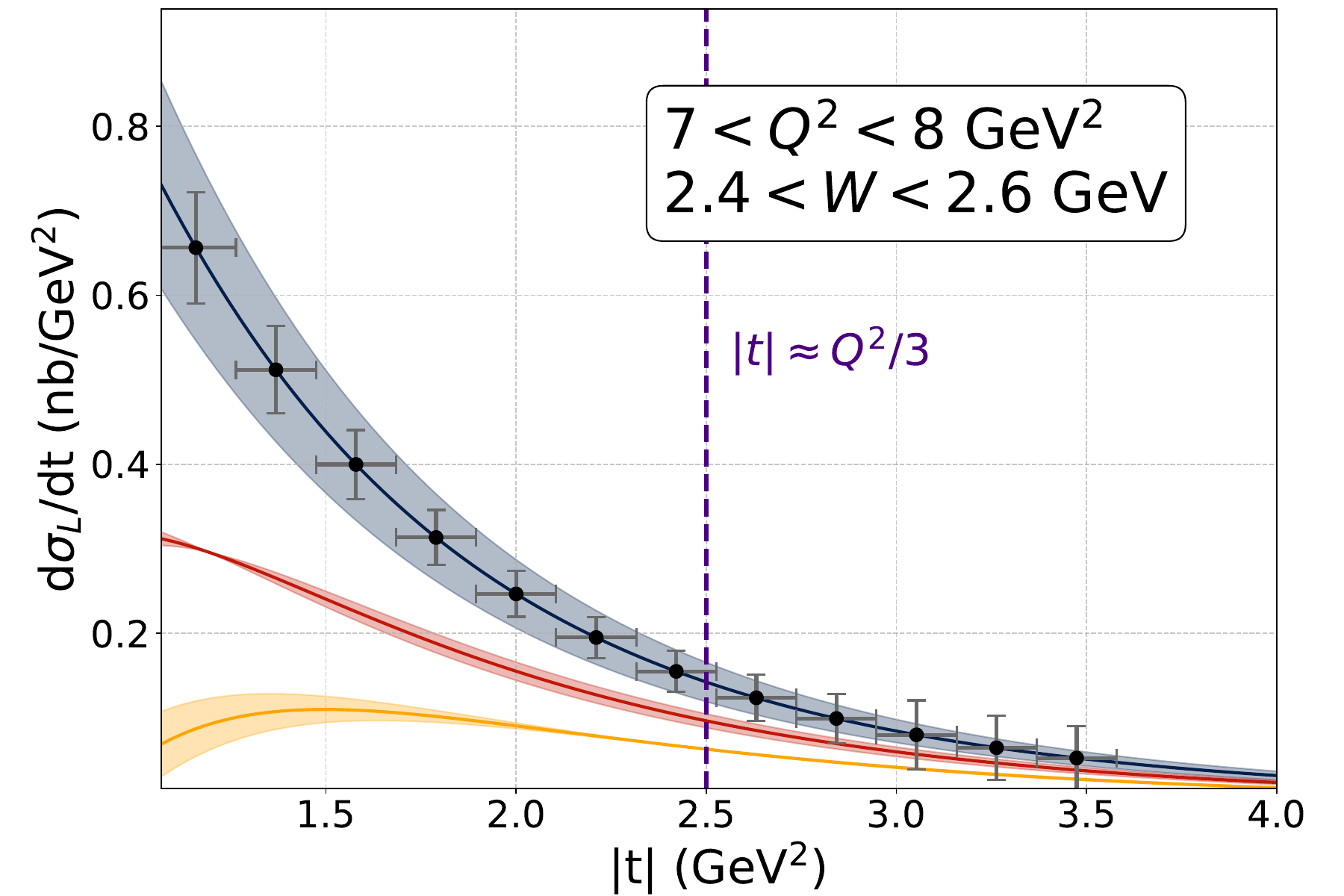}\\
      \includegraphics[width=\textwidth]{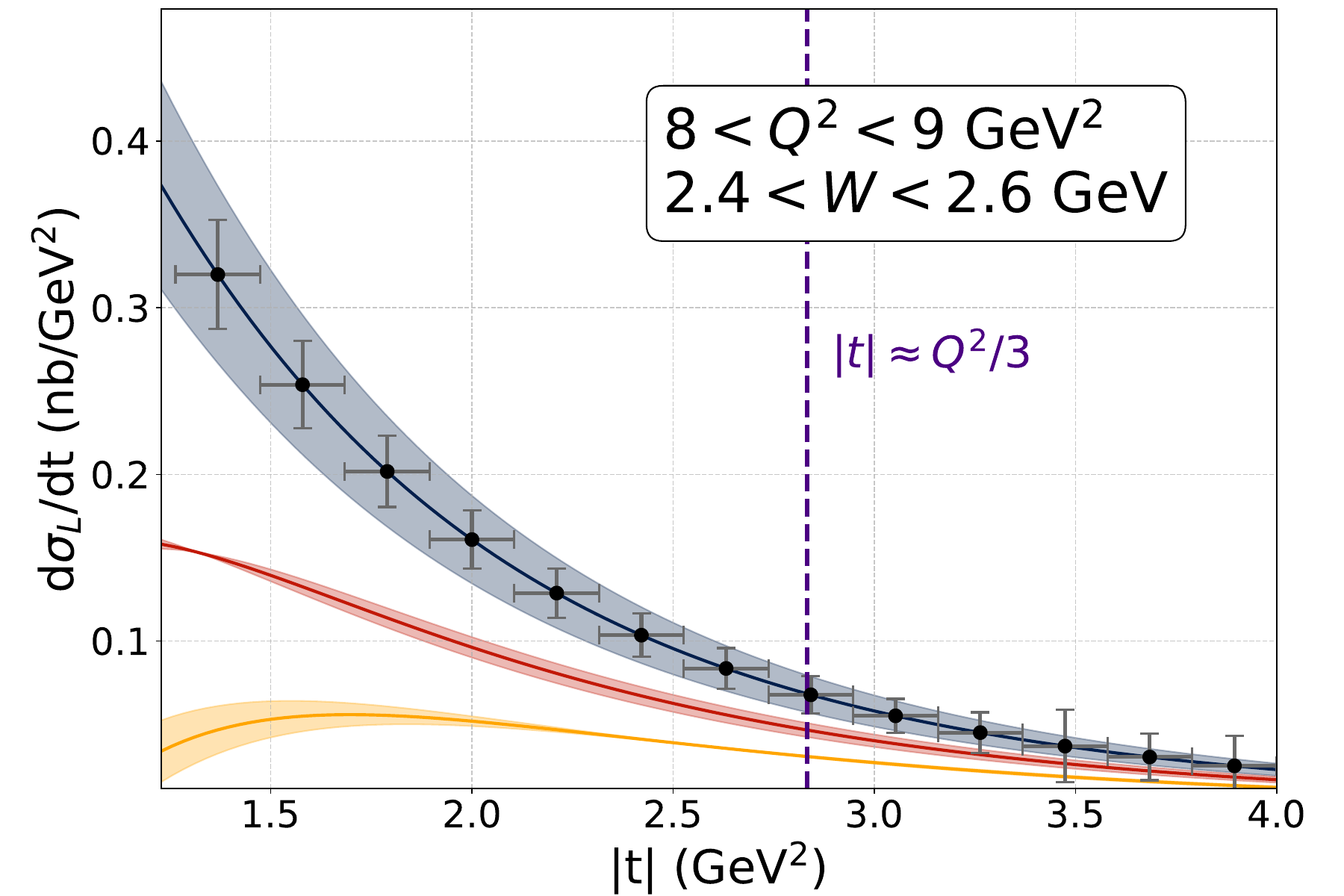}
      \par\phantomsection\label{fig:w25}
    \end{minipage}
    \hfill
    \begin{minipage}[t]{0.23\textwidth}
      \includegraphics[width=\textwidth]{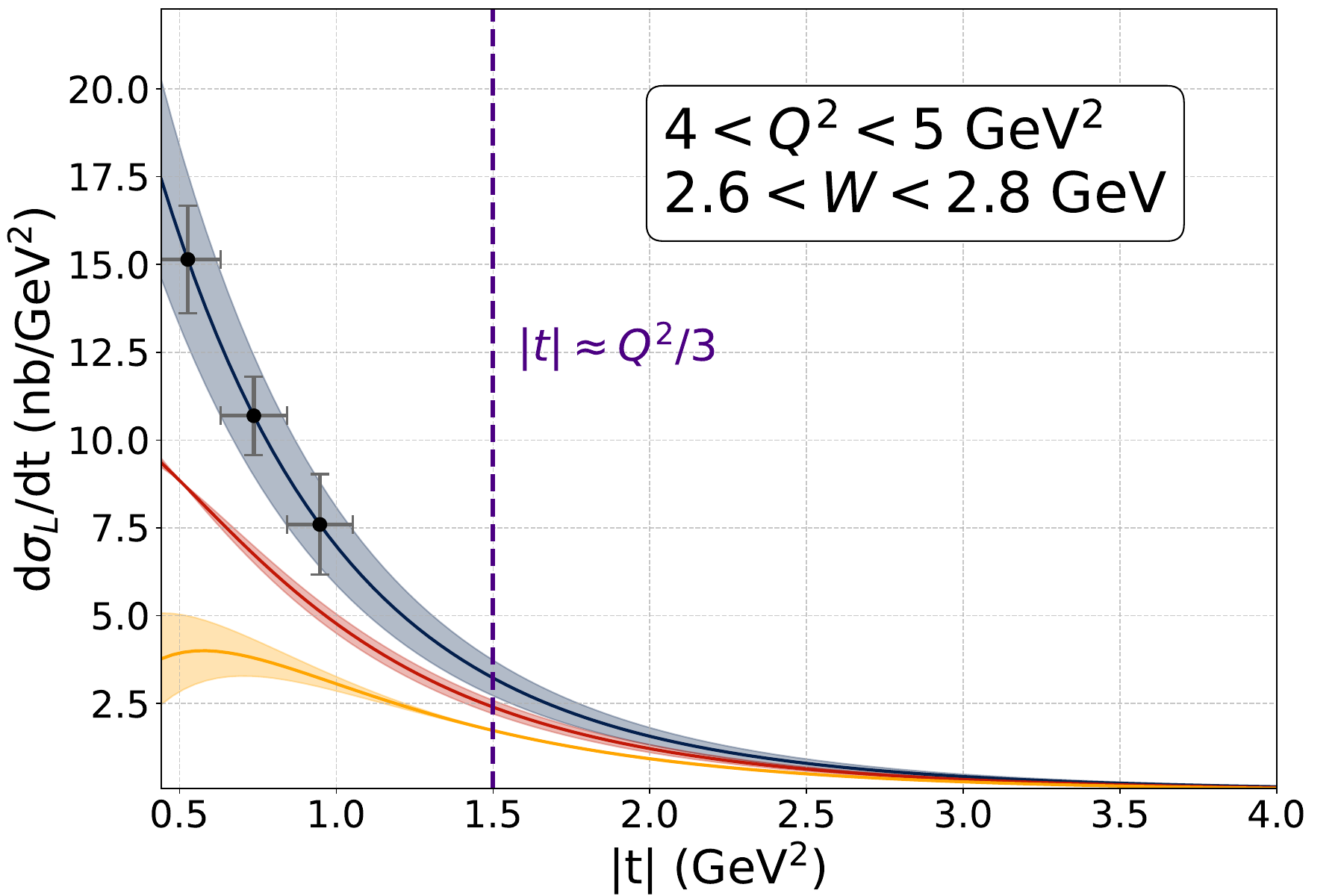}\\
      \includegraphics[width=\textwidth]{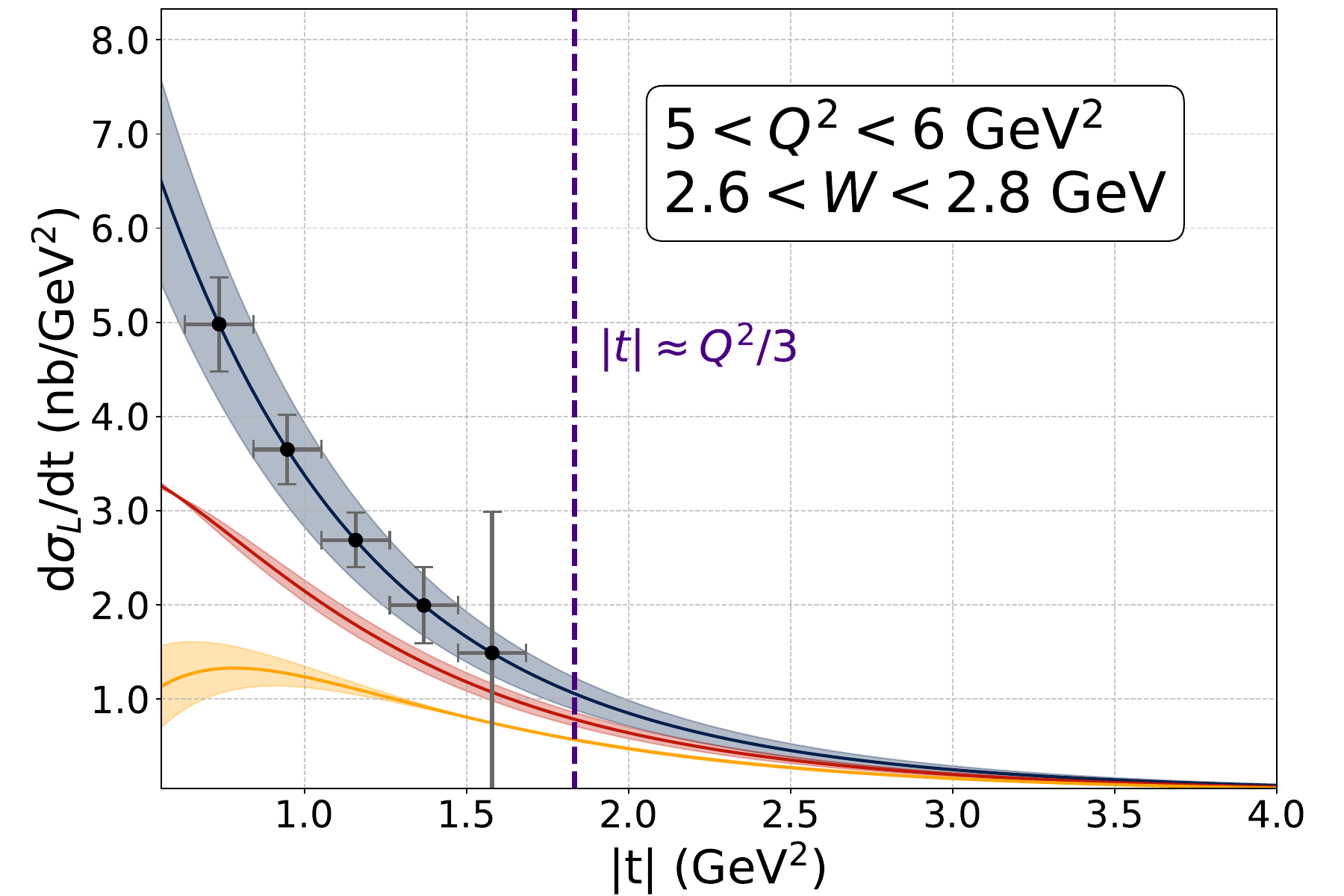}\\
      \includegraphics[width=\textwidth]{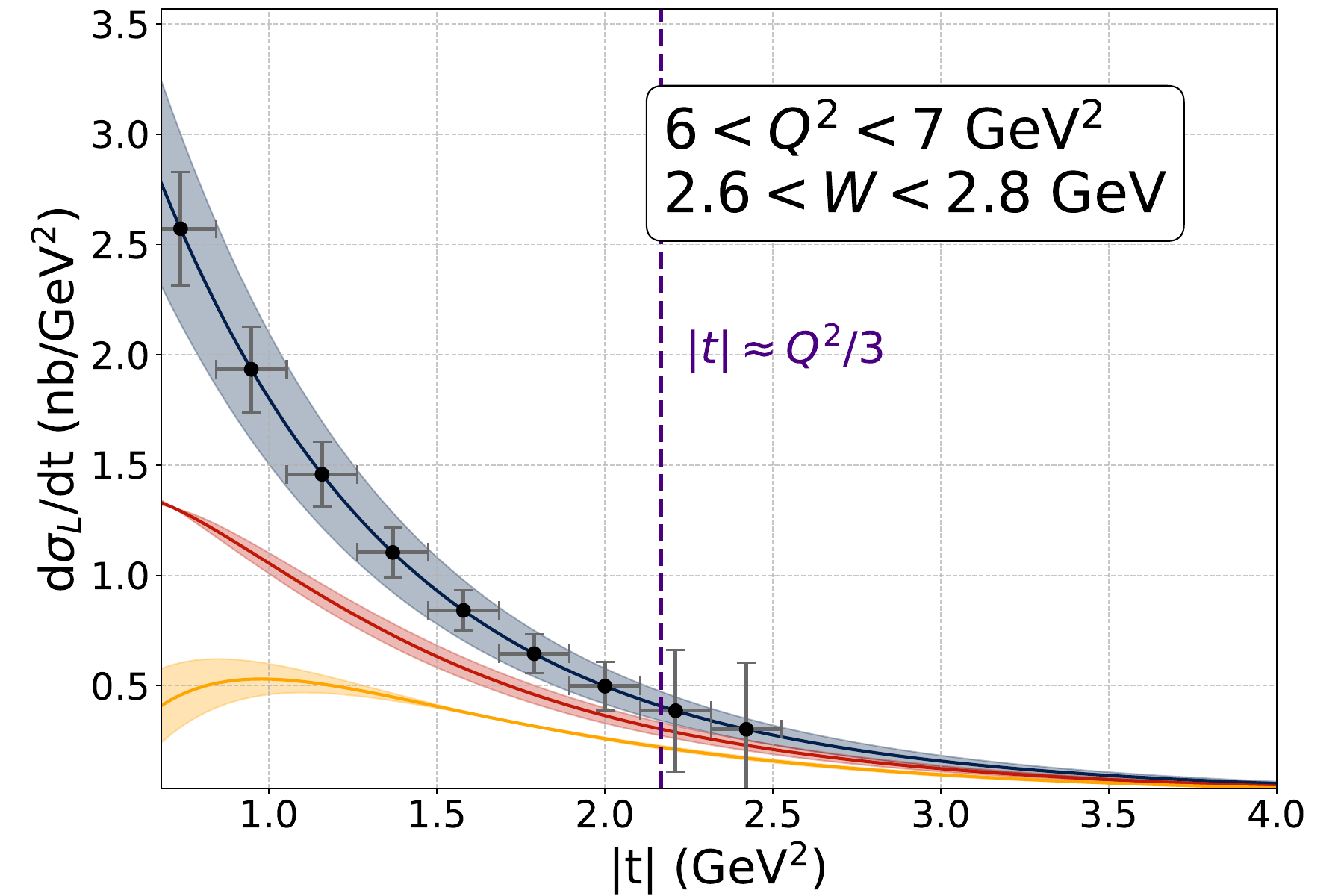}\\
      \includegraphics[width=\textwidth]{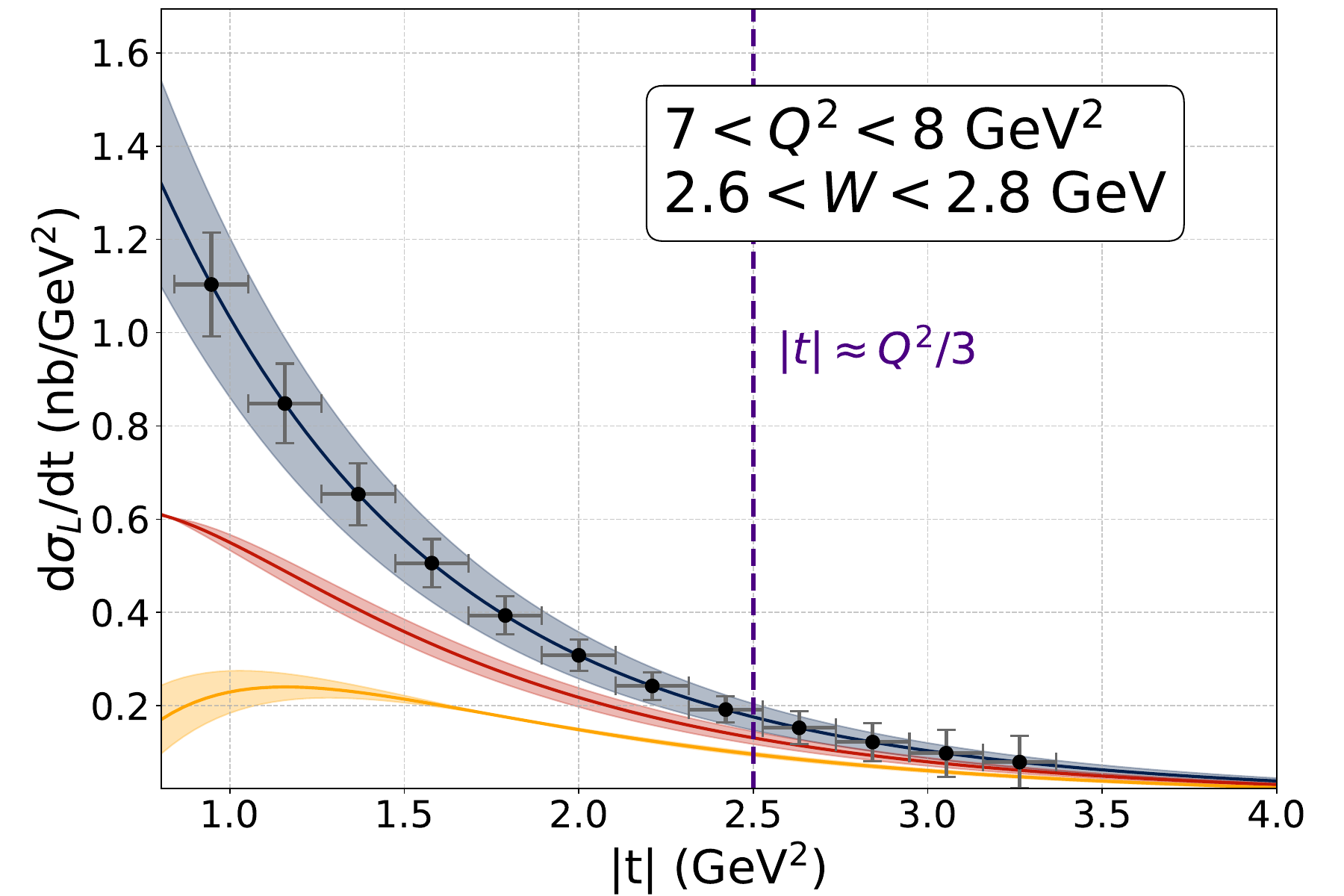}\\
      \includegraphics[width=\textwidth]{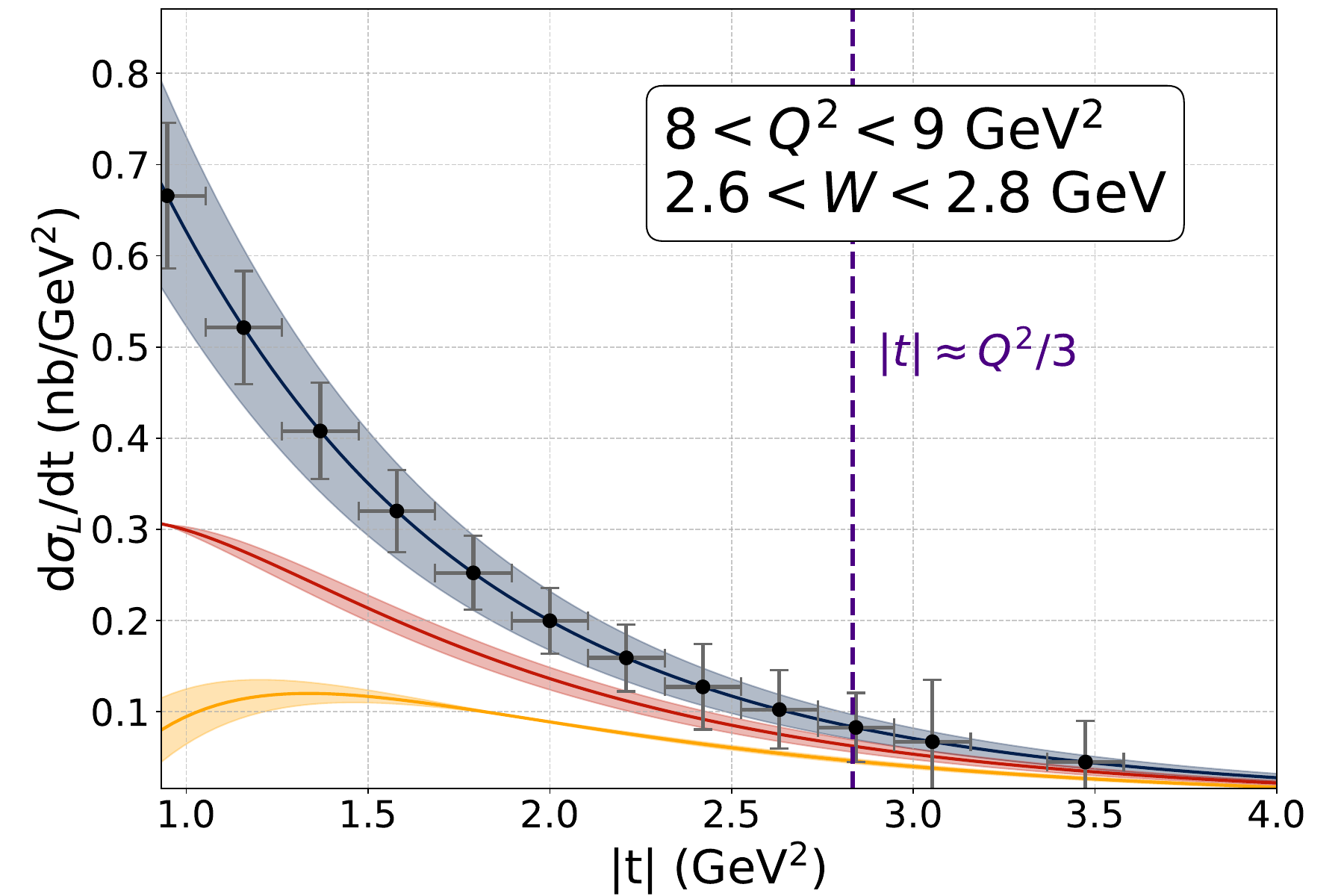}
      \par\phantomsection\label{fig:w27}
    \end{minipage}
    \hfill
    \begin{minipage}[t]{0.23\textwidth}
      \includegraphics[width=\textwidth]{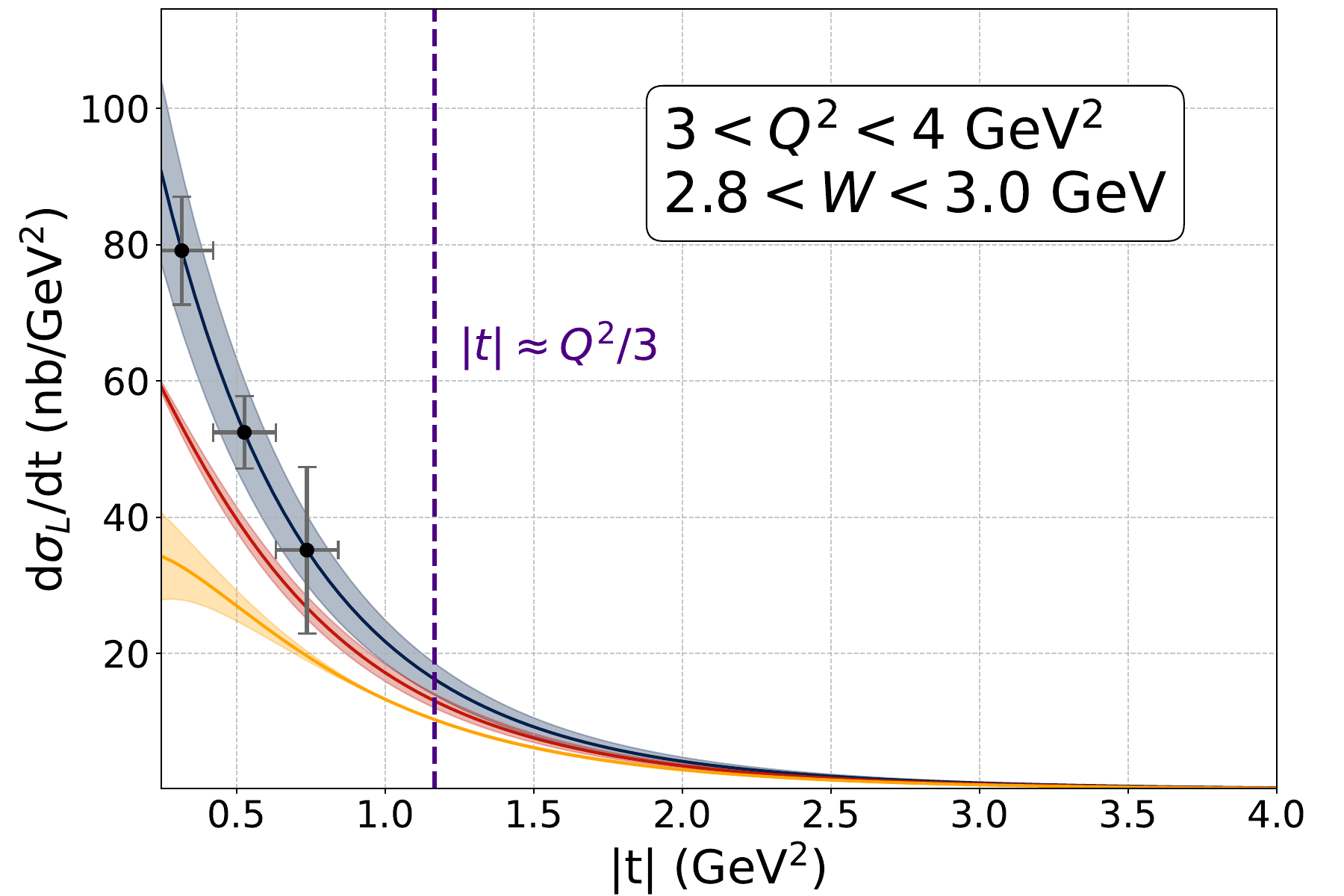}\\
      \includegraphics[width=\textwidth]{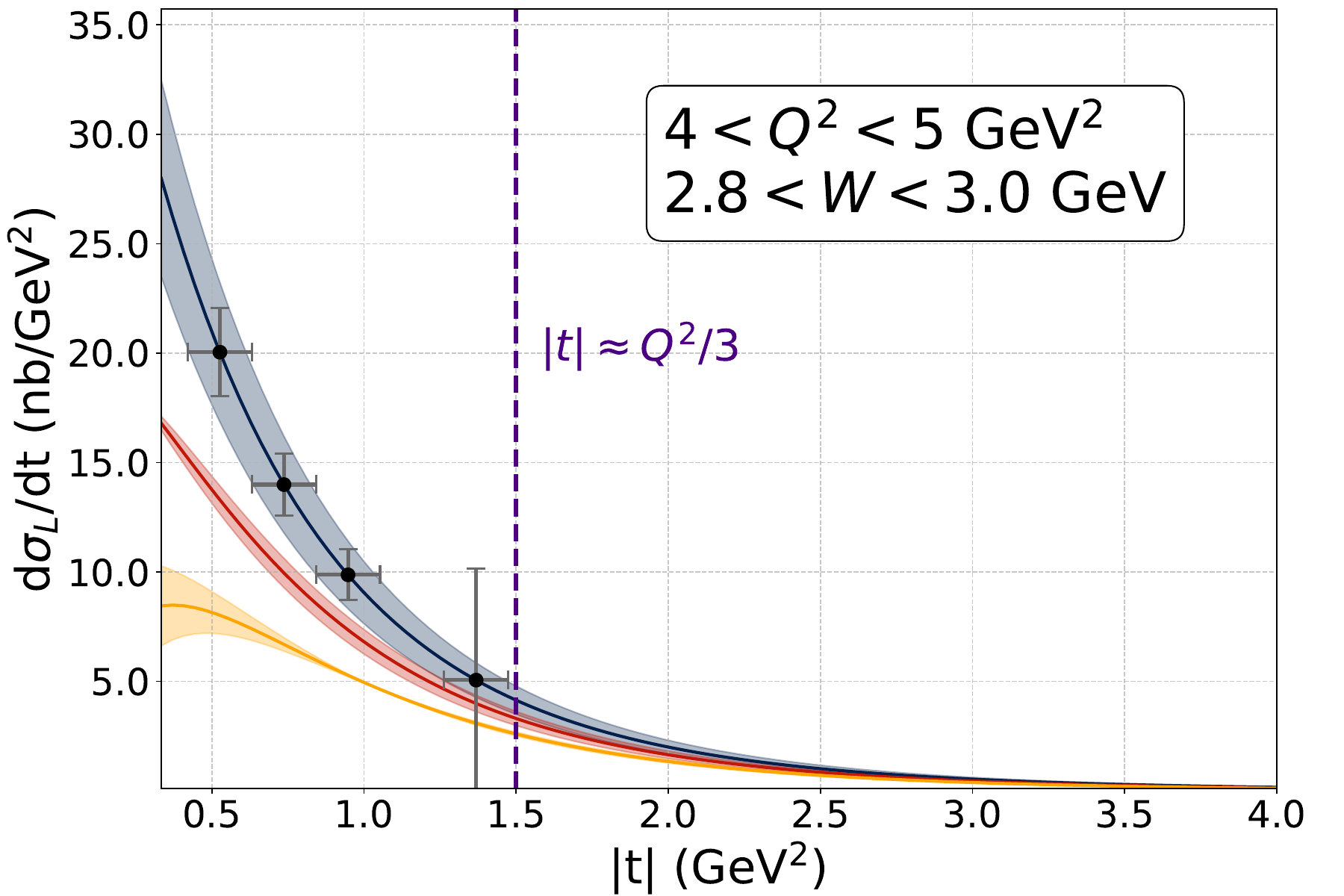}\\
      \includegraphics[width=\textwidth]{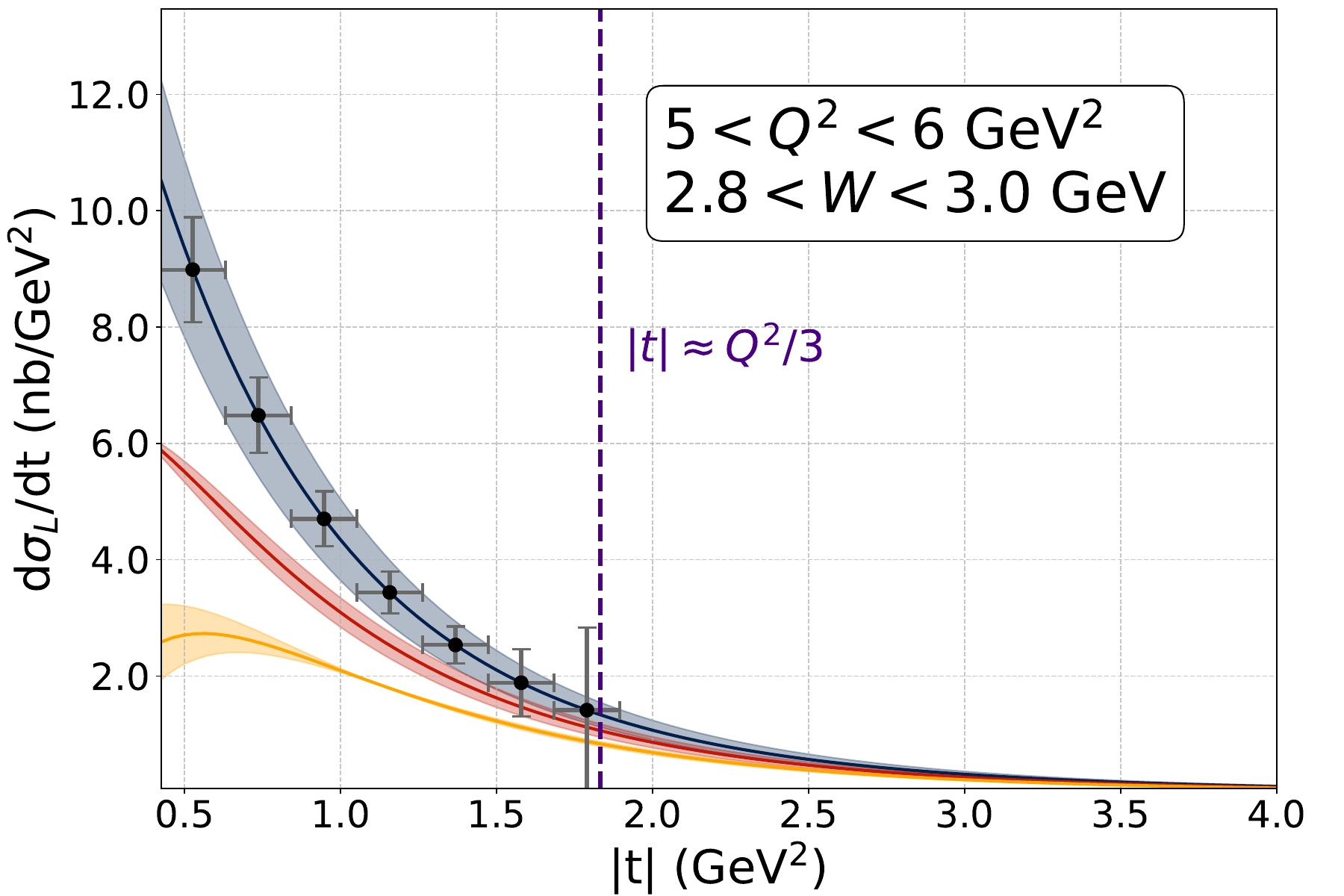}\\
      \includegraphics[width=\textwidth]{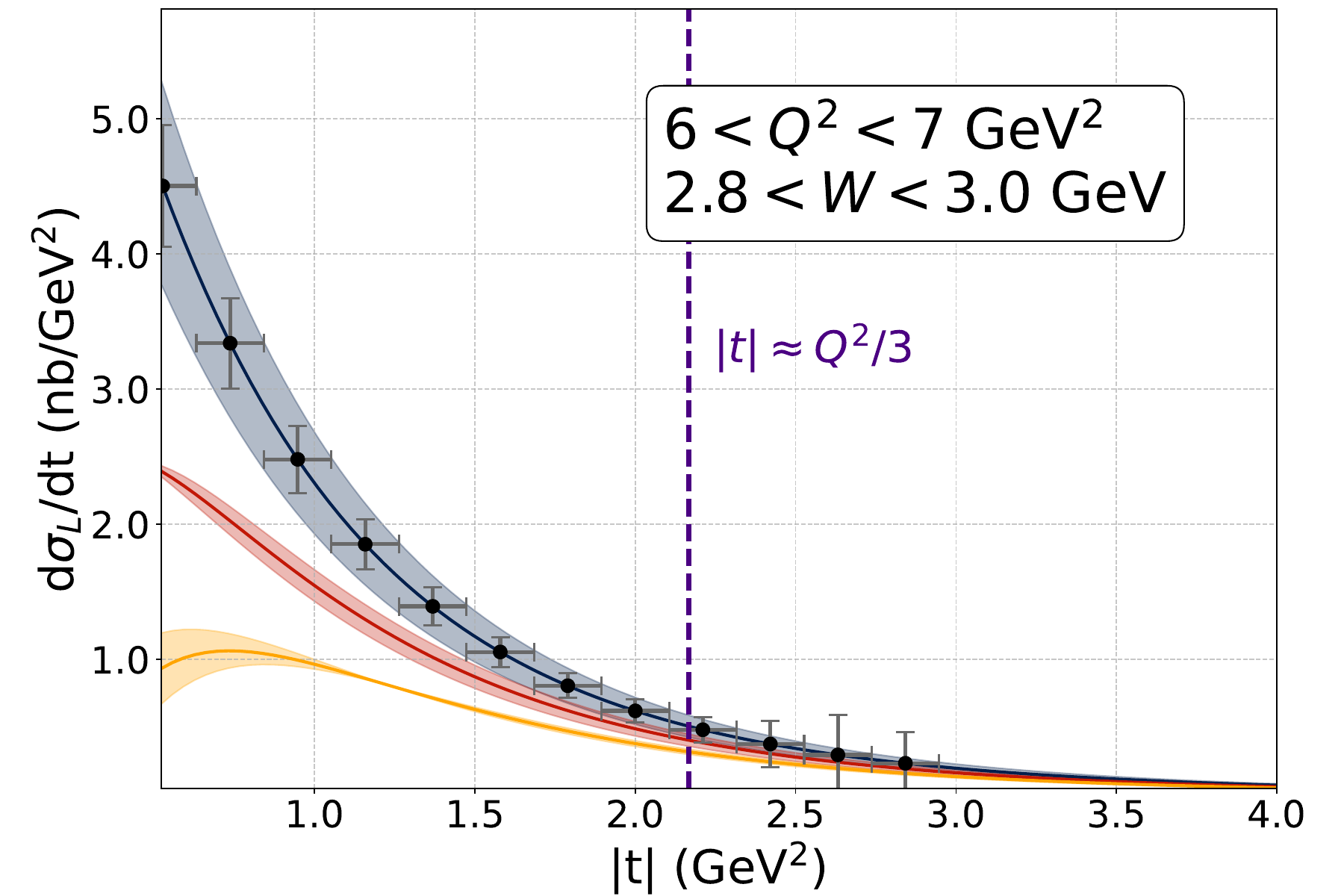}\\
      \includegraphics[width=\textwidth]{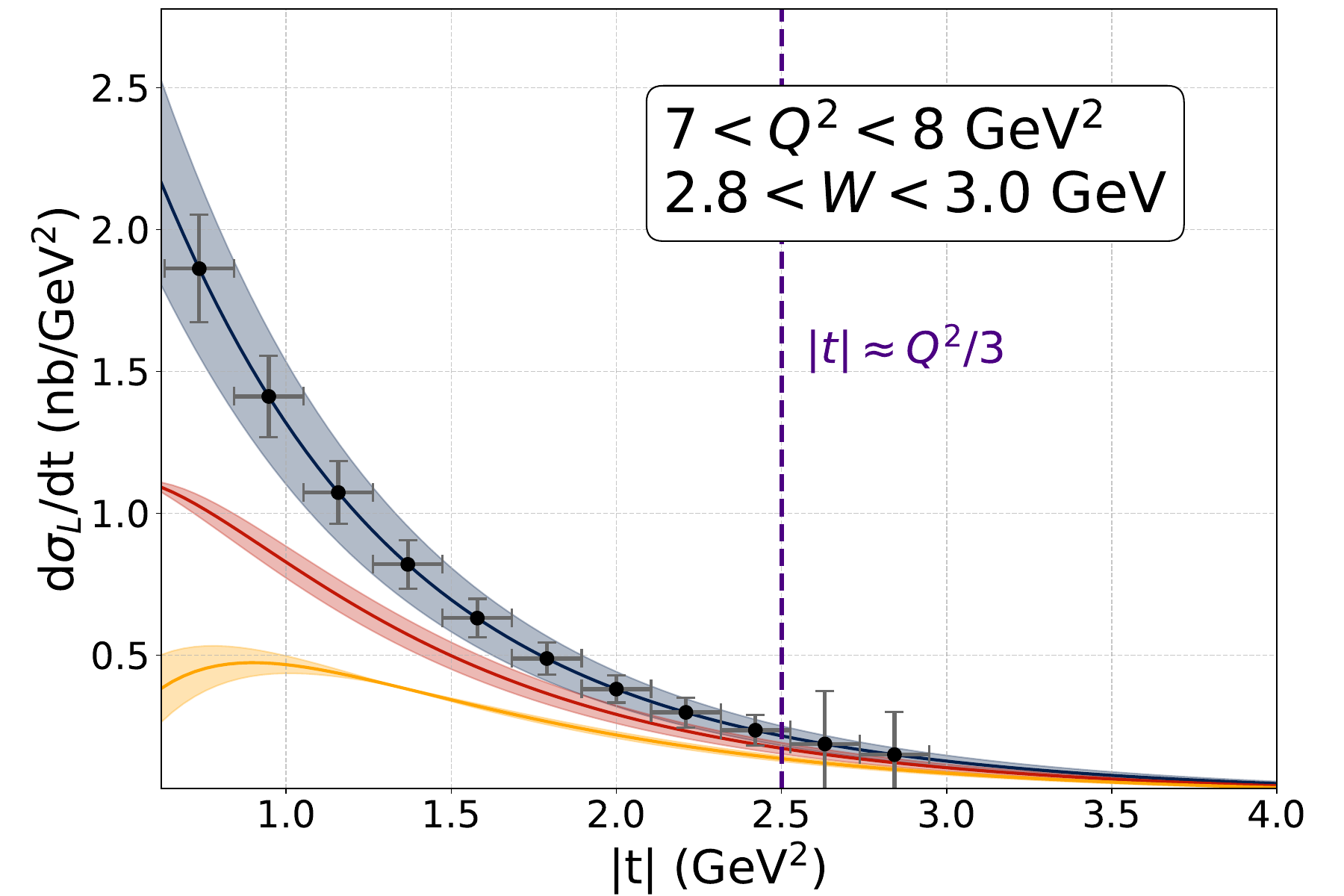}
      \par\phantomsection\label{fig:w29}
    \end{minipage}
    \hfill
    \begin{minipage}[t]{0.15\textwidth}
      \par\phantomsection\label{fig:wx}
    \end{minipage}

    \caption{%
      Projected results at SoLID for the $\gamma^*_{L}+p \rightarrow \phi+p$ cross section, 
      differential in $|t|$ and $Q^2$ in the near-threshold regime. 
      The assumed integrated luminosity is 43.2\,ab$^{-1}$. 
      The first (leftmost) column shows 4 $Q^2$ bins for $W=2.18$, while the subsequent 
      columns each represent a bin in $W$ with five $Q^2$ bins. 
      Note that the accessible range in $Q^2$ changes in each column due to the kinematic dependence on $W$. 
      The purple dashed line is drawn at $|t|\approx Q^2/3$, where 
      $Q^2$ is assumed to be the bin center. 
    }
    \label{fig:SoLIDResults}
\end{figure}

For comparison with the results for the EIC, in the range of $2.4 < W < 2.8$ GeV and $Q^2 > 3$ GeV$^2$, we project that SoLID will reconstruct around 270,000 events with all four final-state particles being detected. The SoLID acceptance in the near-threshold region ($W<2.8$ GeV) depends strongly on $Q^2$, ranging from almost 10\% for events with $7<Q^2<8$ GeV$^2$ to around 0.2\% for $3<Q^2<4$ GeV$^2$. Similarly to the EIC, the kinematic reach and event statistics can be improved beyond what is presented here if only three of the four final-state particles are required to be detected. Another possible way to improve the statistics is to reconstruct the final state $e',p,$ and $K_s^0$ from the $\phi\rightarrow K_s^0K^0_L$ decay and infer the $K^0_L$ from missing mass. As in Fig.~\ref{fig:EICResults}, the theory curves are generated at the bin centers in $Q^2$ and $W$. Due to the form factor of (\ref{FormFactor}), the highest statistics are available at low values of $|t|$. The region $|t|\lesssim Q^2/3$ is observed to be well-populated with statistically precise data points. The acceptance of SoLID increases at higher $Q^2$, due to the increased likelihood that the scattered proton and the $\phi$ decay products have enough transverse momentum to enter the FAD. This effect produces the unintuitive improvement in the number of measurable data points and uncertainties at higher $Q^2$, in spite of the fact that the cross section falls rapidly as a function of $Q$. As can be seen from Fig.~\ref{fig:SoLIDResults} and Fig.~\ref{2}, $\xi>0.4$ for values of $W\lesssim2.6$ at $Q^2=6$ GeV$^2$. Additionally, the range of $W$ for which $\xi>0.4$ grows at higher $Q^2$. Therefore, the region $|t| \lesssim \frac{Q^2}{3}$ and $\xi>0.4$ should be experimentally accessible with a reasonable degree of precision at SoLID. 

Based on these projections, SoLID provides an excellent opportunity to measure near-threshold $\phi$ electroproduction in the region of validity of the GPD factorization discussed in Sec.~\ref{Sec:GPD} due to the combination of high luminosity and acceptance. Together with the precise results on $D_g$ expected from the measurement of $J/\psi$ production in SoLID, a reasonably precise extraction of the strangeness $D$-term appears possible.

\subsection{Other experimental opportunities}

Since one of the primary challenges for measuring near-threshold $\phi$ at the EIC is that the $\phi$ decay products are highly boosted in the forward direction, a lower proton beam energy would likely be advantageous. The Electron-ion collider in China (EicC) plans to operate with beam energy configurations of 5x26, 3.5x20, and 3.5x16 GeV~\cite{Anderle:2021wcy}. Furthermore, the instantaneous luminosity of the EicC at the 3.5x20 energy setting is projected to be 2$\cdot10^{33}$cm$^{-2}$/s, which is 4.5 times higher than the EIC at 5x41. Therefore, it is reasonable to expect that the EicC, with a suitably designed detector, could perform a strong measurement of near-threshold $\phi$ electroproduction and complement measurements at Jefferson Lab and the EIC.

In Hall B at JLab, Run Group A of the CLAS12 experiment is expected to collect a total of approximately 760 fb$^{-1}$ of data on a liquid hydrogen target with a 10.6 GeV electron beam. Projections on exclusive $\phi$ production in CLAS12 are presented in the proposal to PAC 39~\cite{JeffersonLabPR12-12-007}, although the focus was on larger values of $W$ where the cross section is larger. Nevertheless, CLAS12 certainly provides a promising system with which to study $\phi$ production near-threshold, and an analysis of $\phi$ electroproduction is underway. 

A letter-of-intent was submitted to Jefferson Lab PAC52 in Spring 2024 for an experiment to study electroproduction of $\phi$ using the High Momentum Spectrometer (HMS) and Super High Momentum Spectrometer (SHMS)  in Hall C~\cite{Klest:2025rek}. In this experiment, only the scattered electron and proton would be detected and the $\phi$ would be reconstructed in the missing mass spectrum. Using this technique, a reasonable precision could be achieved on the $\gamma^*+p \rightarrow \phi+p$ cross section as a function of $|t|$ in a fixed bin of $Q^2$ and $W$. For comparison of that result with the predictions formulated in the previous sections for the $\gamma^*_{L}+p \rightarrow \phi+p$ cross section, a model based on the existing world data or theoretical considerations would be applied to estimate the value of $R$ at the measured kinematic point. Another option, albeit one likely to require an extended beamtime compared to the original letter-of-intent, is to exploit the excellent resolution of the HMS and SHMS to Rosenbluth separate the cross section.

\section{Conclusion}

In this work, we have presented a new perturbative QCD analysis of near-threshold $\phi$-meson electroproduction in $ep$ scattering. The biggest advantage of being near the threshold is that the skewness variable $\xi$ is order unity. More specifically, we have identified the range  $0.4<\xi<0.6$ as the practical region of interest.  The DVMP  amplitudes are then dominated by the gravitational form factors \cite{Hatta:2021can,Guo:2021ibg}. This can be seen most dramatically  in the conformal partial wave expansion 
where keeping only the $j=1$ moment provides a very good approximation to the entire amplitude \cite{Guo:2023qgu}.  We have demonstrated, for the first time, that this `threshold  approximation' works also at 
next-to-leading order 
in perturbation theory and even in the quark channel. Using the GK model as an example, we estimate that the truncation error can be as good as 5\% (10\%) at the amplitude (cross section) level in the kinematics considered in this paper. This is expected to be smaller than other sources of errors such as higher twist corrections. In principle, the approximation should work even better at higher $Q^2$, although measurements then become challenging due to the fast falloff of the cross section with increasing $Q^2$.

We find that the differential cross section $d\sigma/dt$ depends rather sensitively on the strangeness and gluon D-terms $D_s$, and $D_g$, a finding that consolidates and extends the previous   argument of Ref.~\cite{Hatta:2021can}. Assuming that the longitudinal/transverse separation is possible, we have also carried out realistic event generator  simulations at SoLID and EIC. The results demonstrate that the gravitational form factors likely remain measurable even after taking into account theoretical and experimental uncertainties. In the future, data on $\phi$ production should be combined with those from  $J/\psi$ and $\Upsilon$ photo- or electroproduction  and other processes (see, e.g., \cite{Burkert:2018bqq,Kumericki:2019ddg,Hatta:2023fqc,Hagiwara:2024wqz,Dutrieux:2024bgc}) in a global analysis framework.  On the theoretical side, the threshold approximation can be improved by including various corrections, such as the imaginary part of the amplitude and  higher (or all) moments of the meson distribution amplitude. Furthermore, the contribution from the twist-four gluon condensate, or  equivalently, the $\bar{C}_{q,}$ term in (\ref{gff}) \cite{Hatta:2018sqd}, can also be included   \cite{Boussarie:2020vmu}. Such a global analysis, in conjunction with systematically improvable QCD theory and state-of-the-art experiments, appears to be a practical path toward the precise determination of the nucleon gravitational form factors.

\section*{Note added}

On the same day our paper was posted on arXiv, a preprint \cite{Guo:2025jiz} was released,  where the authors discuss the connection between $J/\psi$ photoproduction near the threshold and the gluon GFFs to next-to-leading order. More recently, the validity of the threshold approximation has been tested also for the pion target \cite{Hatta:2025ryj}. 

\section*{Acknowledgements}

This work was made possible by Institut Pascal at Université Paris-Saclay with the support of the program “Investissements d’avenir” ANR-11-IDEX-0003-01. We extend our gratitude to S. Joosten for helping to implement our new model into \texttt{lAger} and L. DeWitt for thorough editing. 
Y.~H. and J.~S. were supported by the U.S. Department of Energy under Contract No. DE-SC0012704, and also by LDRD funds from Brookhaven Science Associates.
H.~K. was supported by the U.S. Department of Energy under Contract No. DE-AC02-06CH11357.

\bibliography{ref}

\end{document}